# Broadband dielectric behavior of MIL-100 metal-organic framework as a function of structural amorphization

*Arun Singh Babal,[a] Barbara E. Souza,[a] Annika F. Möslein,[a] Mario Gutiérrez,[a] Mark D. Frogley,[b] and Jin-Chong Tan[a,*]*

[a]Multifunctional Materials and Composites (MMC) Laboratory, Department of Engineering Science, University of Oxford, Parks Road, Oxford, OX1 3PJ, United Kingdom

[b]Diamond Light Source, Harwell Campus, Chilton, Oxford, OX11 0DE, United Kingdom

**ABSTRACT:** The performance of modern electronics is associated with the multi-layered interconnects, encouraging the development of a low-*k* dielectrics. Herein, we studied the effects of phase transition from crystalline to amorphous on dielectric, optical and electrical properties of MIL-100 (Fe) and Basolite F300 metal-organic framework (MOF) obtained using different synthesis techniques in both the radio (4-1.5 MHz) and infrared (1.2-150 THz) frequency regimes, which are important for the microelectronics, infrared optical sensors, and high-frequency telecommunications. The impact of amorphization on broadband dielectric response was established based on: (1) The comparison in the dielectric characteristics of a commercially available amorphous Basolite F300 versus a mechanochemically synthesized crystalline MIL-100 in the MHz region. (2) By tracking the frequency shifts in the vibrational modes of MIL-100 structure in the far-IR (phonons) and mid-IR regions. We showed that various parameters such as pelleting pressure, temperature, frequency, density, and degree of amorphization greatly affect the dielectric properties of the framework. We also investigated the influence of temperature (20-100 °C) on the electric and dielectric response in the MHz region, crucial for all electronic devices.

## 1. INTRODUCTION

For 60 years, the development efforts in microelectronics to integrate new dielectric insulator materials in the interconnects of Very large-scale integration (VLSI) along with several billions of components and their interconnects in ever-shrinking chips have paved the way for continuous performance enhancement following the Moore's law.[1] The dimension shrinking increases the resistance and capacitance of the back end of line structures, which causes signal delay, power dissipation, and cross-talk between the devices and becomes a bottleneck for the next generation Ultra large-scale integration (ULSI) devices.[2] Due to low mechanical and chemical stability of both organic and hybrid dielectric materials, they have faced challenges to be integrated into ICs, and later porous organosilicate glasses (OSG) ($\varepsilon'$~3.0) were used as an alternative to the traditional $SiO_2$ materials ($\varepsilon'$~4.0).[3-4]

Metal-organic frameworks (MOFs) are a subset of nanoporous inorganic-organic materials and display the potential for yielding a good combination of chemical and mechanical stability required for interconnects. MOF exhibits a low dielectric constant value caused by its low density[5] and uniform pore distribution (< 2 nm), which can potentially be used between the intermetal spacing (~10 nm).[6] Due to the frequency dependence of dielectric constant, static (DC) values are considered as a good approximation in modern microprocessors, whose operating limit is of the order of $10^9$ Hz (GHz). For developing high-speed communication devices in IR and optical region, understanding of dielectric constant dependency on the atomic and ionic polarization, which are structural dependent, is essential. The effect of various key parameters such as material density, stress-dependent amorphization, operating temperature, etc., on new dielectric materials is also necessary for its implementation in the real-world scenario. Only a handful of studies are available on pressure-induced amorphization of MOF frameworks and its relation to the physical properties.[7-9]

MOF materials have started to gain attention from the scientific community as a promising low-*k* dielectric due to their tunable physical and chemical properties.[10-11] A complete dielectric spectrum over the frequency is required to exploit its full potential, while only a single type of polarization is studied mainly in radio frequency for most of the cases[12-13] and rarely in the higher frequency regions.[14-16] For instance, one study reported the dielectric property of ZIF-based MOF in the IR region[15] using synchrotron specular reflectance spectroscopy, whereas other studies focused on the dielectrics behavior in the kHz-MHz region.[17] Similarly, HKUST-1 films were studied in the optical frequency (visible region) for electronic polarization using the spectroscopic ellipsometry technique.[18] While other studies on HKUST-1 MOF have focused on the radio frequency region.[19] So a complete dielectric study of the system is a must to grasp the full understanding and to gauge the potential of the material as a dielectric interlayer for next-generation electronics, sensors, and telelcommunication devices.

Here, we have studied for the first time the electrical and broadband dielectric (Hz-MHz-THz) properties of both MIL-100 (Fe) and Basolite F300 system under varying temperatures and pelleting pressure. The key feature of this study is to establish the relation between the degree of structural amorphization and how it may affect the electric, dielectric and optical properties of the MOF framework. For this purpose, a comparative measurement was carried out on the MOF samples, which were obtained from two different synthetic routes. Furthermore, the MHz and THz dielectric data obtained from the LCR meter (inductance-capacitance-resistance) and synchrotron infrared specular reflectance spectroscopy, respectively, were combined to give us a deeper understanding of the dielectric and optical response of the MIL-100 (Fe) system, surpassing the static dielectric behavior commonly found in the literature.

## 2. EXPERIMENTAL SECTION

**2.1. Materials.** The Basolite F300, a form of Fe-BTC, along with iron (III) nitrate nonahydrate [$Fe(NO_3)_3 \cdot 9H_2O$], and benzene-1,3,5-tricarboxylic acid linker ($H_3BTC$) were purchased from Sigma Aldrich and used without further purification.

**2.2. Mechanochemistry of MIL-100 (Fe).** 3 mmol of iron (III) nitrate (1.212 g) and 2 mmol of $H_3BTC$ (0.420 g) were ground in the agate mortar for 10 min. The resulting material was heated in the oven for 4 h at 160 °C to complete the annealing process. The product was then washed by centrifugation (8000 rpm for 10 min) with methanol and deionized water to remove the unreacted molecules. It was first activated in the vacuum oven at 150 °C for 12 h and then immersed in deionized water and kept under stirring for the next 12 days at room temperature to perform the framework reconstruction. The obtained MIL-100 (Fe) powder is termed MIL-100-MG, where MG stands for manual grinding. Finally, the sample was recovered and reactivated. The synthesis method of MIL-100-MG was adapted from reference [20].

**2.3. Pellet preparation.** A manual hydraulic press (Specac 15 tons capacity) was used to prepare pellets (150 mg each) from the MIL-100 powder under different mechanical loads (0.5, 1, 2, 3, 5, 7 and 10-tons force). The pellets obtained from commercially available Basolite F300 powder are designated as Basolite F300-*N*-ton, whereas MIL-100-MG-*N*-ton for pellets using the MIL-100 powder obtained from the mechanochemistry technique, where *N* is the applied mechanical load during pelleting.

**2.4. Fourier-transform infrared (FTIR) spectroscopy.** The mid-IR modes of the MOF pellets were characterized at 0.5 $cm^{-1}$ spectral resolution using the Nicolet-iS10 FTIR spectrometer equipped with attenuated total reflection (ATR) accessory. The nano-FTIR spectra were acquired using a neaSNOM instrument (Neaspec GmbH), where an AFM tip in tapping-mode is illuminated by a broadband mid-infrared laser source (Toptica). Recently, the nano-FTIR technique has been successfully employed to characterize the local IR spectra of MOF nanocrystals.[21] To probe the near-field signal without the interference of background contributions, the detector signal is demodulated at higher harmonics of the oscillation frequency of the AFM tip ($\Omega$ = 253 kHz). Each spectrum is an average of more than five individual point spectra probed with a spot size of 20 nm. These were acquired by Fourier transform spectroscopy averaging over 13 individual interferograms with 1024 pixels and an integration time of 11 ms per pixel, normalized by a reference spectrum measured on the silicon substrate.

**2.5. Thermogravimetric analyses (TGA).** The thermogravimetric analysis (TGA) was carried out to examine the thermal stability of the MIL-100 powder and pellets using the TGA-Q50 (TA Instruments) equipped with an induction heater (1000 °C max temperature) under an $N_2$ inert atmosphere. The samples were heated in a platinum sample holder at a rate of 10 °C/min from 40 to 500 °C.

**2.6. X-ray diffraction (XRD).** The pressure-dependent amorphization effect in both Basolite F300 and MIL-100 system was studied using the XRD patterns collected using the Rigaku Miniflex bench-top X-ray diffractometer, at a scan rate of 0.2°/min with a step size of 0.05°.

**2.7. Using the LCR meter to measure Hz-MHz range.** The dielectric response of pellets in the radio frequency range (4 Hz-1.5 MHz) was investigated using the IM3536 LCR meter, which is based on the parallel-plate capacitor principle. The silver coating was applied on both sides of the pellet to simulate the parallel plate electrodes. The dielectric sample holder was placed in the high vacuum oven (pressure = $10^{-3}$ torr) and calibrated for the pellet thickness. Samples were evacuated for 16 h at 20 °C before starting the measurements. The temperature-dependent dielectric response was noted at a step size of 10 °C. The real ($\varepsilon'$) and imaginary ($\varepsilon''$) parts of the dielectric constant as a function of frequency ($\omega$) were calculated from the following standard equations:

$$\varepsilon'(\omega) = \frac{C(\omega)d}{\varepsilon_0 A} \quad \text{............(1)}$$

$$\varepsilon''(\omega) = \varepsilon'(\omega)\tan\delta \quad \text{............(2)}$$

where $C$ is the capacitance, $d$ is the distance between the pair of parallel plate electrodes, $A$ is the area of the electrode, $\varepsilon_0$ is vacuum permittivity and $\tan\delta$ is the loss tangent.

**2.8. Using synchrotron infrared (IR) specular reflectance spectroscopy to measure the far-IR (THz), mid-IR and near-IR frequencies.** The dielectric response of pellets in the IR region (1.2-150 THz) was obtained by transforming the reflectivity spectrum, collected from the Bruker Vertex 80V FTIR interferometer equipped with the Pike Technologies VeeMAX II variable angle specular reflectance accessory at Beamline B22 MIRIAM in the Diamond Light Source (Harwell, UK).

Two beam splitters were used in the IR region - (i) 6-μm thick Mylar broadband multi-layer coated beam splitter for far-IR and (ii) KBr beam splitter for mid-IR. The background and pellet spectra were collected for both regions at an angle of 30° from the normal axis of the pellet surface at 2 $cm^{-1}$ resolution and 512 scans per spectral scan, which was kept constant for all the pellets for comparison. The dielectric and refractive indices were evaluated by applying the Kramers-Kronig Transformation (KKT)[22] on the reflectivity data R($\omega$) using the following relations:

$$\phi(\omega_a) = \frac{2\omega_a}{\pi}\int_0^{\infty}\frac{\log(\sqrt{R(\omega)})}{\omega^2-\omega_a^2}d\omega \quad \ldots\ldots\ldots\ldots(3)$$

$$n(\omega) = \frac{1-R(\omega)}{1+R(\omega)-2\sqrt{R}\cos(\phi(\omega))} \quad \ldots\ldots\ldots\ldots(4)$$

$$\kappa(\omega) = \frac{-2\sqrt{R}\sin(\phi(\omega))}{1+R(\omega)-2\sqrt{R}\cos(\phi(\omega))} \quad \ldots\ldots\ldots\ldots(5)$$

where $\phi$ is the phase angle, $\omega_a$ is the arbitrary wavenumber, $n$ is the real part of refractive index and $\kappa$ is the imaginary part of the refractive index.

$$\tilde{\varepsilon}(\omega) = \tilde{n}(\omega)^2 \ldots\ldots\ldots\ldots(6)$$

$$\varepsilon'(\omega) = n(\omega)^2 - \kappa(\omega)^2 \ldots\ldots\ldots\ldots(7)$$

$$\varepsilon''(\omega) = 2n(\omega)\kappa(\omega) \ldots\ldots\ldots\ldots(8)$$

where $\varepsilon'$ and $\varepsilon''$ are the real and imaginary parts of the dielectric constant, respectively.

## 3. RESULTS AND DISCUSSION

In this study, a series of 13-mm diameter disk pellets were prepared using the manual hydraulic press by applying 0.5, 1, 2, 3, 5, 7 to 10 tons of force (see Figure 1(a), which summarizes the MIL-100 powder synthesis and the pellet preparation). The nominal density curve was plotted for both types of pellet: commercially available (Basolite F300) and in-house synthesized MIL-100 (mechanochemistry technique, termed as MIL-100-MG), which were used in dielectric experiment, confirm that similar density pellets were used to compare and obtain an understanding of the effect of amorphization on material dielectrics (see Figure 1(b)). In contrast to the Basolite F300 pellets, the MIL-100-MG pellets are slightly denser due to the presence of residual linker molecules in the framework cavity, which is a common occurrence in MOFs synthesized using the mechanochemistry technique.[20] The estimated yield strength of the compressed powder: Basolite F300 and MIL-100-MG were 310 MPa and 366 MPa, respectively, indicating that confinement of guests in the framework cavity increases its mechanical stability under compression (see inset of Figure 1(b)). A systematic shift in the color of the pellets from light brown to dark brown was also observed, which hints toward the increase in the refractive index with pellet density.

The MIL-100 structure has a large pore (~3 nm mesocage), thus it is highly susceptible to mechanical pressure-induced amorphization. Herein, we track the structural evolution upon pelleting using the powder X-ray diffraction (PXRD). The XRD spectra of Basolite F300 powder show a significant broadness in its characteristic Bragg peaks followed by diminishing of the long-range periodicity over increased pelleting pressure (see Figure S1(a)), whereas MIL-100-MG powder demonstrates high crystallinity; and with increased pelleting pressure an amorphous hump arises at $2\theta = 8°$ arises starting from 37 MPa (0.5t) (see Figure S1(b)). The XRD spectra of MIL-100-MG pellets prepared at higher pelleting pressure (>221.76 MPa) is mainly dominated by the amorphous peak, which is obvious from the Figure S1 (c). Both the broadness in characteristic peaks and the appearance of an amorphous hump in the spectra can be attributed to the complete pore collapse and structural amorphization in the MIL-100 framework. The FTIR spectrum of the Basolite F300 system also shows broadness in its characteristic peaks and a frequency-dependent redshift under compression, which follows the density pattern (see Figure S2). As the pelleting pressure was increased to 10t, the nano-FTIR spectra in Figure 1(c-d) revealed that the main IR absorption bands (except carboxylate stretch) remained unchanged with some broadening followed by a redshift in the peak positions. In contrast, the carboxylate asymmetric stretch at 1578 $cm^{-1}$ showed significant broadening and the appearance of a shoulder band at ~1535 $cm^{-1}$, which is the characteristic asymmetric stretch of monodentate coordinated carboxylates indicating the broken bonds of $Fe\text{-}O_{COO}$.[7] The result supports the notion that bond breakage is responsible for amorphization of MIL-100 subject to mechanical stress. The Basolite F300 also shows higher thermal stability of up to 300 °C, which increases slightly with pelleting pressure, even after amorphization (see Figure S3).

Figure 2 contrasts the amorphization effect on the dielectric properties in both types of MOF pellets as a function of pelleting pressure and temperature at 0.01, 0.1, and 1 MHz frequency. For both Basolite F300 and MIL-100-MG pellets, the real part of dielectric constant ($\varepsilon'$) shows a systematic increase under compression. Interestingly, for individual pressure pellet, the $\varepsilon'$ value decreases as a function of frequency, which can be associated with the decline in the orientational polarizability (inherent dipole moment inertia cannot keep up with the oscillating field) caused by the higher alternating field.

In Figure 2(a), the $\varepsilon'$ shows a sudden surge from 1t pellet to the 2t pellet of Basolite F300 sample, which can be associated with the loss of the long-range periodicity confirmed by XRD. Similarly, in Figure 2(b) of MIL-100-MG pellets, the $\varepsilon'$ value shows a step increase from 1-2, 2-3, and 3-5t pellets, which follows the XRD trend revealing the amorphization of the open framework. The rising temperature also assists in better alignment of the dipole moment of the amorphized structures, resulting in a higher $\varepsilon'$ value in both pellet types. Additionally, the MIL-100-MG pellets show a higher dielectric value over its counterpart Basolite F300 pellets due to pore blockage (resulted from the mechanochemical grinding method).

For all the pressure pellets, the increase in the imaginary part of dielectric constant ($\varepsilon''$) with frequency is associated with the higher energy dissipation resulted from the dipole motion. As evident from Figure 2(c) and (d), MIL-100-MG pellets showed slightly higher energy dissipation over its counterpart Basolite F300 pellets, resulting from the pore blockage and higher degree of amorphization (Figure S1). Figure 2(e) summarizes the $\varepsilon'$ values for both Basolite F300 and MIL-100-MG pellets. The complete datasets of the temperature- and frequency-dependent MHz dielectrics for the individual pellets of Basolite F300 and MIL-100-MG MOFs are presented in the SI, see Figures S4-S10.

The MIL-100-MG pellets were chosen to study the optical properties, due to their phase transition from crystalline to amorphous phase under compression, whereas the as-received Basolite F300 already had some amorphous phase associated with it (see Figure S1). The dielectric and refractive properties in the broadband infrared (IR) frequencies (40-5000 $cm^{-1}$) were evaluated by implementing KKT[22] on the high-resolution reflectance spectra (Figure S11). The optical properties of the framework in terms of refractive index mainly depend upon its optical density, its surrounding, temperature, and the frequency of the incidental light. The components of complex refractive index, the real ($n$) and ($\kappa$) imaginary parts associated with the phase velocity and incidental light beam absorption on the framework, respectively, show a rise with pellets optical density resulted from the pelleting pressure and decrease as a function of higher frequency (see Figure S12).

The calculated dielectric properties of MIL-100-MG pellets in the IR region were presented in conjunction with the Hz-MHz data for a better illustration of the broadband dielectric spectrum (4 Hz to 150 THz frequency range) of the MIL-100 framework (see Figure 3). Its overall dielectric constant can be expressed in terms of the different polarization mechanisms, as follows:

$$\varepsilon'_{\text{Total}} = \varepsilon'_{\text{Space charge (below 100 Hz)}} + \varepsilon'_{\text{Dipole/Orientational (MHz region)}} + \varepsilon'_{\text{Atomic/Vibrational (THz region)}} + \varepsilon'_{\text{Optical/Electronic (above 100 THz)}} \quad \ldots\ldots\ldots\ldots (9)$$

Switching the frequency from MHz to THz causes a step decrease in dielectric values, correlated with the loss of dipole-based polarization. The step increase in the dielectric values of individual pellets in the IR region is proportional to the pressure-induced framework amorphization and pelleting density, which follows the low-frequency (Hz-MHz) trend.

In the far-IR region (<20 THz), the dielectric properties of a material depend upon the bond polarity and the phonons of the structure. The distinctive transitions evidenced in the individual pellet $\varepsilon'$ curve (Figure 3(b)) is associated with the collective vibrational modes, i.e., metal-linker and linker deformations of the MIL-100 system, and accompanied by an absorption peak (dissipation) in the $\varepsilon''$ plot associated with the loss of energy during these collective vibrations.[23-24] Our data suggest that pelleting pressure-dependent amorphization also yields chemical bond polarization in the framework, thereby contributing to a higher $\varepsilon'$ value observed under compression. Similarly, the dielectric polarization associated with the vibrational modes of molecular moieties, i.e., torsional, bending, and stretching, dominate the mid-IR (~20-100 THz) region.

The absence of transition modes in the near-IR region (100-150 THz) shows the insensitivity of the dielectric response of MIL-100 towards its structural deformation. Due to the dependency of electronic polarization on atomic volume and its density, the effect of pelleting pressure-dependent amorphization can be evidenced in this region, thus limiting the $\varepsilon'$ values lying within the dielectric band to be 1.3-2.1. The individual contribution of all these mechanisms can be further broken down as summarized in Table 1.

Table 1: Contributions of the main polarization mechanisms to the total dielectric response of MIL-100-MG. The contribution of space charge is not considered in this analysis.

| Polarization mechanisms → | $\varepsilon'_{Dipole}$ / Orientational polarization | $\varepsilon'_{Atomic}$ / Vibrational polarization | $\varepsilon'_{Optical}$ / Electronic polarization | $\varepsilon'_{Total}$ (MIL-100-MG) | $\varepsilon'_{Total}$ (Basolite F300) |
|---|---|---|---|---|---|
| Pelleting pressure (MPa) ↓ | MHz (4 Hz - 1.5 MHz) | Far-IR to Mid IR (1.5 - 100 THz) | Near-IR (100 - 150 THz) | Full-spectrum (100 Hz) | Full-spectrum (100 Hz) |
| 36.96 | 1.83 | 1.18 | 1.29 | 4.30 | 2.96 |
| 73.92 | 2.06 | 1.26 | 1.39 | 4.71 | 3.20 |
| 147.84 | 2.37 | 1.38 | 1.6 | 5.35 | 4.63 |
| 221.76 | 2.53 | 1.60 | 1.84 | 5.97 | 4.87 |
| 369.60 | 2.97 | 2.31 | 1.91 | 7.19 | 5.64 |
| 517.44 | 3.45 | 2.37 | 2.00 | 7.82 | 5.75 |
| 739.20 | 3.36 | 2.50 | 2.10 | 7.96 | 6.13 |

A summary of the $\varepsilon'$ trend of MIL-100-MG pellets was shown in Figure 4(a) to reveal the effect of transition from crystalline to amorphous phase on the broadband dielectric behavior. We found that the mechanical deformation and structural densification of the framework cause phonon modes softening. Due to the structural dependency of the atomic polarization, the decline rate of $\varepsilon'$ in the far-IR to mid-IR range is much greater as a function of pelleting pressure when compared to the plateau observed in the near-IR region (mainly associated with the electron density). Interestingly, for both far- and mid-IR regions, a systematic pelleting pressure-dependent red shift was observed for a few of the $\varepsilon''$ absorption peaks of the pellets, from which the amplitude of two majorly shifted peaks are plotted against the nominal pressure as shown in Figure 4(b) and Figure

S13. These step changes in the redshifts indicate the softening in the phonon modes within THz frequencies, and the results can be explained by the amorphization trend underpinning the pellets.

In addition to the dielectric and optical properties, the pelleting pressure-dependent electrical conductivity measurements were also carried out on both types of pellets in the frequency range of 4 Hz to 1.5 MHz. Figure 5(a) and (b) show the pelleting pressure and temperature-dependent AC conductivity values for Basolite F300 and MIL-100-MG pellets at 1 MHz frequency, respectively. Similar to $\varepsilon'$, the increase in conductivity values also follows the amorphization pattern, suggesting that framework distortion may facilitate charge hopping (by decreasing the percolation threshold in the pellets) resulting in a sudden increase in conductivity values from 1-2 and 3-5t pellets for Basolite F300 pellets, and 1-2, 2-3 and 3-5t for MIL-100-MG pellets at 20 °C. The MIL-100-MG pellets show overall higher conductivity values in addition to a higher temperature-dependent conductivity rate over its Basolite F300 counterpart, which was facilitated by the occupied guest molecules in the framework cavity. The temperature rise will ease the electron movement to upper bands and decrease the resistance of the insulating MIL-100 pellet. The temperature and frequency-dependent conductivity plots for the individual Basolite F300 and MIL-100-MG pellets are shown in Figures S14 and S15, respectively.

## 4. CONCLUSIONS

In addition to the surrounding temperature and operating frequency, both the pelleting pressure-dependent amorphization and the pore occupancy greatly affect the dielectric, optical and electrical properties of the MOF framework. Herein we contrasted properties of both Basolite F300 and MIL-100 (Fe) framework obtained via different processing techniques and studied their framework properties at different instances from crystalline to amorphous phase. The MIL-100 pellets obtained using mechanochemistry was chosen to further explore the dependence of dielectric constant on dipole orientation, phonon, and electronic polarization, due to its clear phase transition from crystalline to amorphous as a function of pelleting pressure. To this end, we studied its effect on dielectric properties in different frequency regimes: (i) radio frequency region (4-1.5 MHz) for dipole polarization in the framework, (ii) IR region (1.2-150 THz) - framework collective modes in far-IR (<20 THz), atomic vibrations in mid-IR (20-100 THz) and electronic response in near-IR (>100 THz). Interestingly, only redshifts were observed for the few transition modes in optical and dielectric spectra due to bond softening. A greater understanding of broadband dielectrics under the influence of structural amorphization provides a cornerstone to new technologies bridging the domains of microelectronics and high-frequency telecommunications.

## ASSOCIATED CONTENT

### Supporting Information

Material characterization (XRD, FTIR and TGA), effect of pelleting pressure and temperature on the properties of dielectric (real, imaginary and loss), optical (reflectivity, real and imaginary part of refractive index) and electrical conductivity for Basolite F300 and MIL-100-MG pellets.

## AUTHOR INFORMATION

### Corresponding Author

Jin-Chong Tan - Multifunctional Materials and Composites (MMC) Laboratory, Department of Engineering Science, University of Oxford, Parks Road, Oxford, OX1 3PJ, United Kingdom

Email: jin-chong.tan@eng.ox.ac.uk

**ACKNOWLEDGMENTS**

ASB is grateful to the Engineering Science (EPSRC DTP – Samsung) Studentship that supports this DPhil research. JCT, AFM, MG acknowledge the European Union's Horizon 2020 research and innovation programme (ERC Consolidator Grant agreement No. 771575 - PROMOFS) for supporting this research. JCT thanks the Samsung GRO Award (DFR00230) for funding this research. We acknowledge the Diamond Light Source for the provision of beamtime SM21472 at B22 MIRIAM. We thank the Research Complex at Harwell (RCaH) for the provision of TGA and FTIR.

## REFERENCES

1. Moore, G. E., Solid-State Circuits Society Newsletter. *IEEE* **2006,** *20*, 33-35.
2. Bohr, M. T. In *Interconnect scaling-the real limiter to high performance ULSI*, Proceedings of International Electron Devices Meeting, IEEE: 1995; pp 241-244.
3. Grill, A.; Perraud, L.; Patel, V.; Jahnes, C.; Cohen, S., Low dielectric constant SiCOH films as potential candidates for interconnect dielectrics. *Low-Dielectric Constant Materials V* **1999,** *565*, 107-116.
4. Thompson, S.; Anand, N.; Armstrong, M.; Auth, C.; Arcot, B.; Alavi, M.; Bai, P.; Bielefeld, J.; Bigwood, R.; Brandenburg, J. In *A 90 nm logic technology featuring 50 nm strained silicon channel transistors, 7 layers of Cu interconnects, low-k ILD, and 1 µm SRAM cell*, Digest. International Electron Devices Meeting, IEEE: 2002; pp 61-64.
5. Zagorodniy, K.; Seifert, G.; Hermann, H., Metal-organic frameworks as promising candidates for future ultralow-κ dielectrics. *Appl. Phys. Lett.* **2010,** *97*, 251905.
6. Farrell, R.; Goshal, T.; Cvelbar, U.; Petkov, N.; Morris, M. A., Advances in ultra low dielectric constant ordered porous materials. *Electrochem Soc Interface* **2011,** *20*, 39.
7. Su, Z.; Miao, Y. R.; Zhang, G. H.; Miller, J. T.; Suslick, K. S., Bond breakage under pressure in a metal-organic framework. *Chem. Sci.* **2017,** *8*, 8004-8011.
8. Bennett, T. D.; Saines, P. J.; Keen, D. A.; Tan, J. C.; Cheetham, A. K., Ball-Milling-Induced Amorphization of Zeolitic Imidazolate Frameworks (ZIFs) for the Irreversible Trapping of Iodine. *Chem. Eur. J.* **2013,** *19*, 7049-7055.
9. Cao, S.; Bennett, T. D.; Keen, D. A.; Goodwin, A. L.; Cheetham, A. K., Amorphization of the prototypical zeolitic imidazolate framework ZIF-8 by ball-milling. *Chem. Commun.* **2012,** *48*, 7805-7807.
10. Stassen, I.; Burtch, N.; Talin, A.; Falcaro, P.; Allendorf, M.; Ameloot, R., An updated roadmap for the integration of metal–organic frameworks with electronic devices and chemical sensors. *Chem. Soc. Rev.* **2017,** *46*, 3185-3241.
11. Mendiratta, S.; Usman, M.; Lu, K. L., Expanding the dimensions of metal-organic framework research towards dielectrics. *Coord. Chem. Rev.* **2018,** *360*, 77-91.
12. Li, W. J.; Liu, J.; Sun, Z. H.; Liu, T. F.; Lu, J.; Gao, S. Y.; He, C.; Cao, R.; Luo, J. H., Integration of metal-organic frameworks into an electrochemical dielectric thin film for electronic applications. *Nat. Commun.* **2016,** *7*.
13. Krishtab, M.; Stassen, I.; Stassin, T.; Cruz, A. J.; Okudur, O. O.; Armini, S.; Wilson, C.; De Gendt, S.; Ameloot, R., Vapor-deposited zeolitic imidazolate frameworks as gap-filling ultra-low-κ dielectrics. *Nat. Commun.* **2019,** *10*, 1-9.
14. Titov, K.; Zeng, Z.; Ryder, M. R.; Chaudhari, A. K.; Civalleri, B.; Kelley, C. S.; Frogley, M. D.; Cinque, G.; Tan, J. C., Probing dielectric properties of metal-organic frameworks: MIL-53(Al) as a model system for theoretical predictions and experimental measurements via synchrotron far- and mid-infrared spectroscopy. *J. Phys. Chem. Lett.* **2017,** *8*, 5035-5040.
15. Ryder, M. R.; Zeng, Z. X.; Titov, K.; Sun, Y. T.; Mandi, E. M.; Flyagina, I.; Bennett, T. D.; Civalleri, B.; Kelley, C. S.; Frogley, M. D.; Cinque, G.; Tan, J. C., Dielectric properties of zeolitic imidazolate frameworks in the broad-band infrared regime. *J. Phys. Chem. Lett.* **2018,** *9*, 2678-2684.
16. Babal, A. S.; Dona, L.; Ryder, M. R.; Titov, K.; Chaudhari, A. K.; Zeng, Z.; Kelley, C. S.; Frogley, M. D.; Cinque, G.; Civalleri, B., Impact of pressure and temperature on the broadband dielectric response of the HKUST-1 metal–organic framework. *J. Phys. Chem. C* **2019,** *123*, 29427-29435.
17. Eslava, S.; Zhang, L.; Esconjauregui, S.; Yang, J.; Vanstreels, K.; Baklanov, M. R.; Saiz, E., Metal-organic framework ZIF-8 films as low-κ dielectrics in microelectronics. *Chem. Mater.* **2012,** *25*, 27-33.
18. Redel, E.; Wang, Z. B.; Walheim, S.; Liu, J. X.; Gliemann, H.; Woll, C., On the dielectric and optical properties of surface-anchored metal-organic frameworks: A study on epitaxially grown thin films. *Appl. Phys. Lett.* **2013,** *103*, 091903.
19. Scatena, R.; Guntern, Y. T.; Macchi, P., Electron density and dielectric properties of highly porous MOFs: binding and mobility of guest molecules in $Cu_3(BTC)_2$ and $Zn_3(BTC)_2$. *J. Am. Chem. Soc.* **2019,** *141*, 9382-9390.
20. Souza, B. E.; Moslein, A. F.; Titov, K.; Taylor, J. D.; Rudic, S.; Tan, J.-C., Green reconstruction of MIL-100 (Fe) in water for high crystallinity and enhanced guest encapsulation. *ACS Sustain. Chem. Eng.* **2020,** *8*, 8247-8255.
21. Moslein, A. F.; Gutierrez, M.; Cohen, B.; Tan, J.-C., Near-field infrared nanospectroscopy reveals guest confinement in metal–organic framework single crystals. *Nano Lett.* **2020,** *20*, 7446-7454.
22. Roessler, D. M., Kramers-Kronig analysis of reflection data. *Br. J. Appl. Phys.* **1965,** *16*, 1119.
23. Ryder, M. R.; Civalleri, B.; Bennett, T. D.; Henke, S.; Rudic, S.; Cinque, G.; Fernandez-Alonso, F.; Tan, J. C., Identifying the role of terahertz vibrations in metal-organic frameworks: from gate-opening phenomenon to shear-driven structural destabilization. *Phys. Rev. Lett.* **2014,** *113*, 215502.
24. Ryder, M. R.; Civalleri, B.; Cinque, G.; Tan, J.-C., Discovering connections between terahertz vibrations and elasticity underpinning the collective dynamics of the HKUST-1 metal–organic framework. *CrystEngComm* **2016,** *18*, 4303-4312.

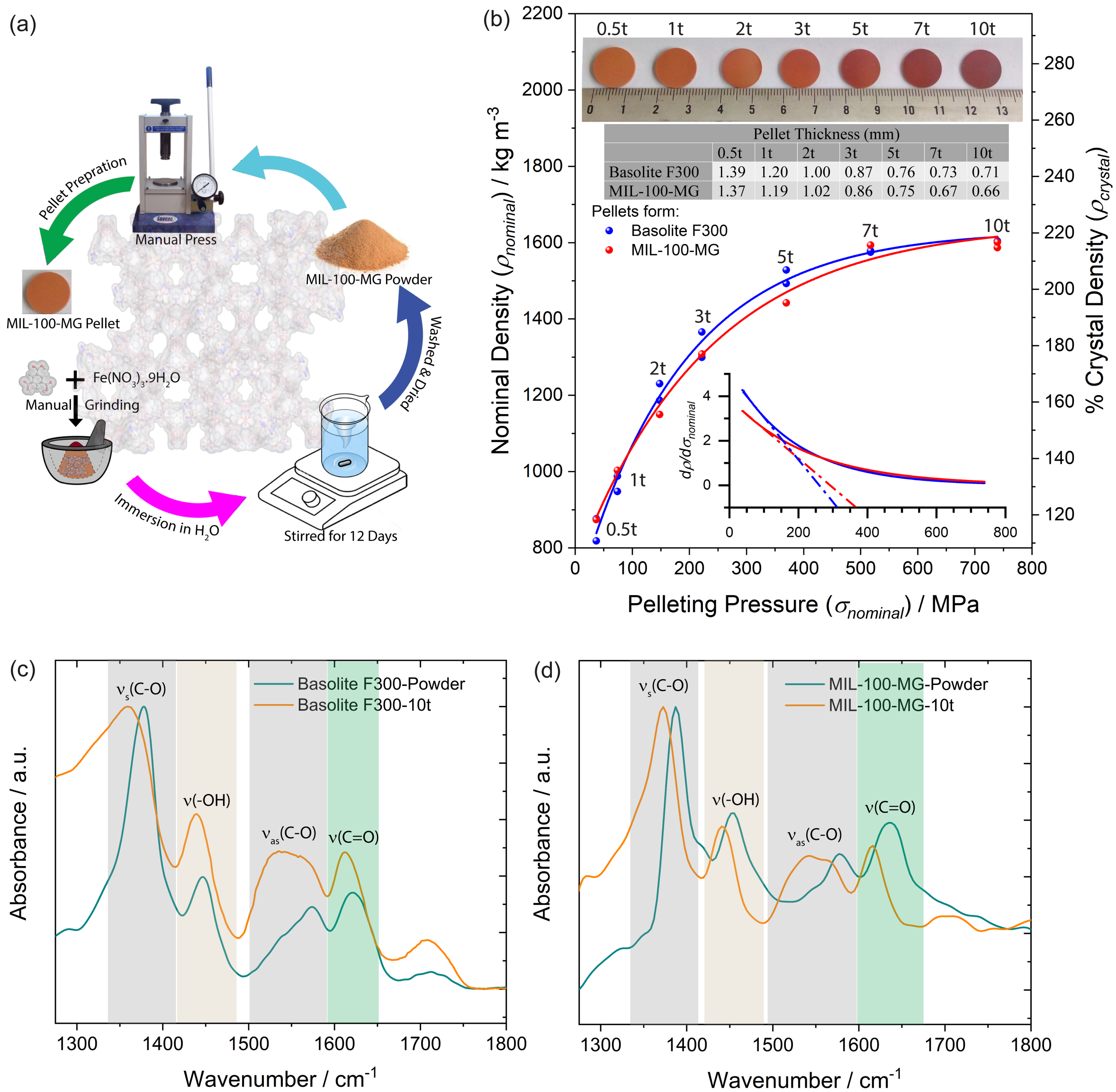

**Figure 1.** (a) MIL-100-MG powder synthesized by mechanochemistry route and its subsequent pellet preparation process. (b) Pelleting pressure-dependent nominal density plot for Basolite F300 and MIL-100-MG pellets used in the experiments. The pellets were labeled in terms of the applied uniaxial force, e.g., 0.5t for pellet pressed at 0.5-ton force. The photograph of pellets used in the experiments is also shown here. The inset shows the plot for the derivative of density over pelleting pressure and was used to estimate the yield strength of the compressed powder of Basolite F300 and MIL-100-MG as 310 MPa and 366 MPa, respectively. Table inset shows the average pellet thickness determined from 3 samples. (c)-(d) nano-FTIR spectra for Basolite F300 and MIL-100-MG samples (powder and 10-ton pellet), respectively.

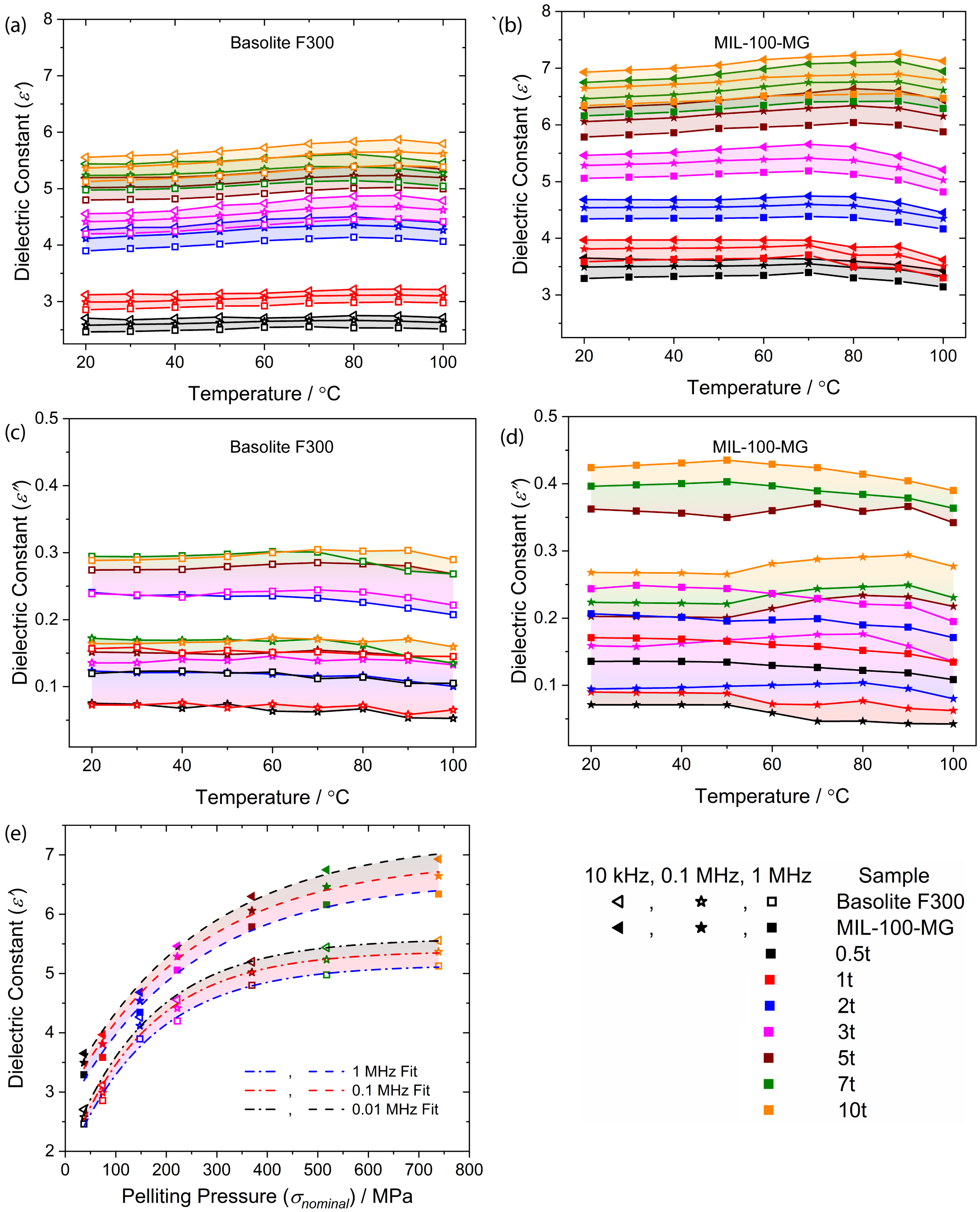

**Figure 2.** Dielectric properties of MOF pellets as a function of temperature: Real part of dielectric constant $\varepsilon'$ - (a) Basolite F300 and, (b) MIL-100-MG pellets. The imaginary part of dielectric constant $\varepsilon''$ - (c) Basolite F300 and, (d) MIL-100-MG pellets. (e) $\varepsilon'$ as a function of pelleting pressure for the three specific frequencies of 0.01, 0.1, and 1 MHz.

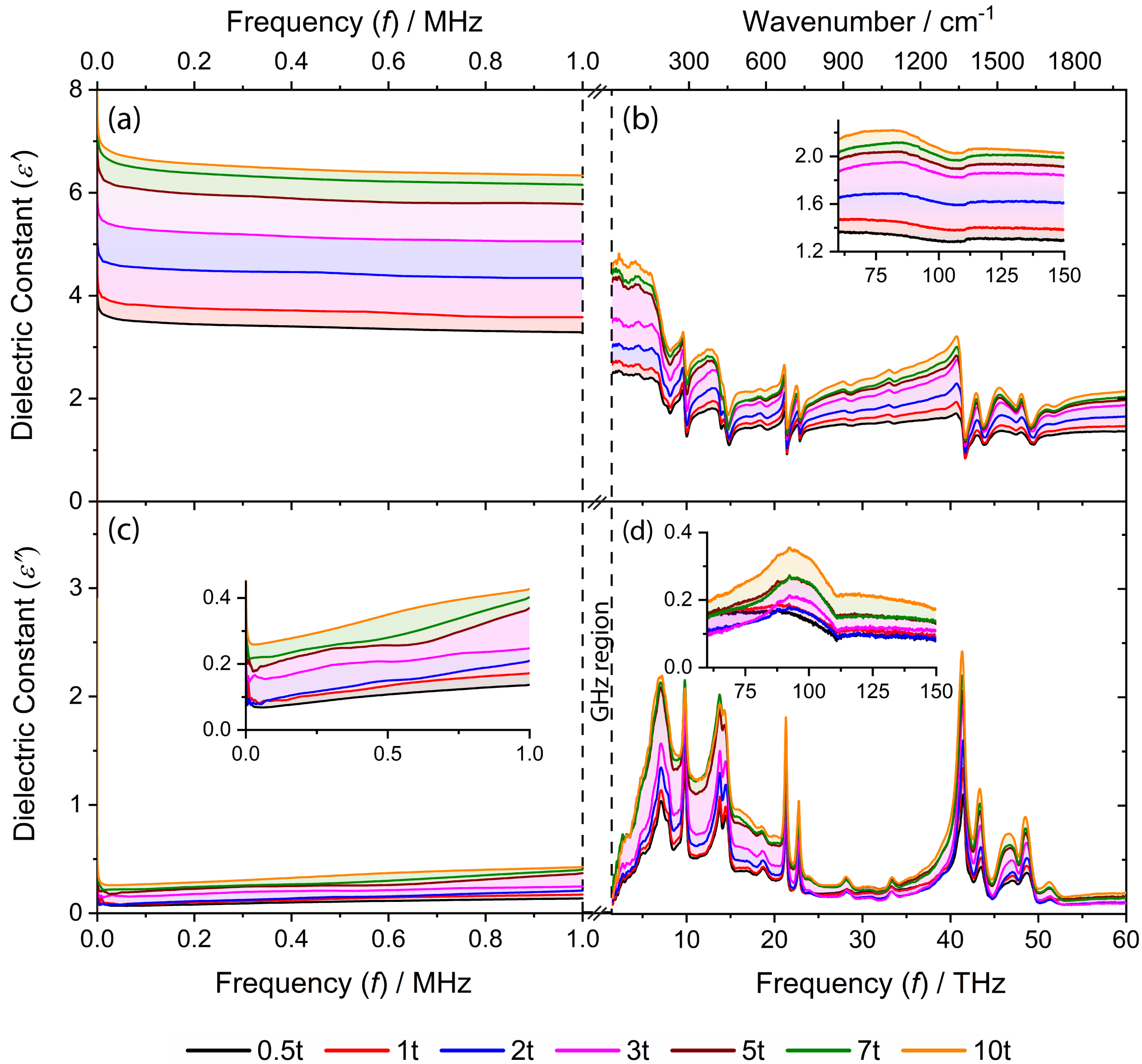

**Figure 3.** The pressure-dependent broadband dielectric spectrum for MIL-100 MOF in the frequency range of 4 Hz to 150 THz. (a, b) The real part of the dielectric constant ε' and (c, d) imaginary part of the dielectric constant $\varepsilon''$. In the radio frequency (Hz-MHz) range, the dielectric measurements are based on the LCR parallel-plate capacitor technique, whereas synchrotron specular reflectance spectroscopy was used to study the optical properties in the IR region (2-150 THz), which were converted into the dielectric parameters. The pellets studied for dielectrics are 0.5t, 1t, 2t, 3t, 5t, 7t, and 10t pellets.

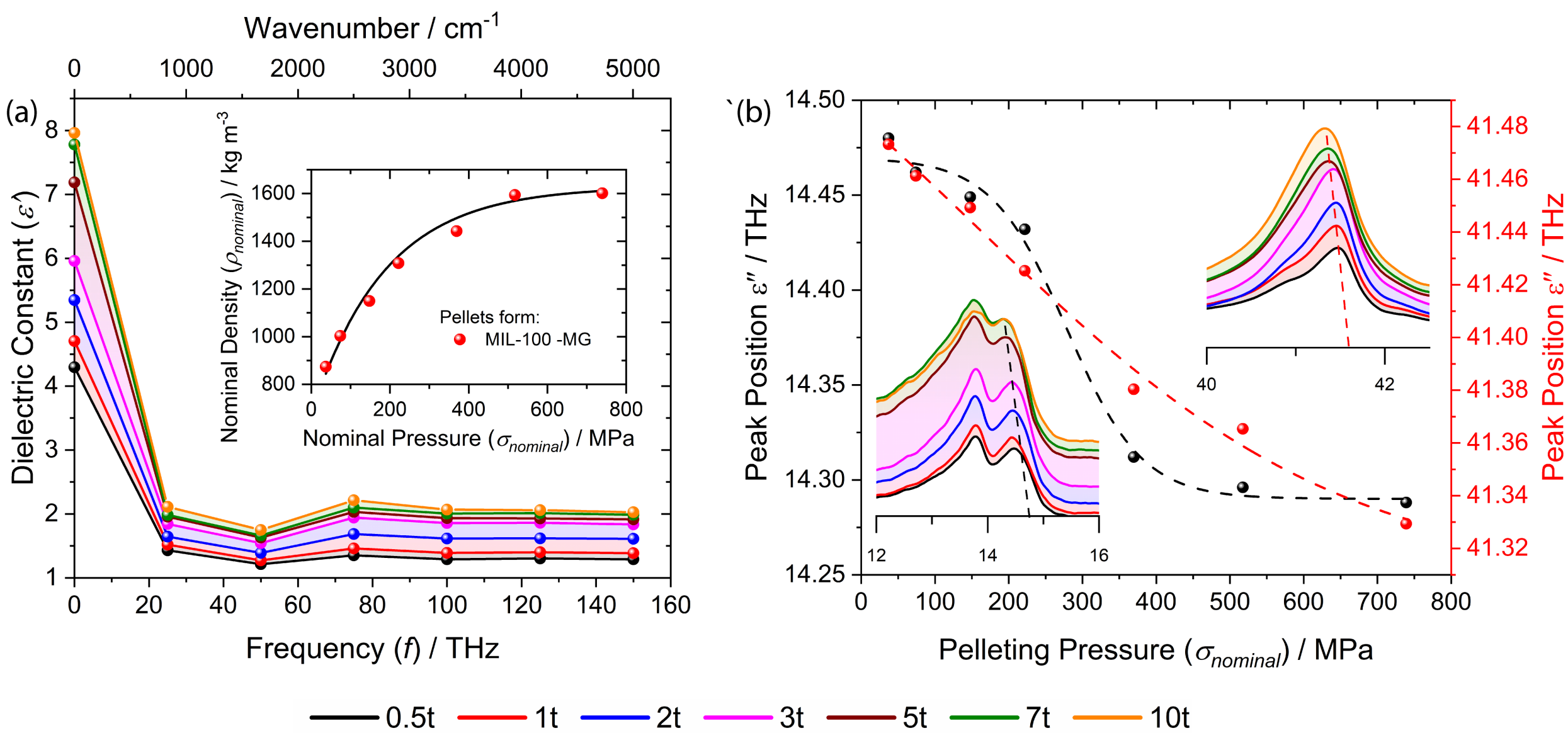

**Figure 4.** (a) The effect of crystalline to the amorphous transition of MIL-100-MG pellets on the real part of dielectric constants at specific broadband frequencies comprising the far-, mid-, and near-IR regions. (b) Pelleting pressure-dependent redshifts in the THz peaks associated with the phonon modes of the molecular moieties.

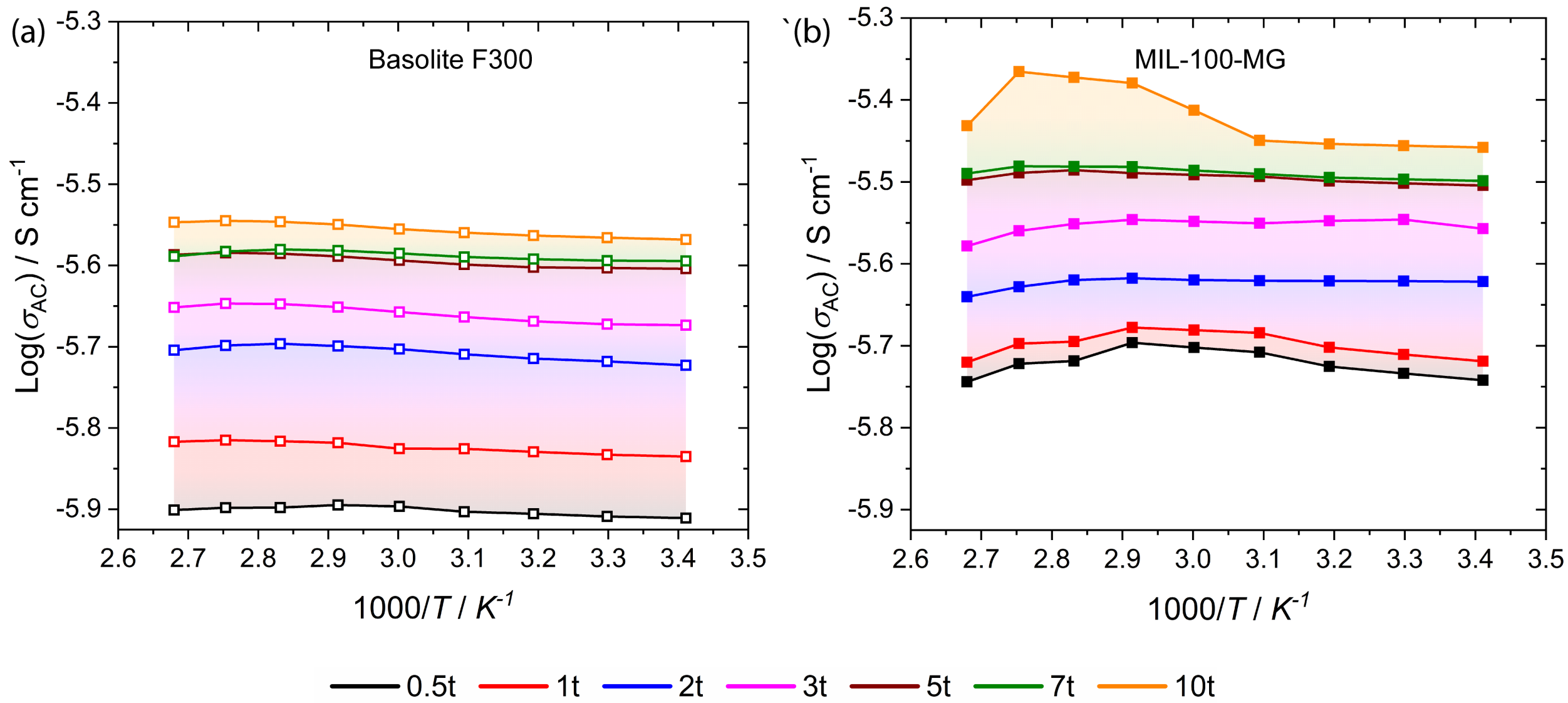

**Figure 5.** AC conductivity of MIL-100 pellets: (a) Basolite F300 and, (b) MIL-100-MG as a function of residual pressure and temperature.

# Broadband dielectric behavior of MIL-100 metal-organic framework as a function of structural amorphization

Arun Singh Babal,[a] Barbara E. Souza,[a] Annika F. Möslein,[a] Mario Gutiérrez,[a]

Mark D. Frogley,[b] and Jin-Chong Tan[a,*]

[a]Multifunctional Materials and Composites (MMC) Laboratory, Department of Engineering Science, University of Oxford, Parks Road, Oxford, OX1 3PJ, United Kingdom

[b]Diamond Light Source, Harwell Campus, Chilton, Oxford, OX11 0DE, United Kingdom

*E-mail: jin-chong.tan@eng.ox.ac.uk

**Contents**

## 1. Powder X-ray diffraction (XRD)

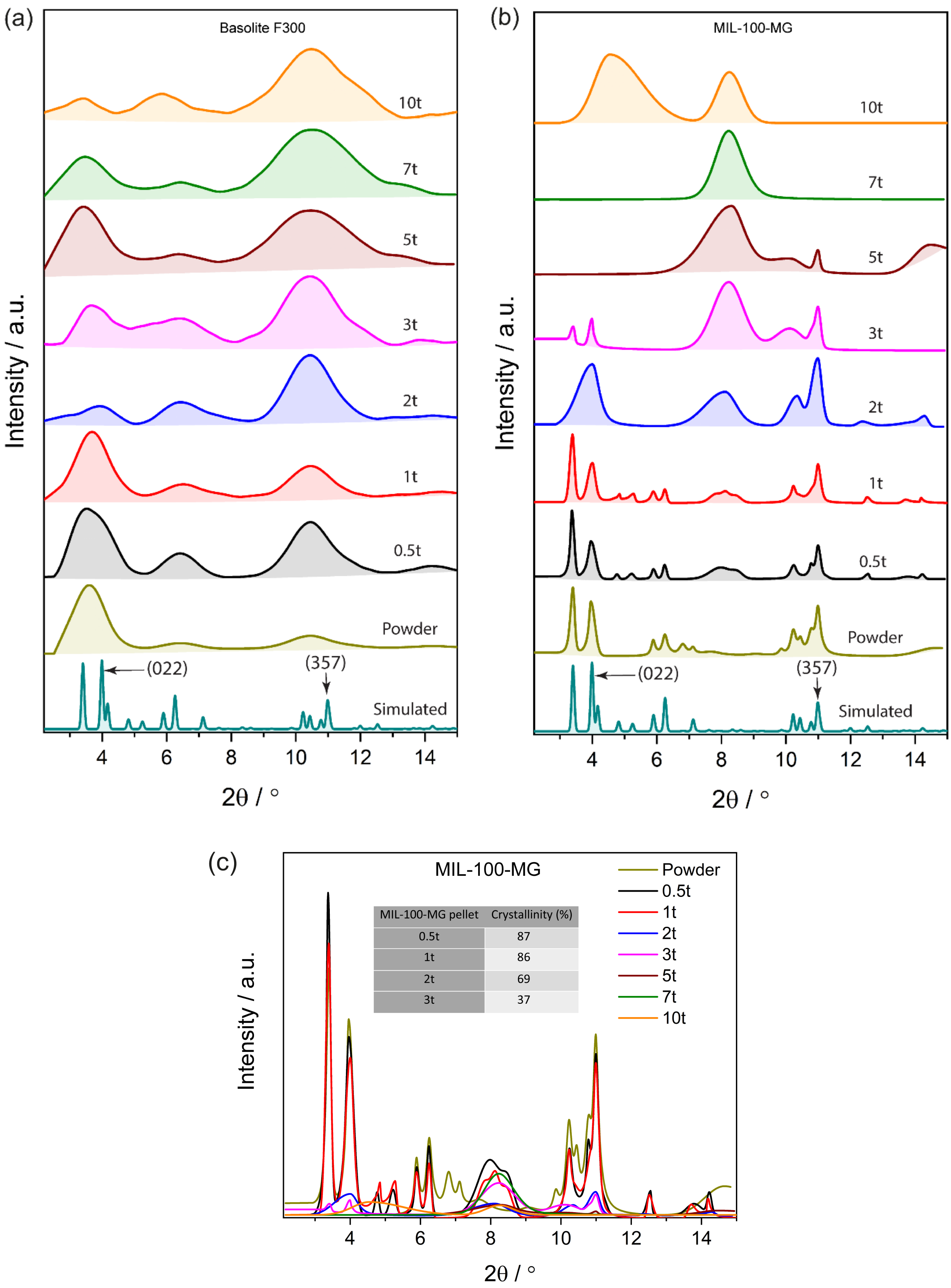

| MIL-100-MG pellet | Crystallinity (%) |
|---|---|
| 0.5t | 87 |
| 1t | 86 |
| 2t | 69 |
| 3t | 37 |

Figure S1: The XRD patterns for: (a) Basolite F300 and (b) MIL-100-MG pellets, both normalized with respect to the highest data point. (c) XRD patterns in absolute intensities for MIL-100-MG pellets. Inset of table shows the pellet crystallinity (%), estimated from the area ratio of the crystalline peaks to the total area found under the XRD pattern.

## 2. Fourier-transform infrared spectroscopy (ATR-FTIR)

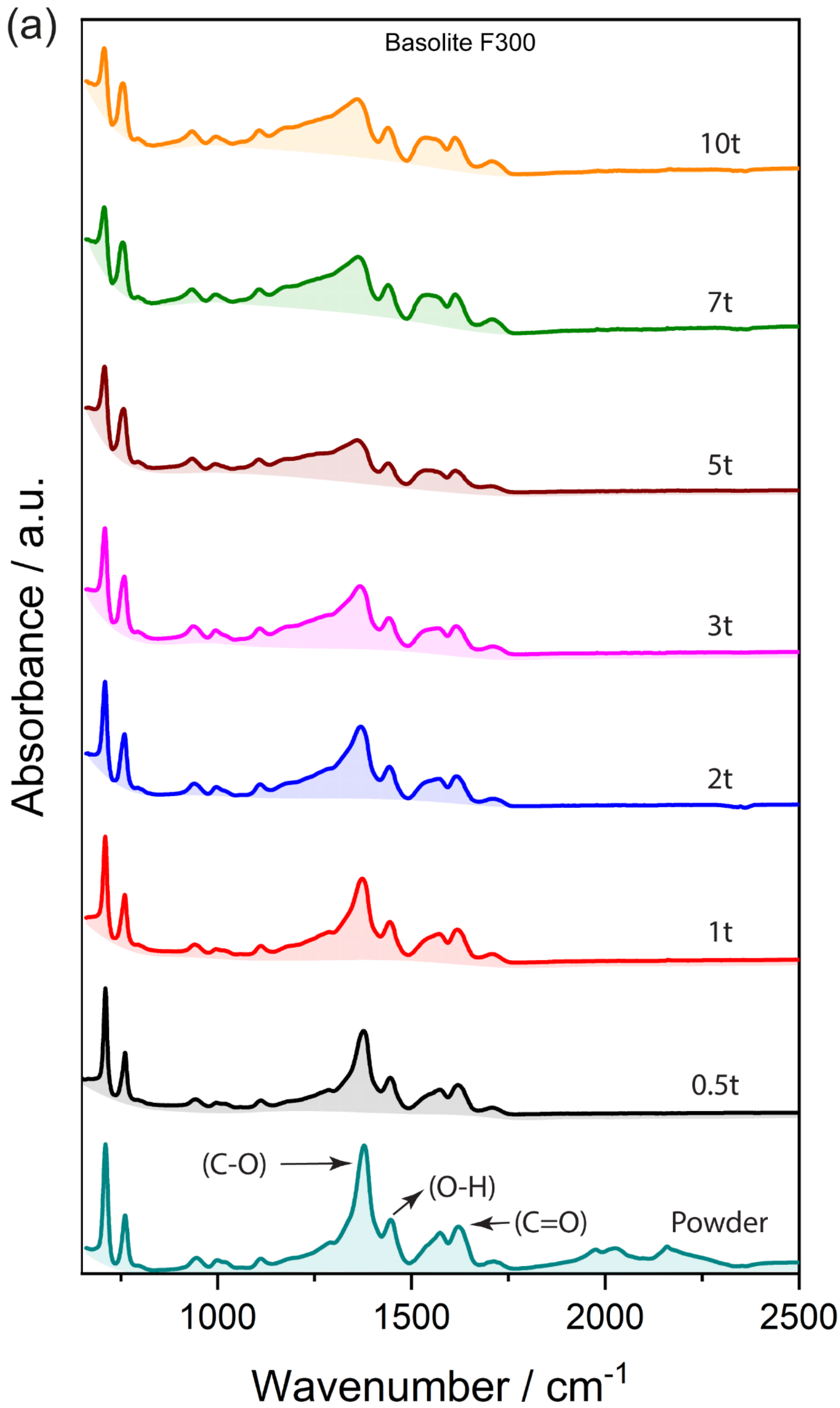

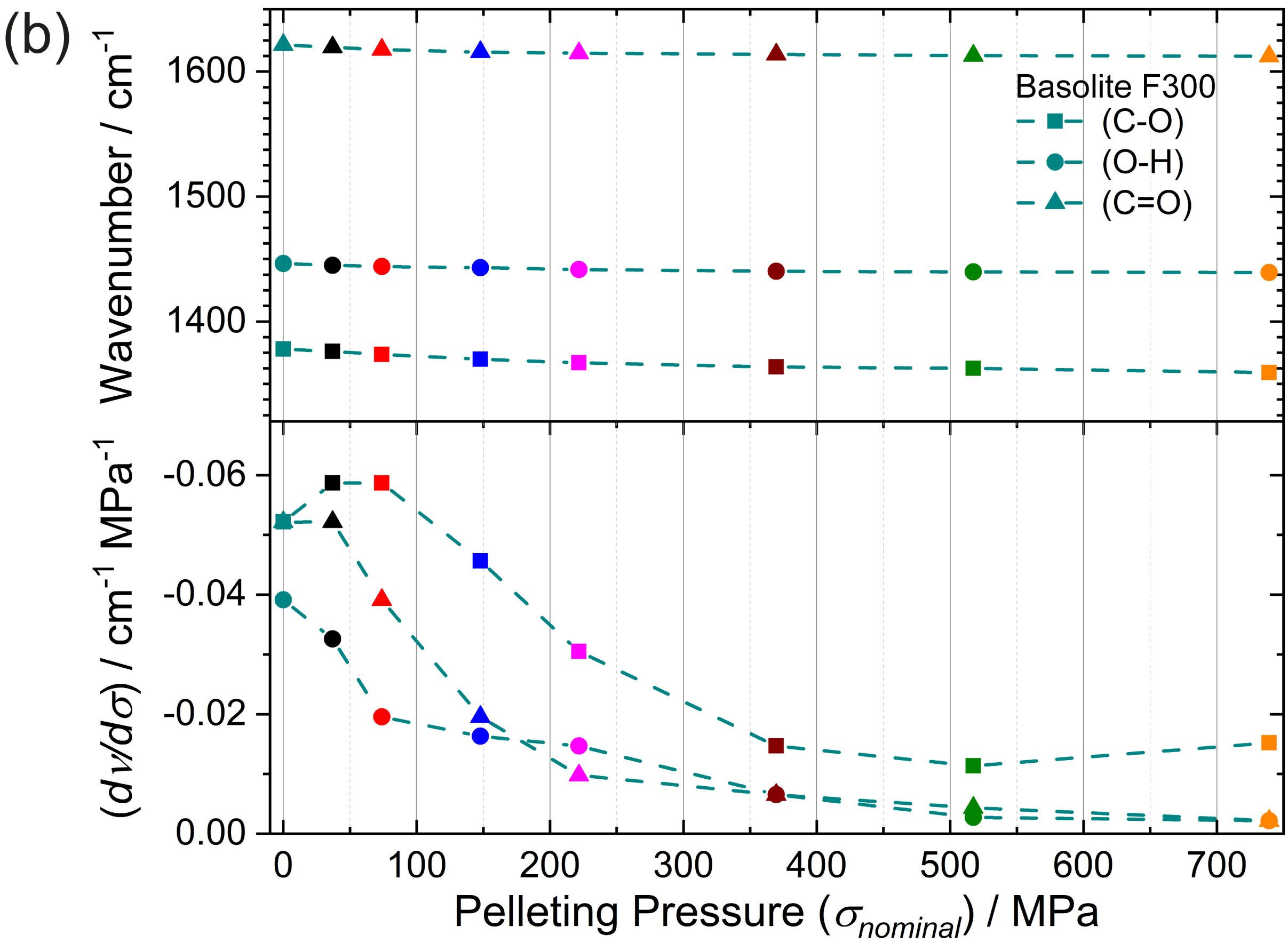

**Figure S2:** (a) The pressure-dependent ATR-FTIR spectra for Basolite F300 pellets, (b) Derivative of peak shift over pelleting pressure.

### 3. Thermogravimetric analysis (TGA)

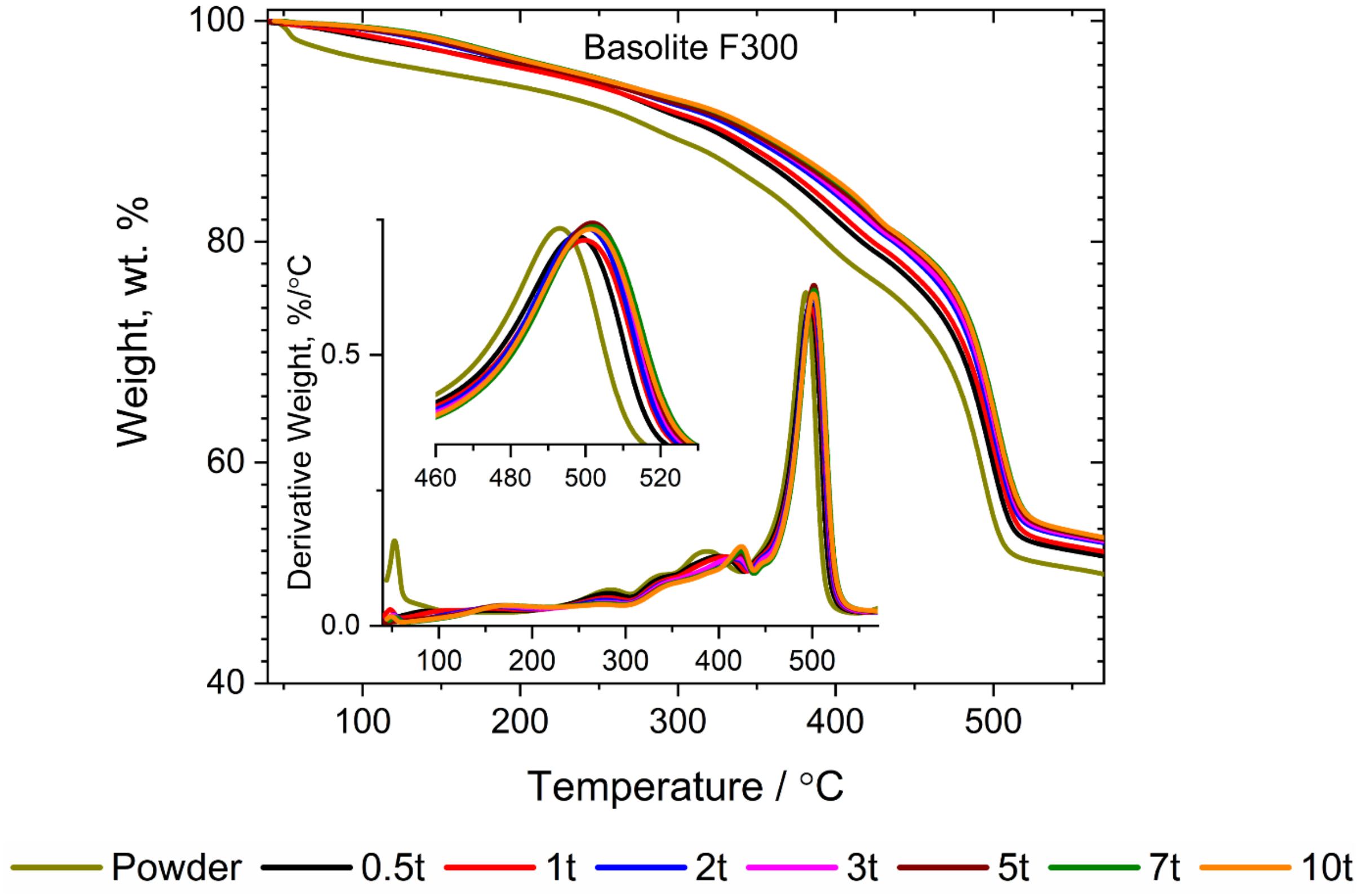

**Figure S3:** The pressure-dependent thermal stability measurement (TGA) for Basolite F300 pellets.

# 4. Dielectric properties

## 4.1 Basolite F300

### 4.1.1 Real Part of Dielectric Constant ($\varepsilon'$)

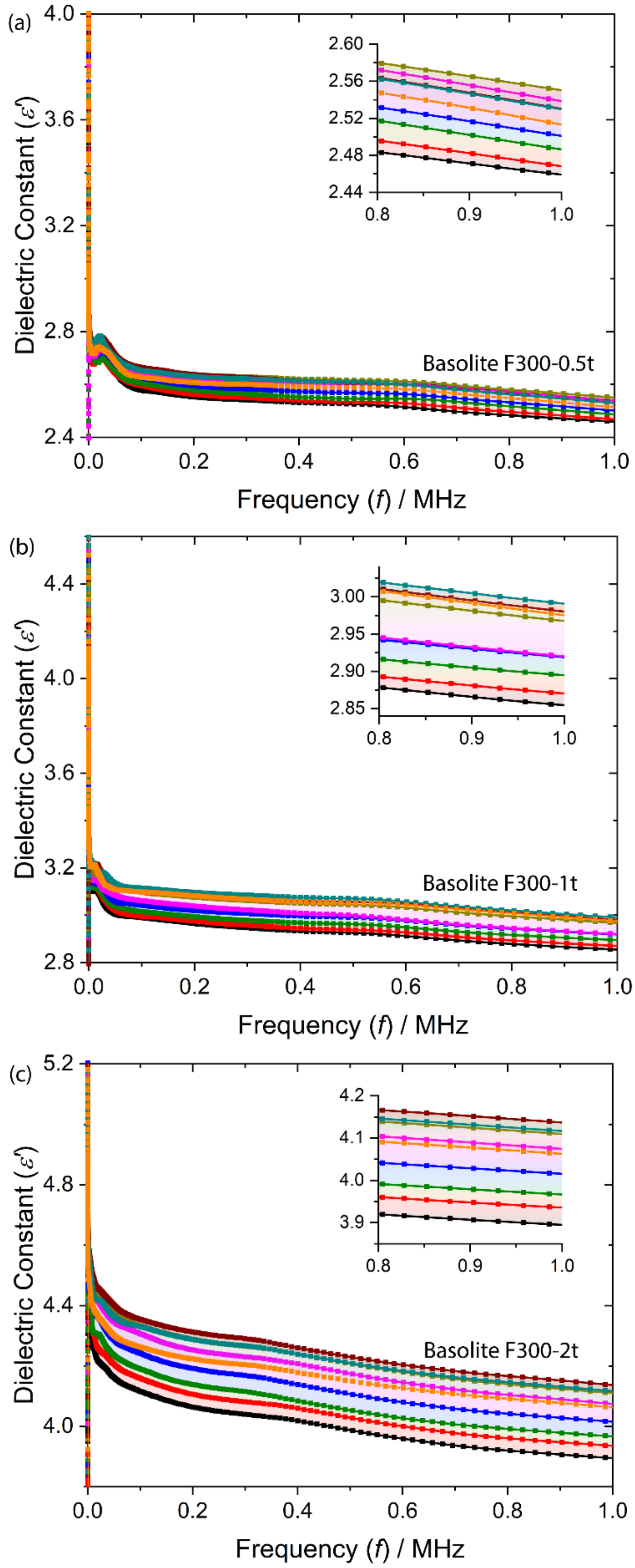

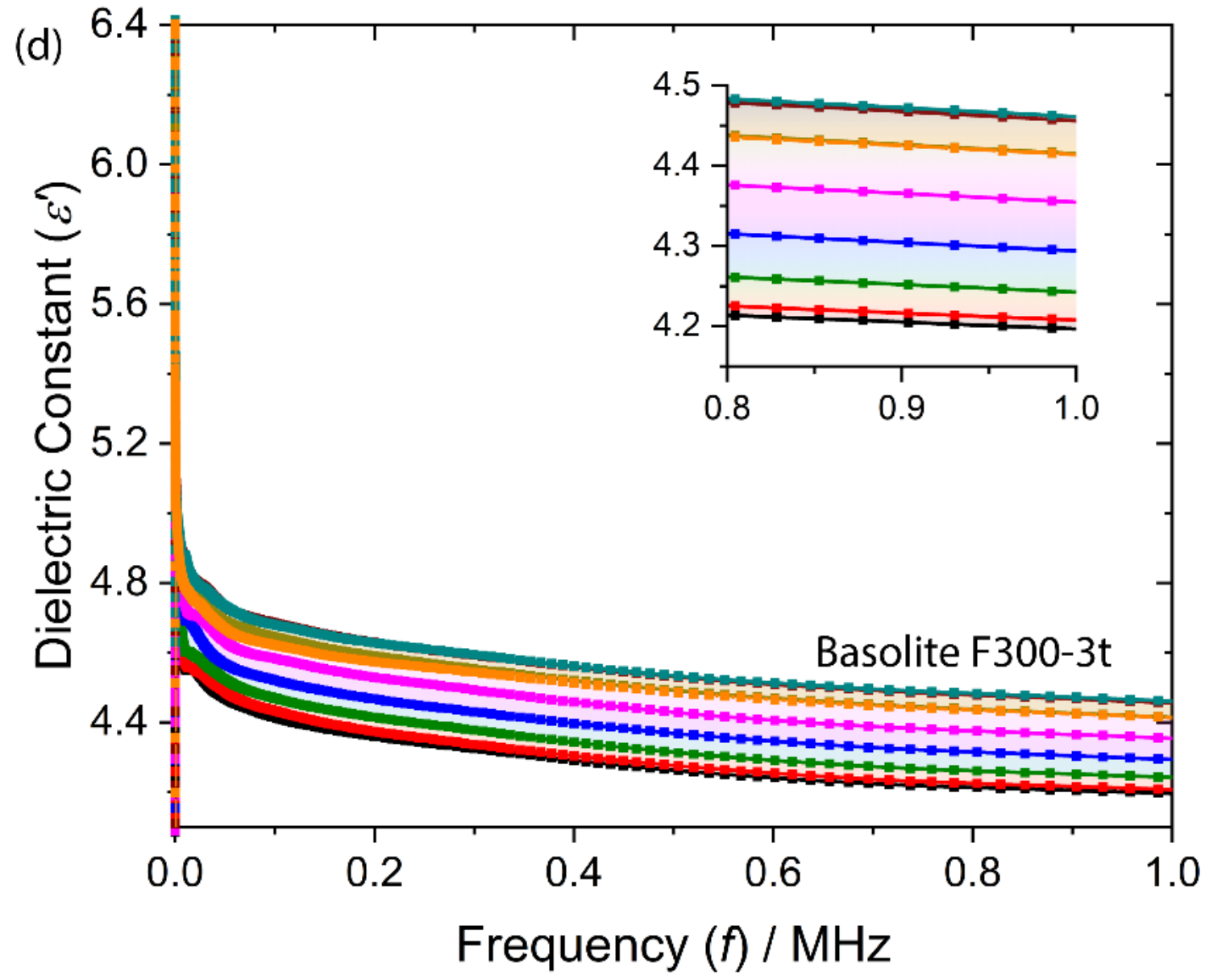
(d)
Dielectric Constant (ε')
Frequency (f) / MHz
Basolite F300-3t

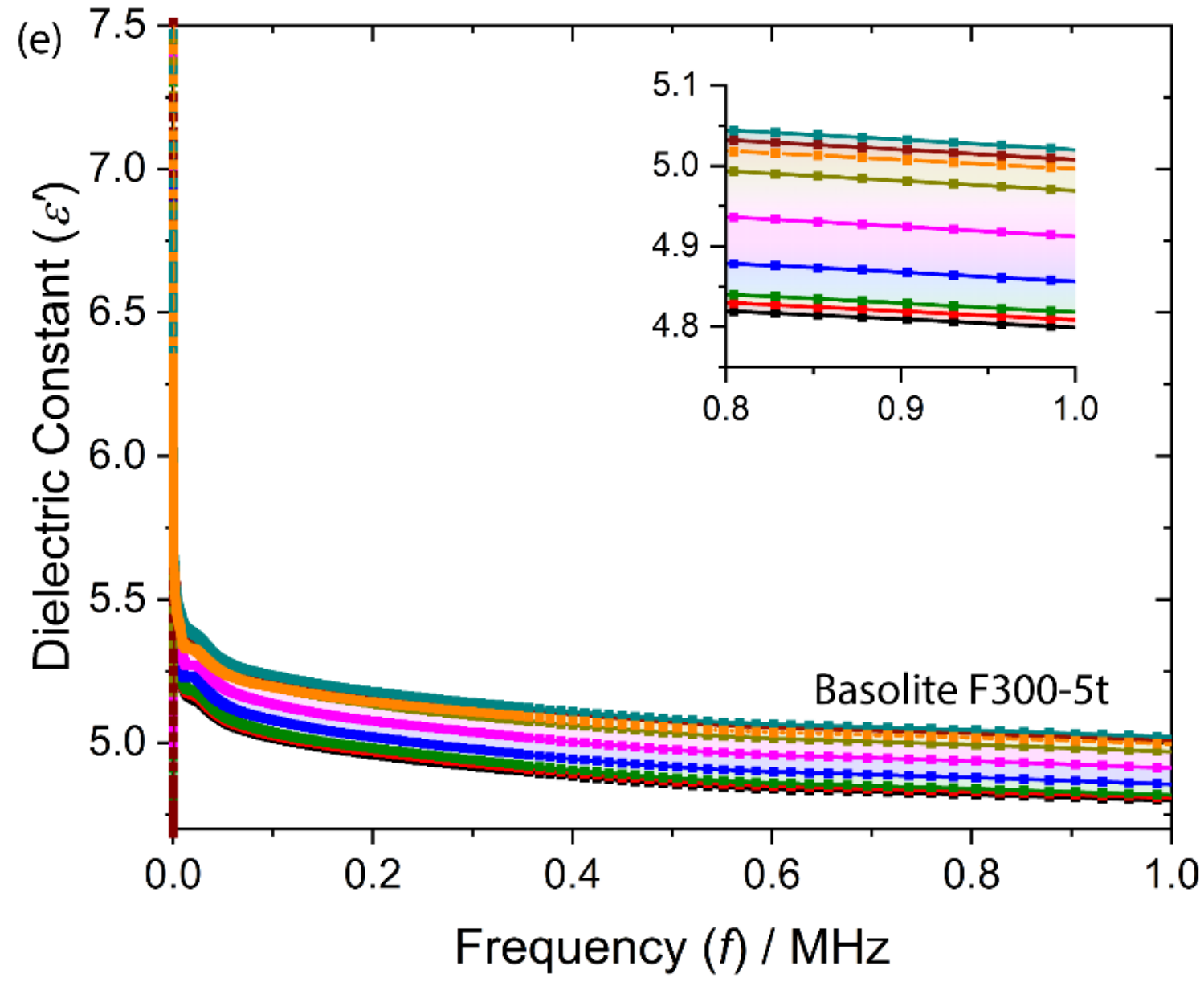
(e)
Dielectric Constant (ε')
Frequency (f) / MHz
Basolite F300-5t

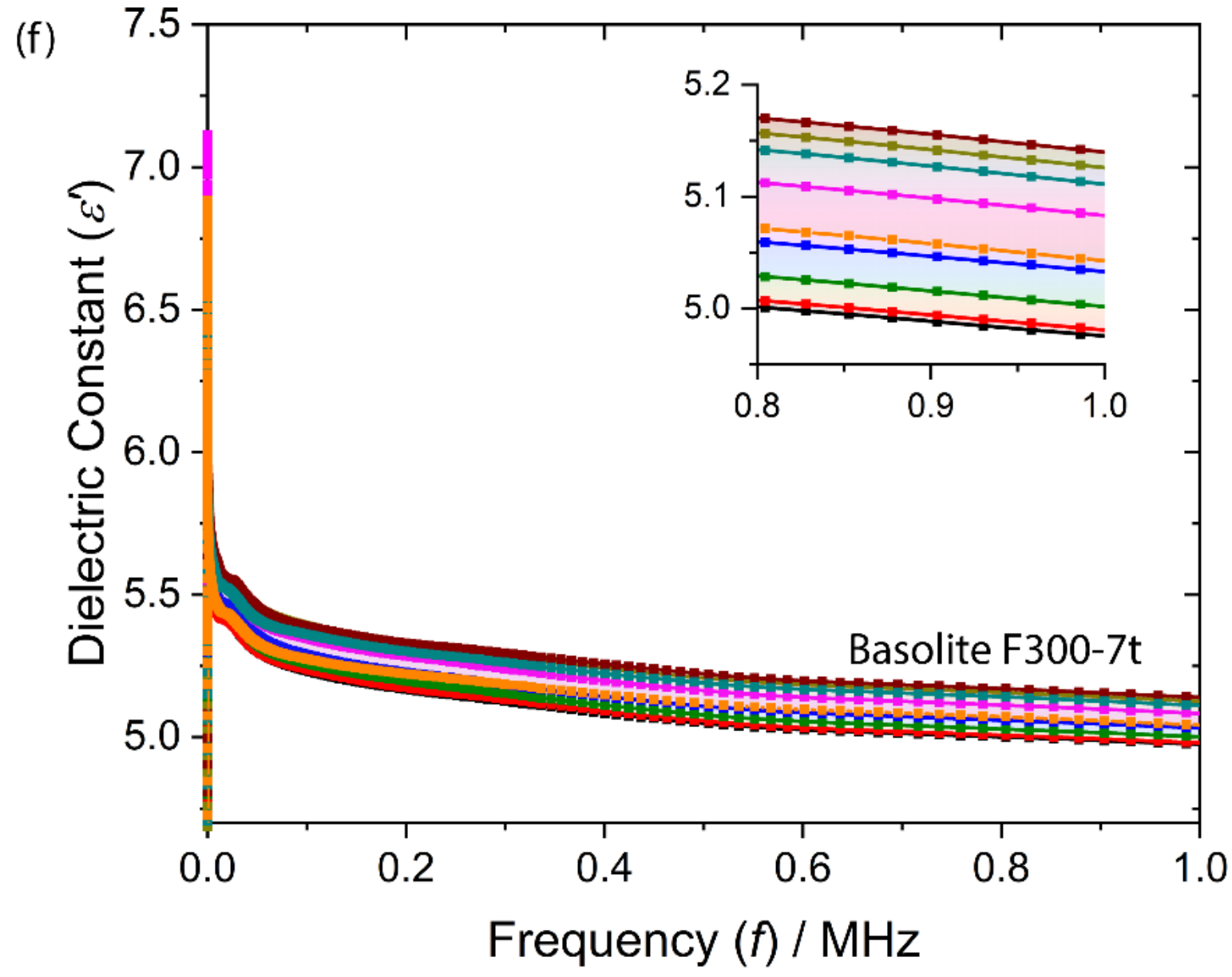
(f)
Dielectric Constant (ε')
Frequency (f) / MHz
Basolite F300-7t

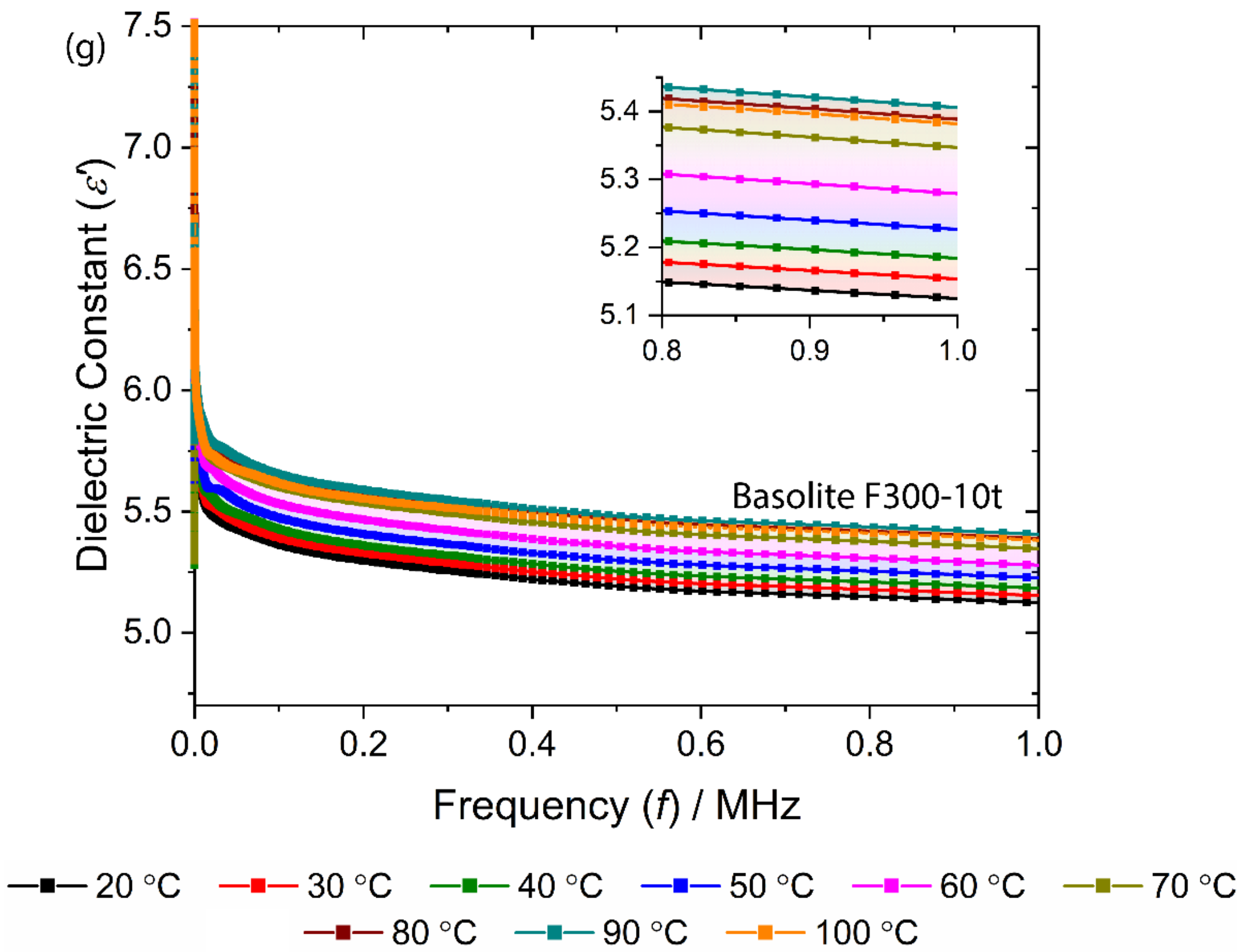

**Figure S4:** The real part of dielectric constant for Basolite F300 pellets prepared under a compression load of: (a) 0.5-ton, (b) 1-ton, (c) 2-ton, (d) 3-ton (e) 5-ton (f) 7-ton, and (g) 10-ton, corresponding to the pressure of 36.96, 73.92, 147.84, 221.76, 369.6, 517.44 and 739.20 MPa, respectively.

### 4.1.2 Imaginary Part of Dielectric Constant ($\varepsilon''$)

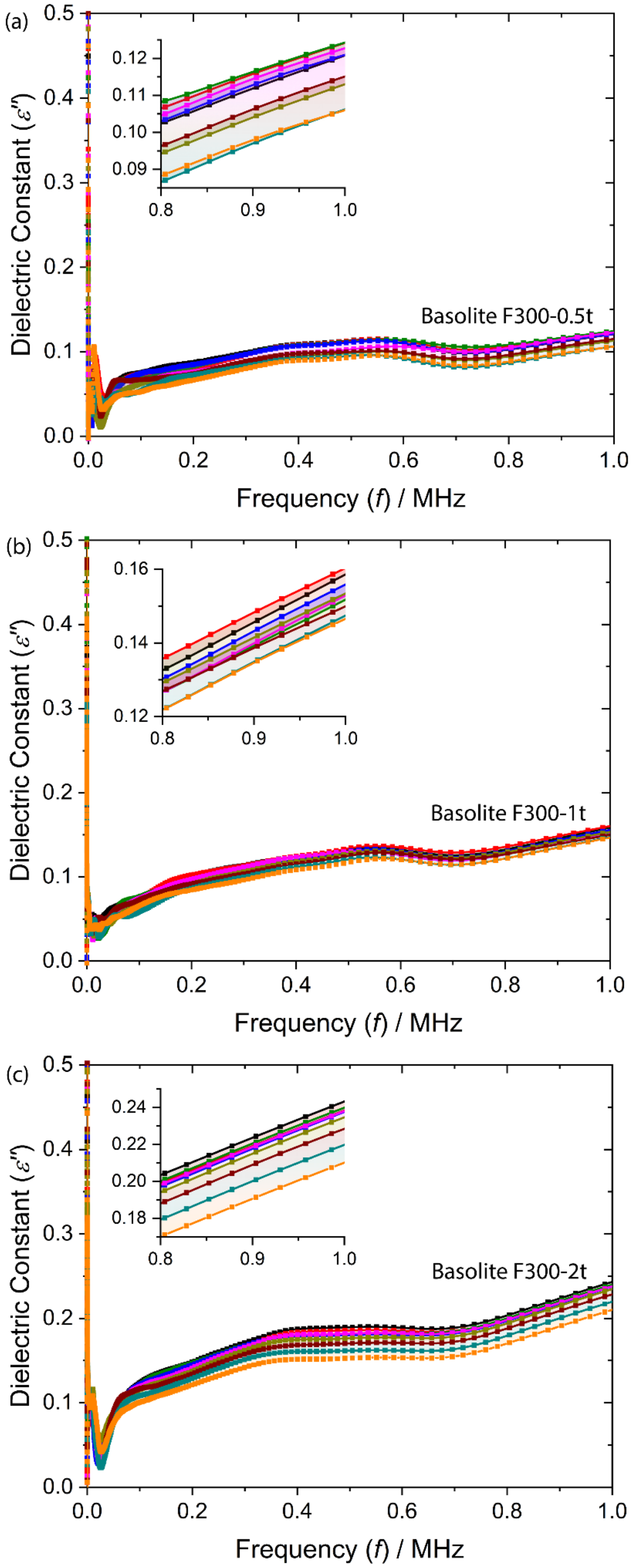

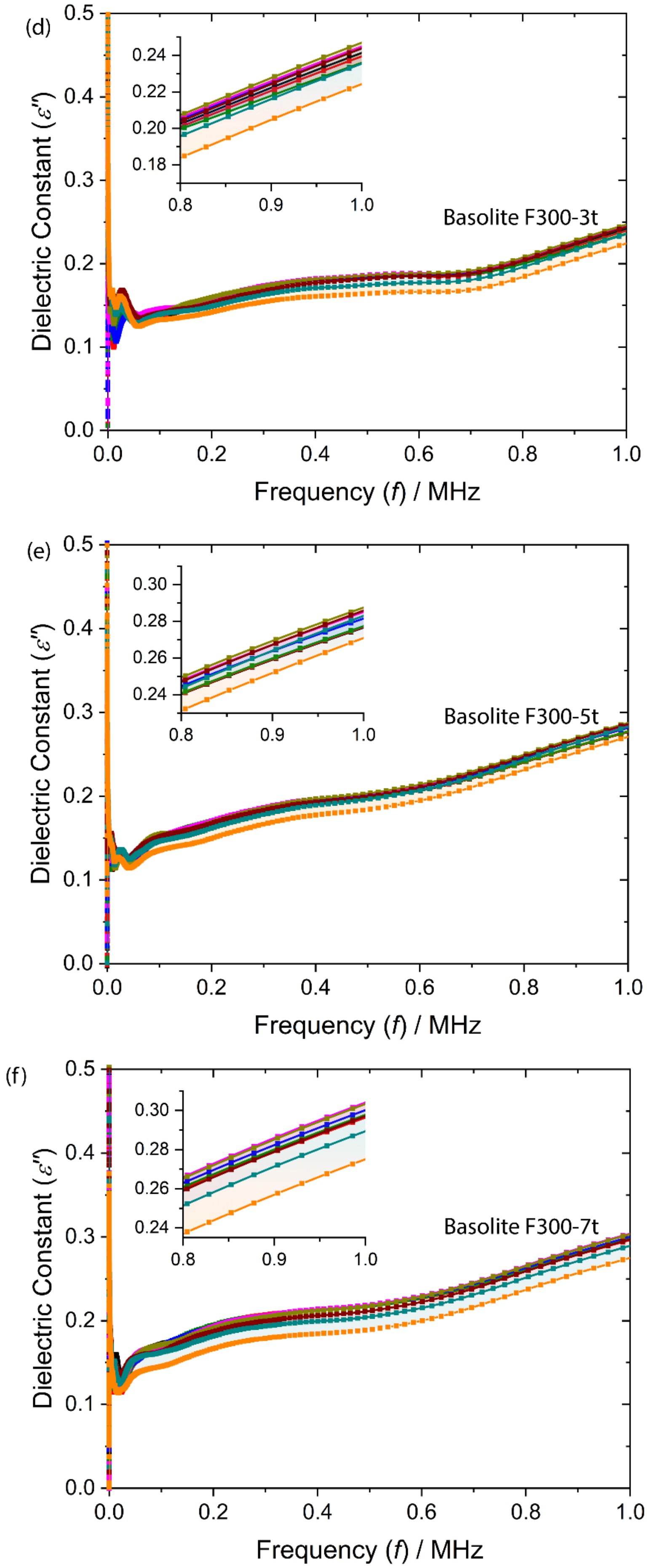
(d)
Basolite F300-3t
Dielectric Constant (ε'')
Frequency (f) / MHz
(e)
Basolite F300-5t
Dielectric Constant (ε'')
Frequency (f) / MHz
(f)
Basolite F300-7t
Dielectric Constant (ε'')
Frequency (f) / MHz

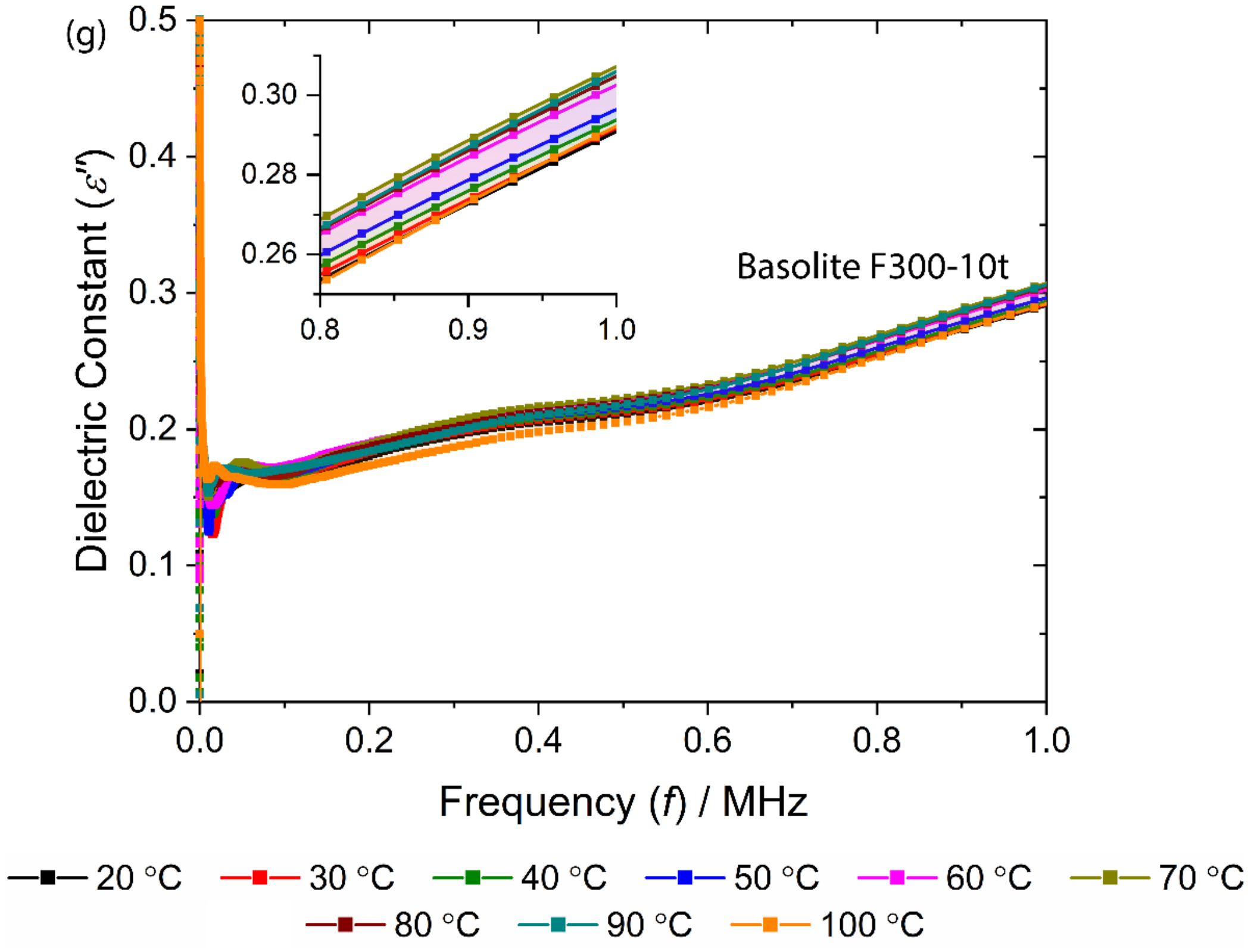

**Figure S5:** The imaginary part of dielectric constant for Basolite F300 pellets prepared under a compression load of: (a) 0.5-ton, (b) 1-ton, (c) 2-ton, (d) 3-ton (e) 5-ton (f) 7-ton, and (g) 10-ton, respectively.

### 4.1.3 Dielectric Loss (tan *δ*)

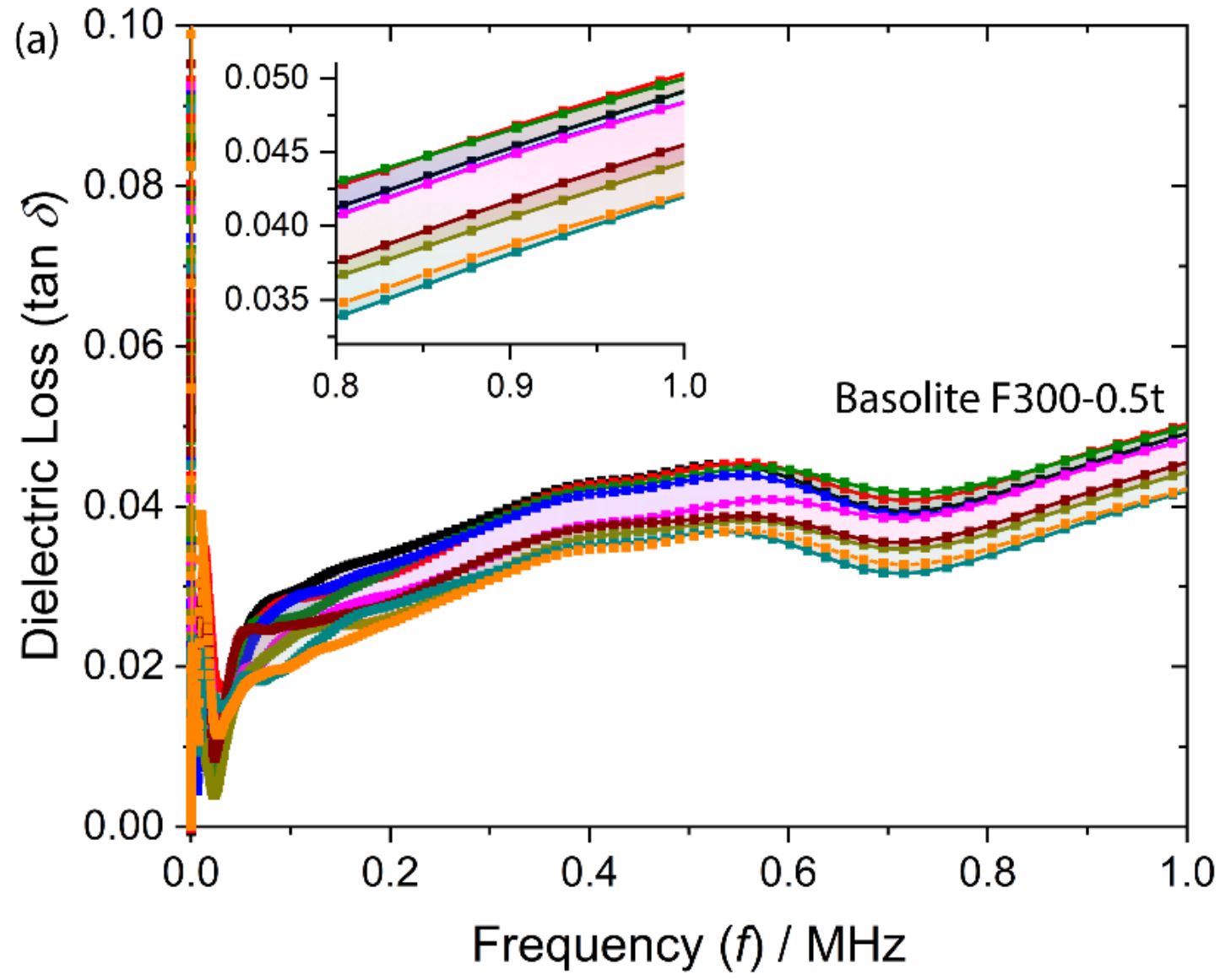

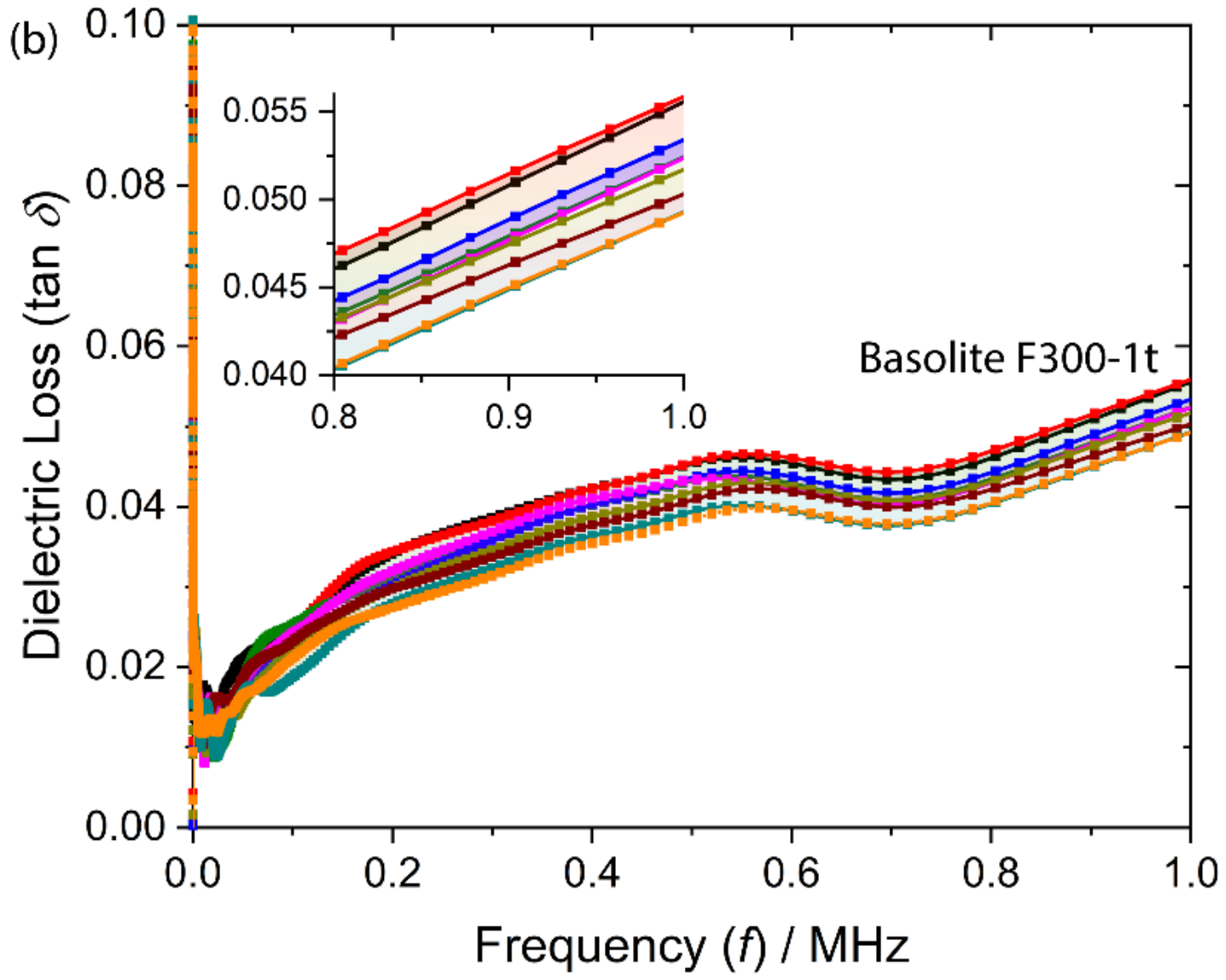

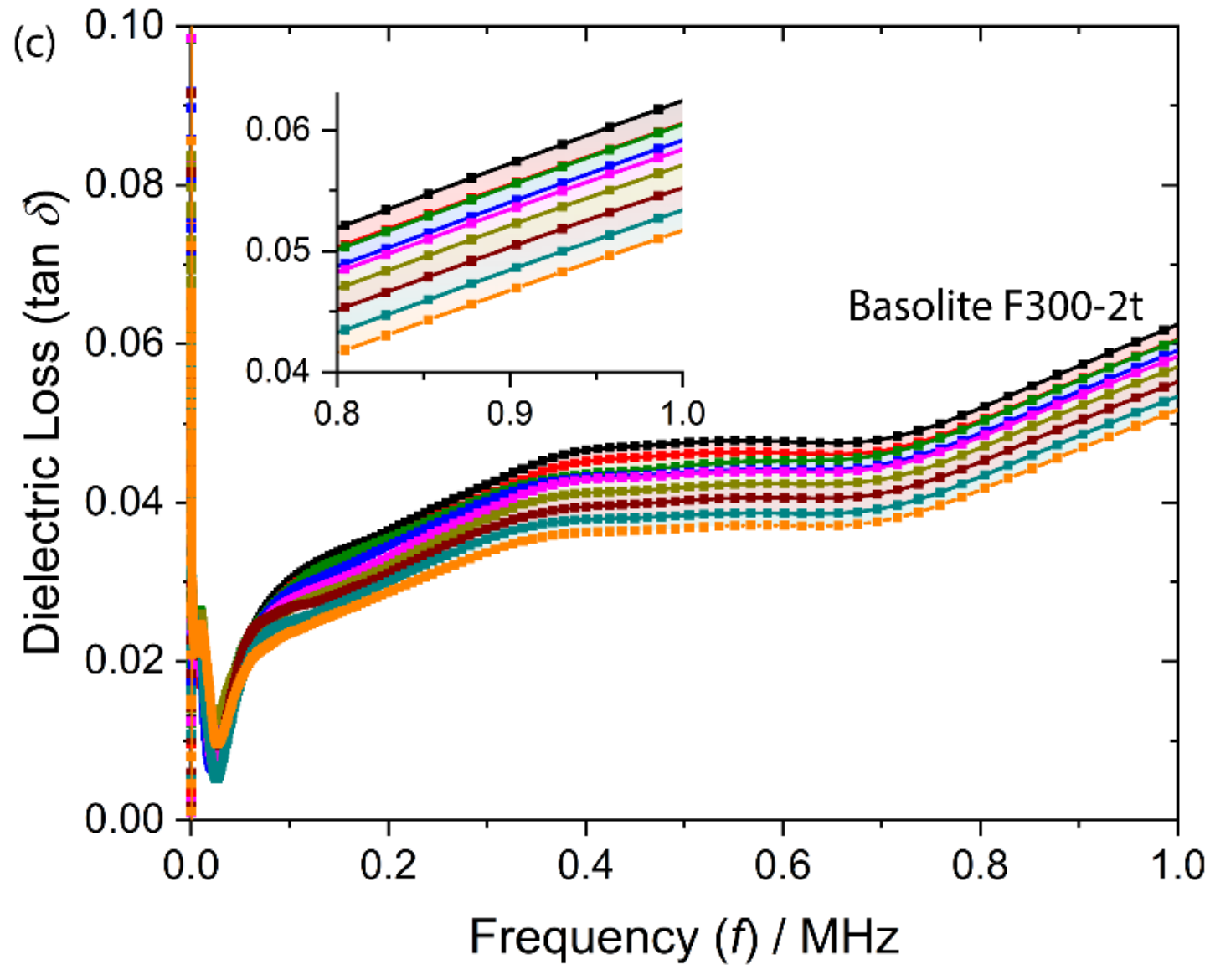

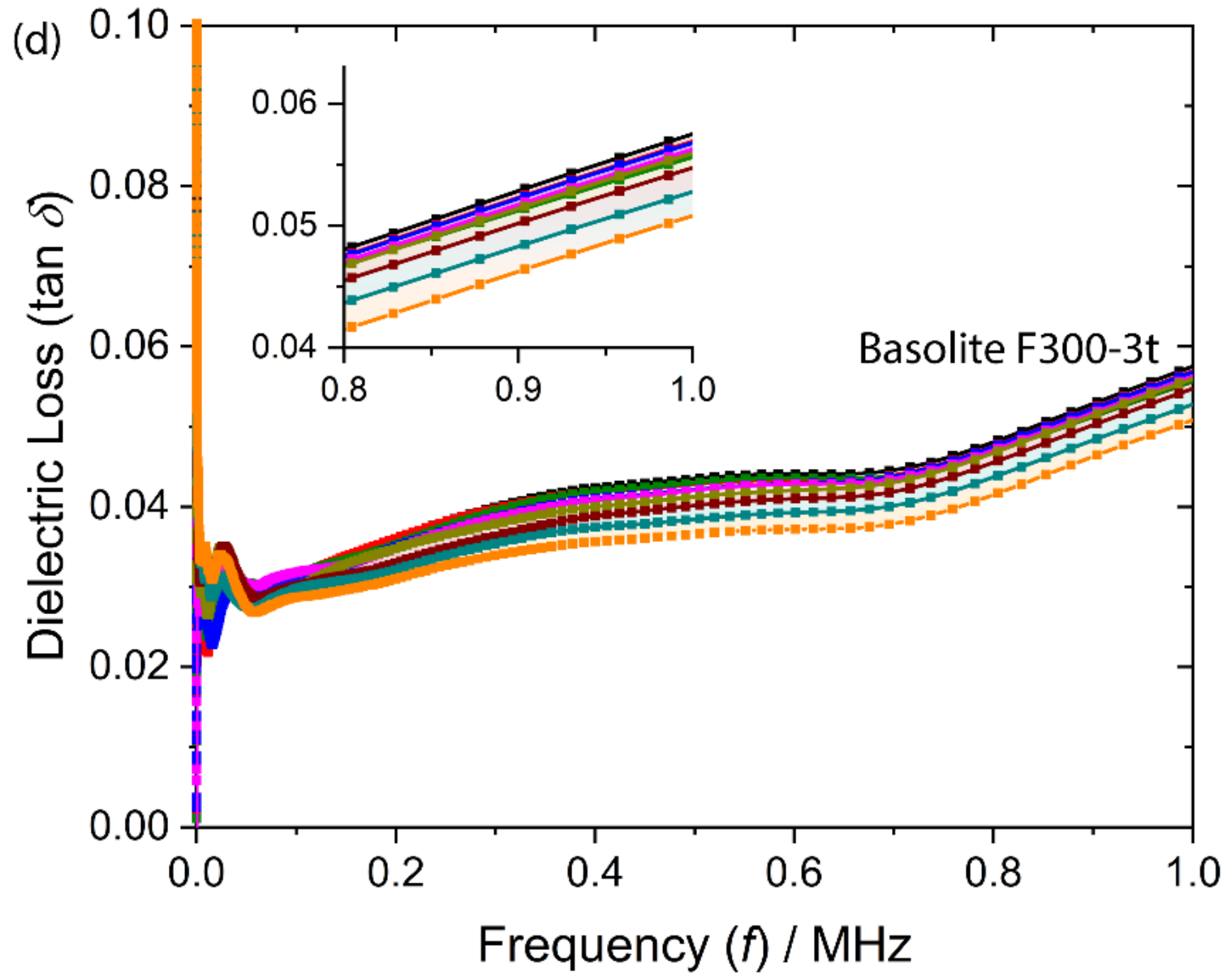
(d)
Dielectric Loss (tan δ)
Frequency (f) / MHz
Basolite F300-3t

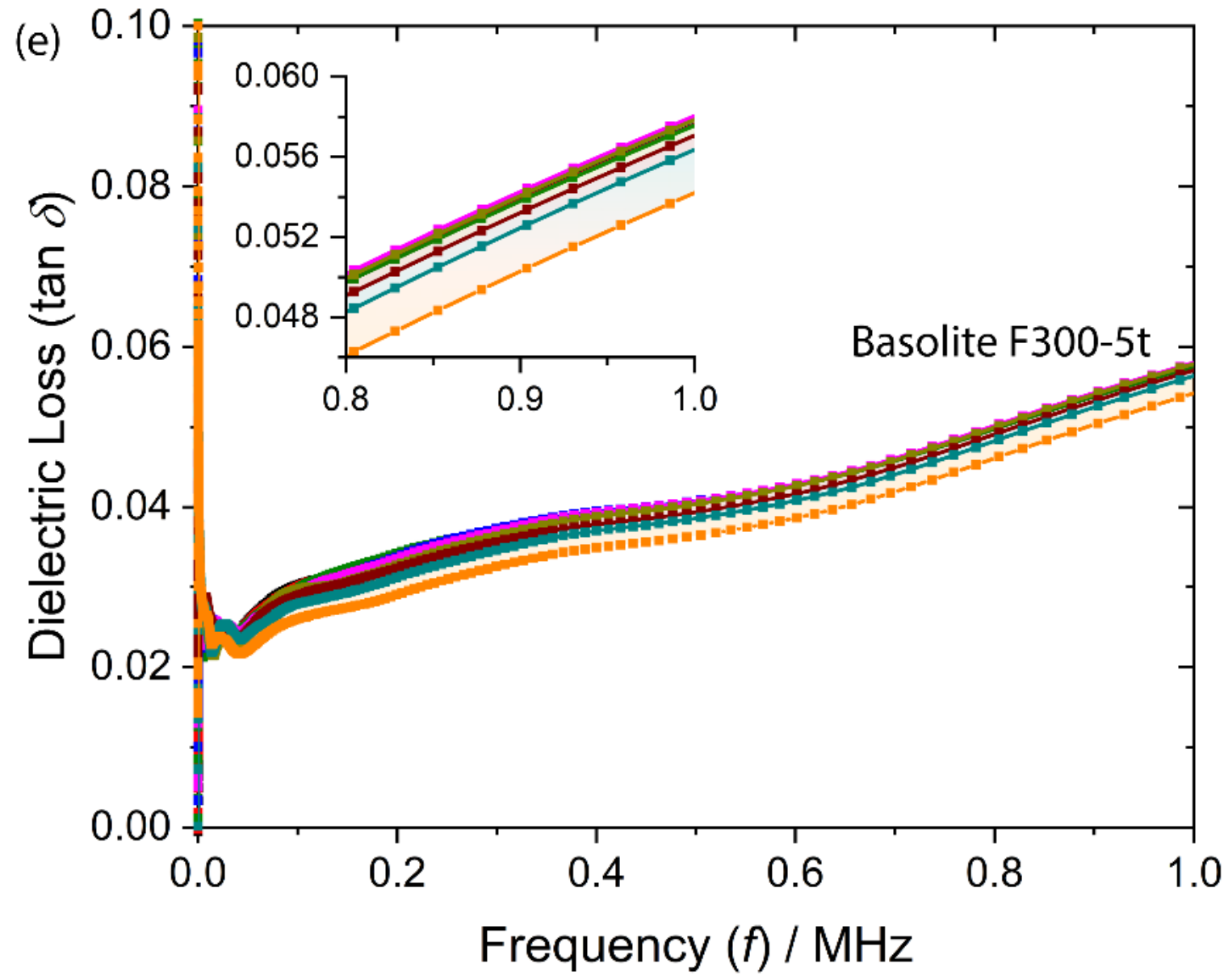
(e)
Dielectric Loss (tan δ)
Frequency (f) / MHz
Basolite F300-5t

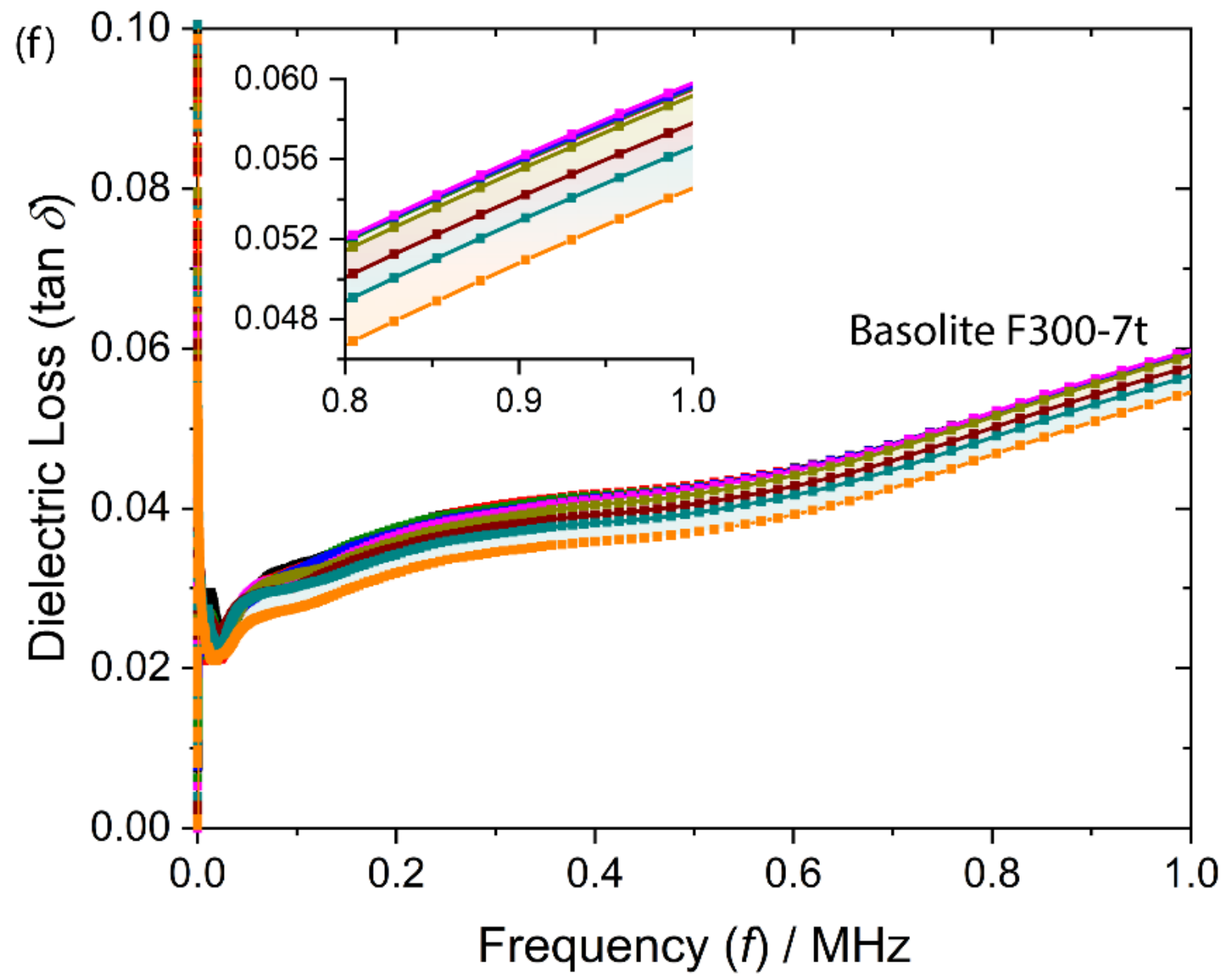
(f)
Dielectric Loss (tan δ)
Frequency (f) / MHz
Basolite F300-7t

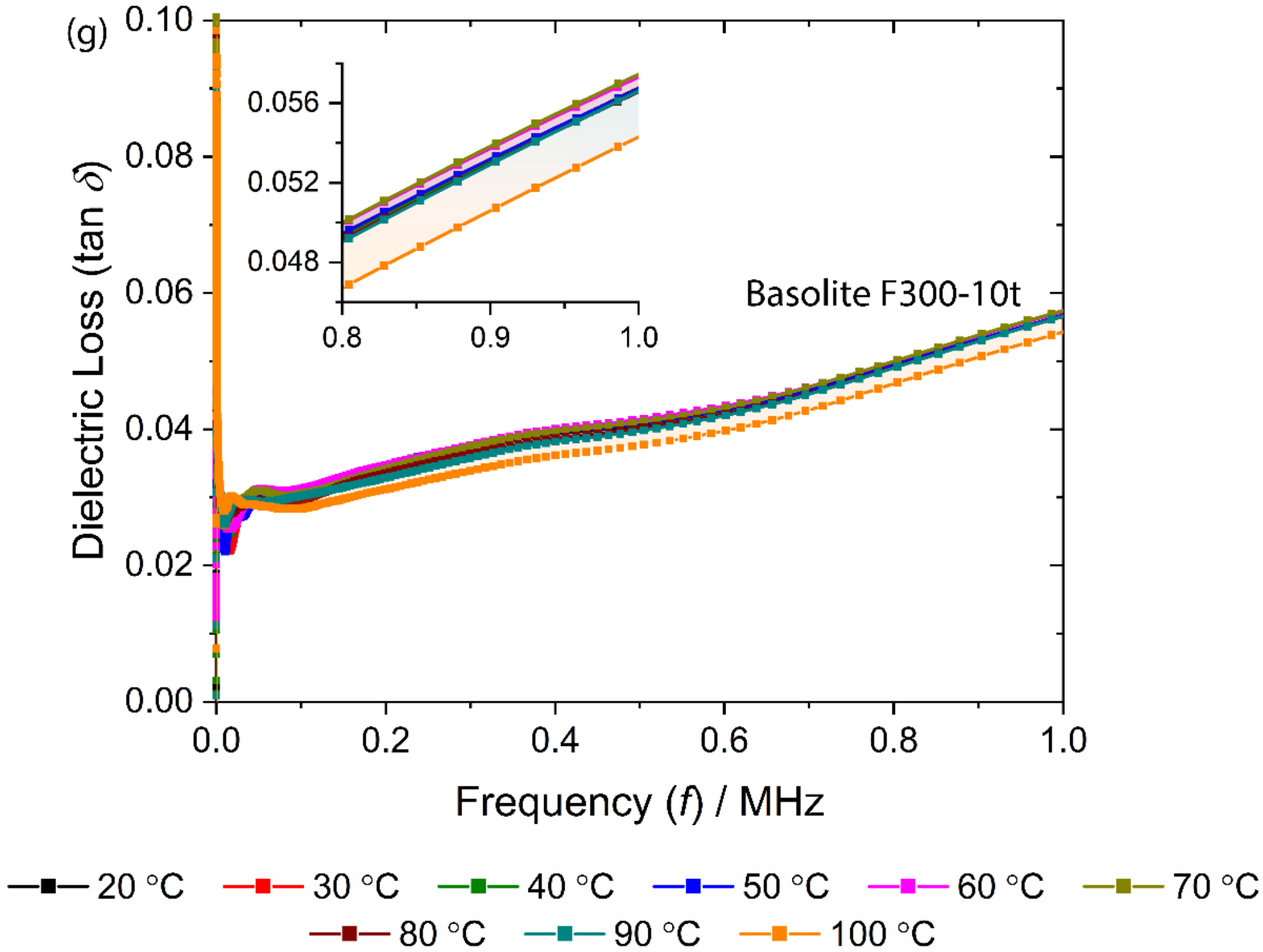

**Figure S6:** The dielectric loss for Basolite F300 pellets prepared under a compression load of: (a) 0.5-ton, (b) 1-ton, (c) 2-ton, (d) 3-ton (e) 5-ton (f) 7-ton, and (g) 10-ton, respectively.

## 4.2 MIL-100-MG

### 4.2.1 Real Part of Dielectric Constant

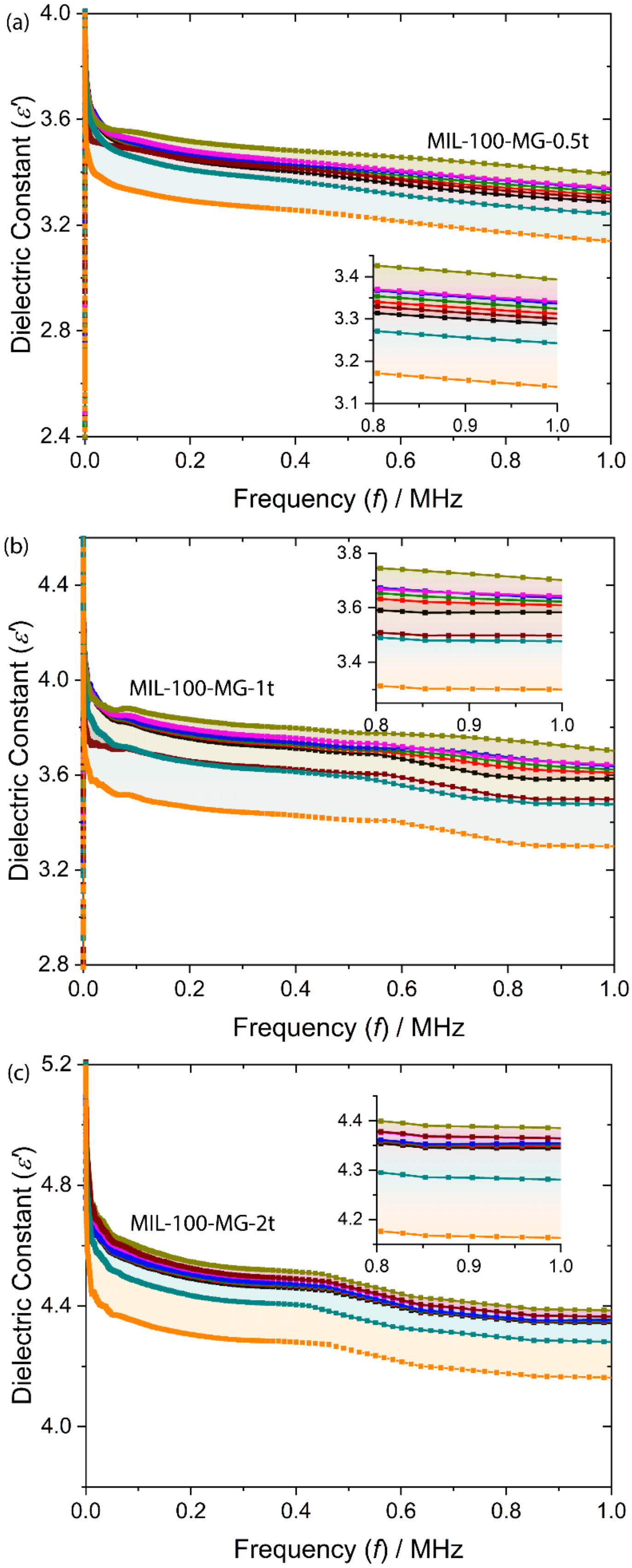

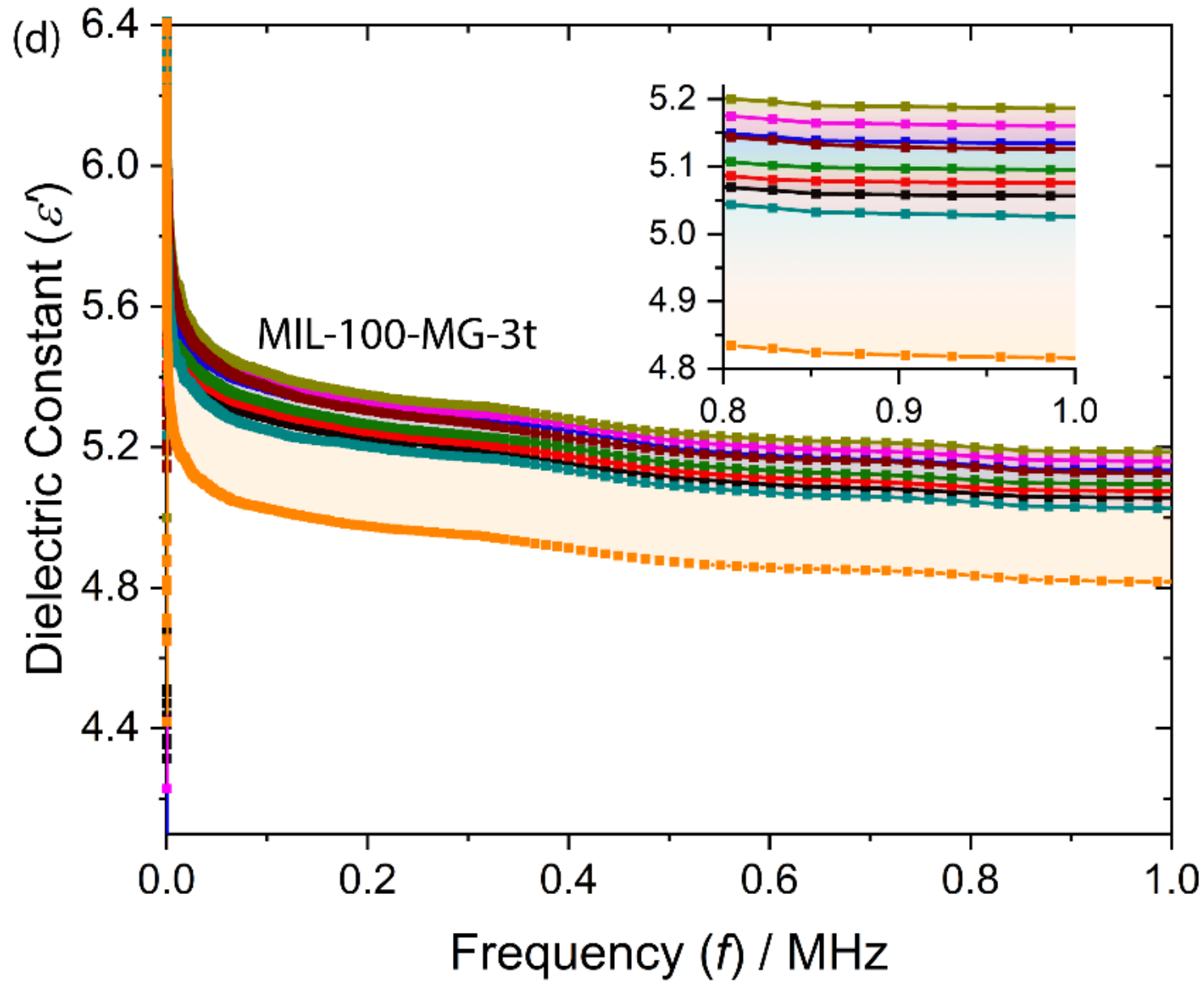
(d)
MIL-100-MG-3t
Dielectric Constant (ε')
Frequency (f) / MHz

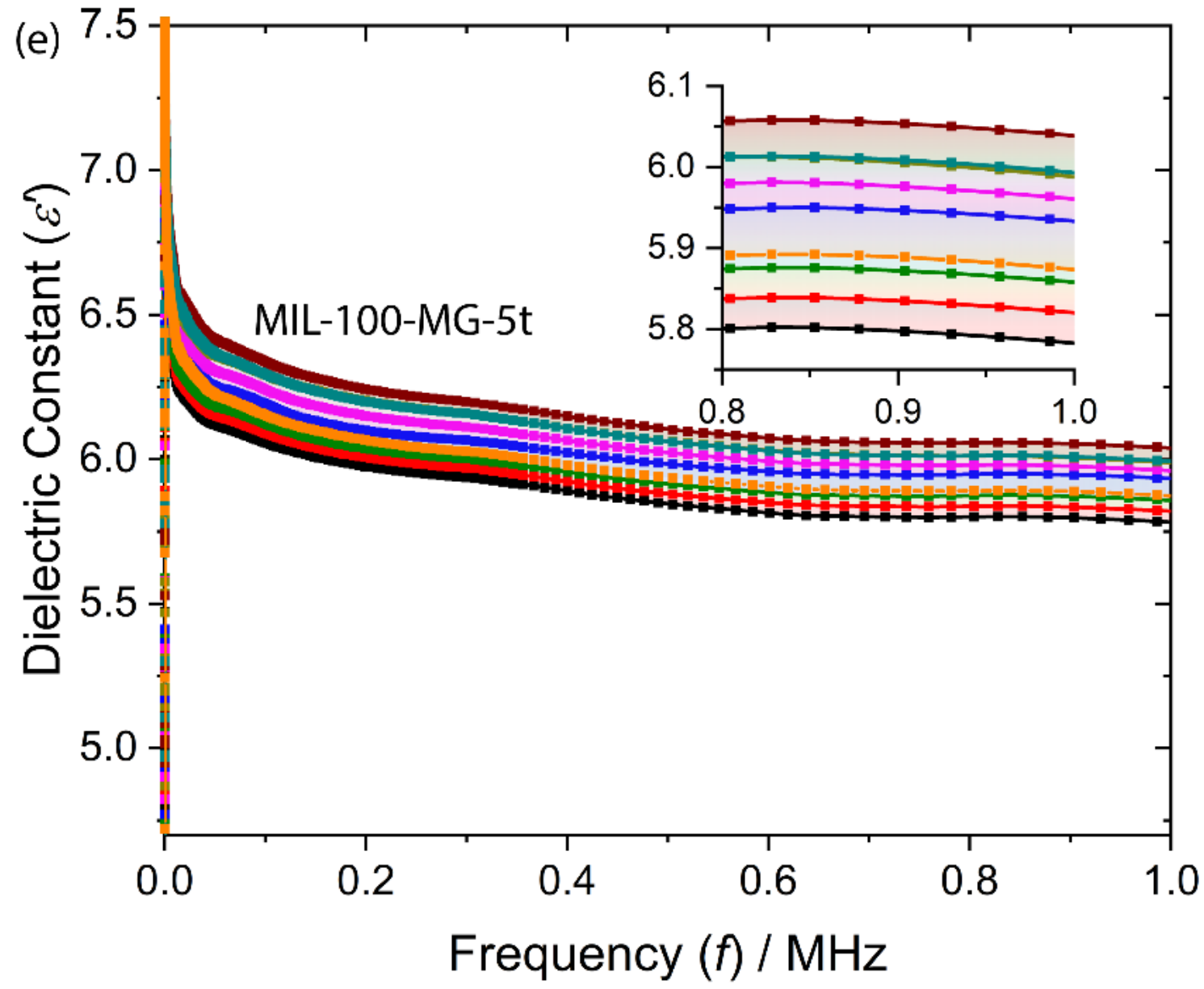
(e)
MIL-100-MG-5t
Dielectric Constant (ε')
Frequency (f) / MHz

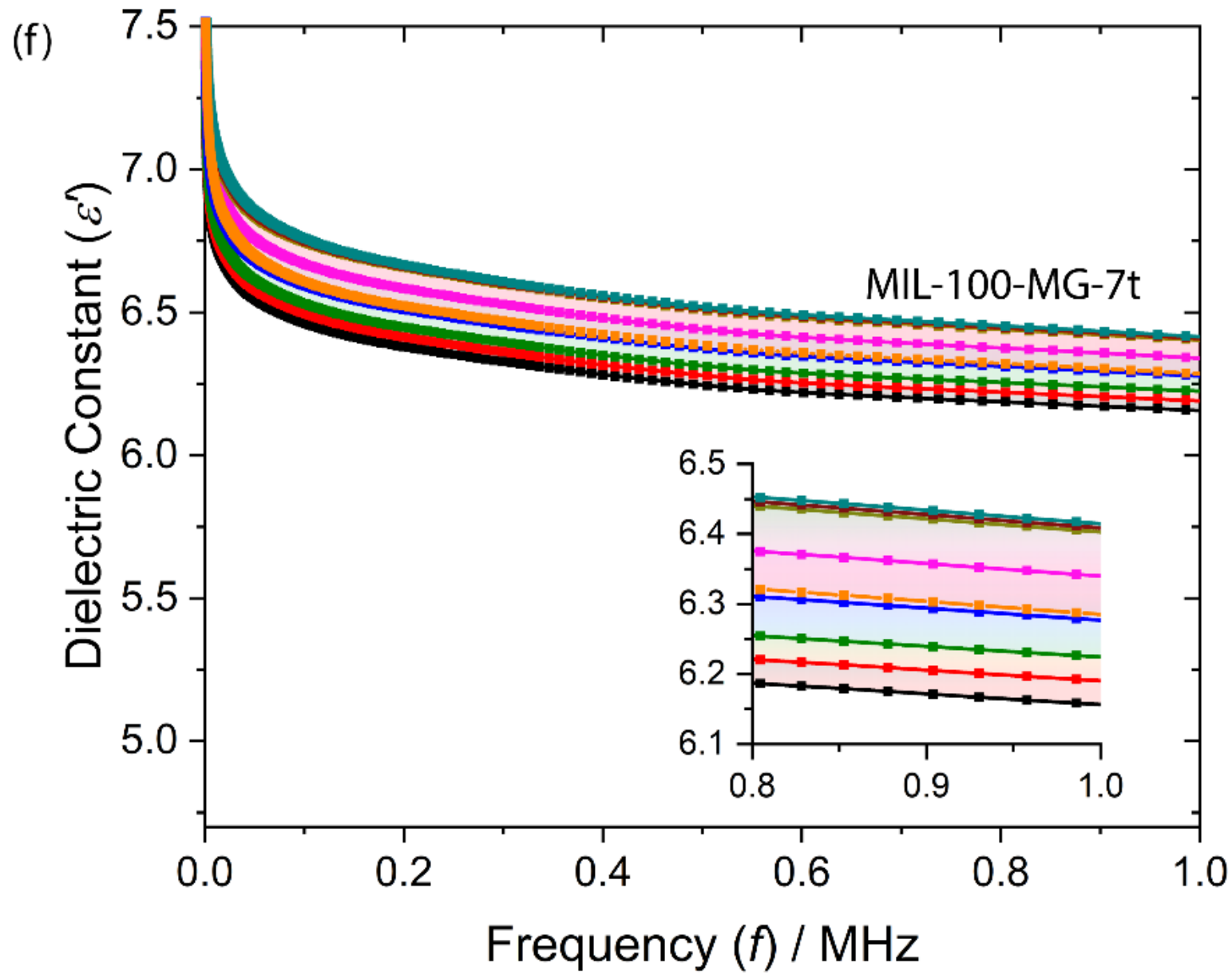
(f)
MIL-100-MG-7t
Dielectric Constant (ε')
Frequency (f) / MHz

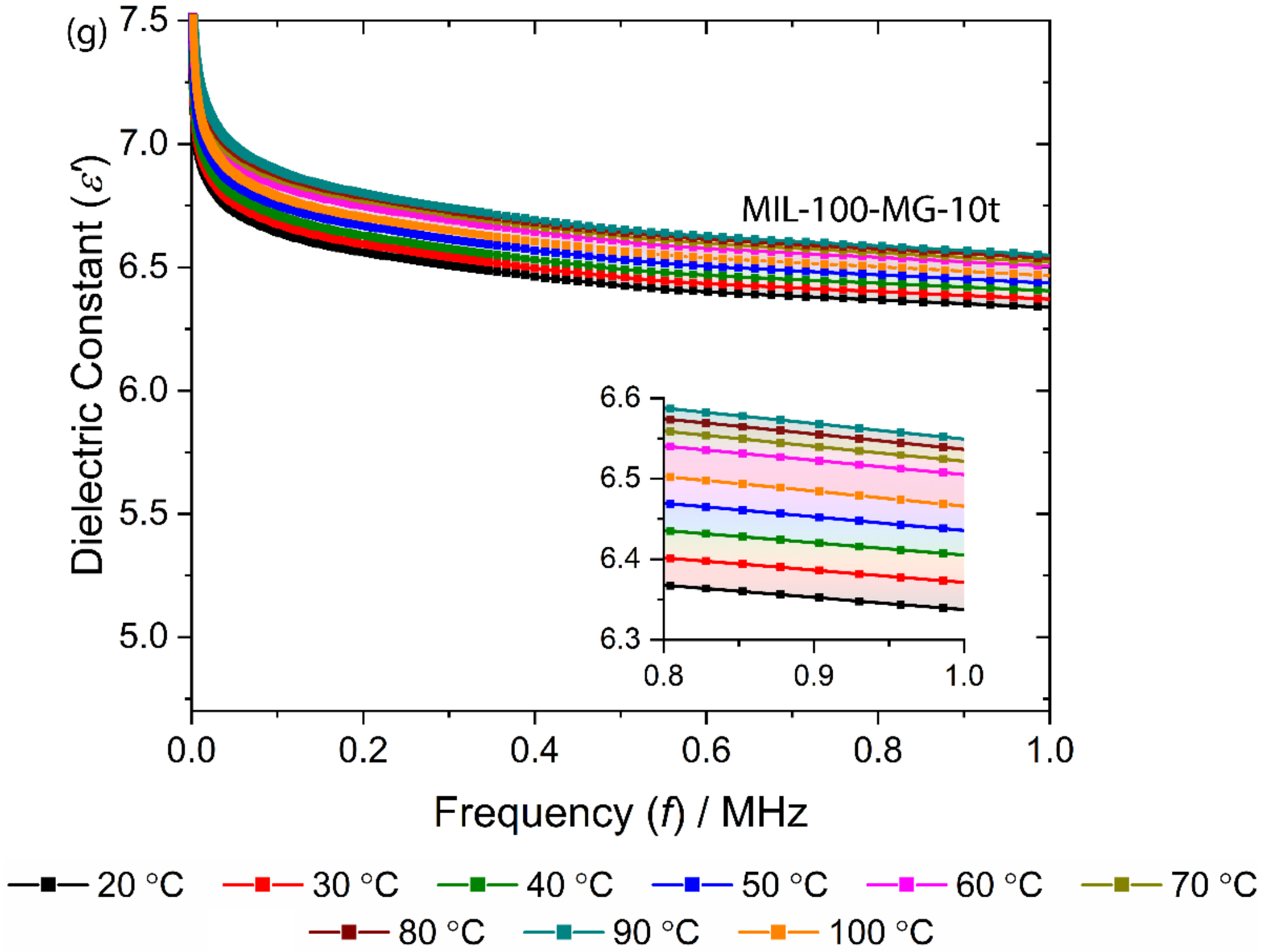

**Figure S7:** The real part of dielectric constant for MIL-100-MG pellets prepared under a compression load of: (a) 0.5-ton, (b) 1-ton, (c) 2-ton, (d) 3-ton (e) 5-ton (f) 7-ton, and (g) 10-ton, corresponding to the pressure of 36.96, 73.92, 147.84, 221.76, 369.6, 517.44 and 739.20 MPa, respectively.

### 4.2.2 Imaginary Part of Dielectric Constant

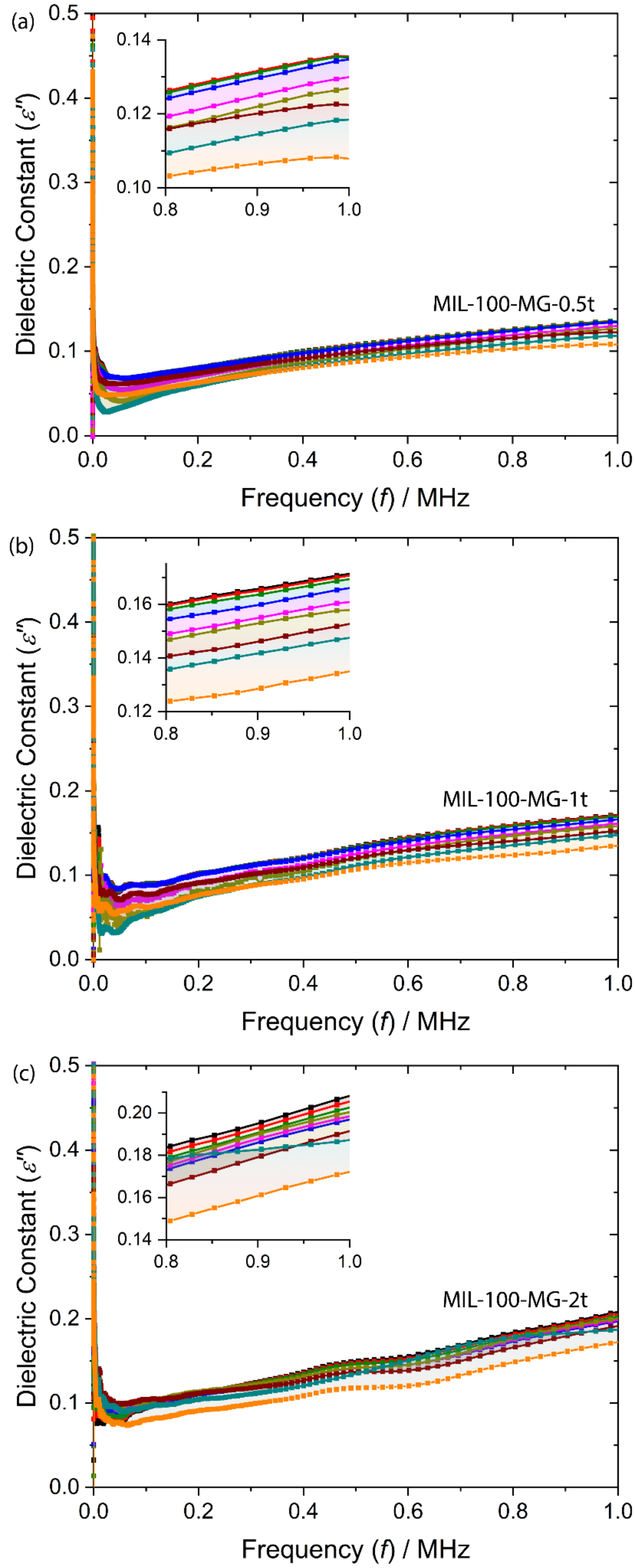

(d)

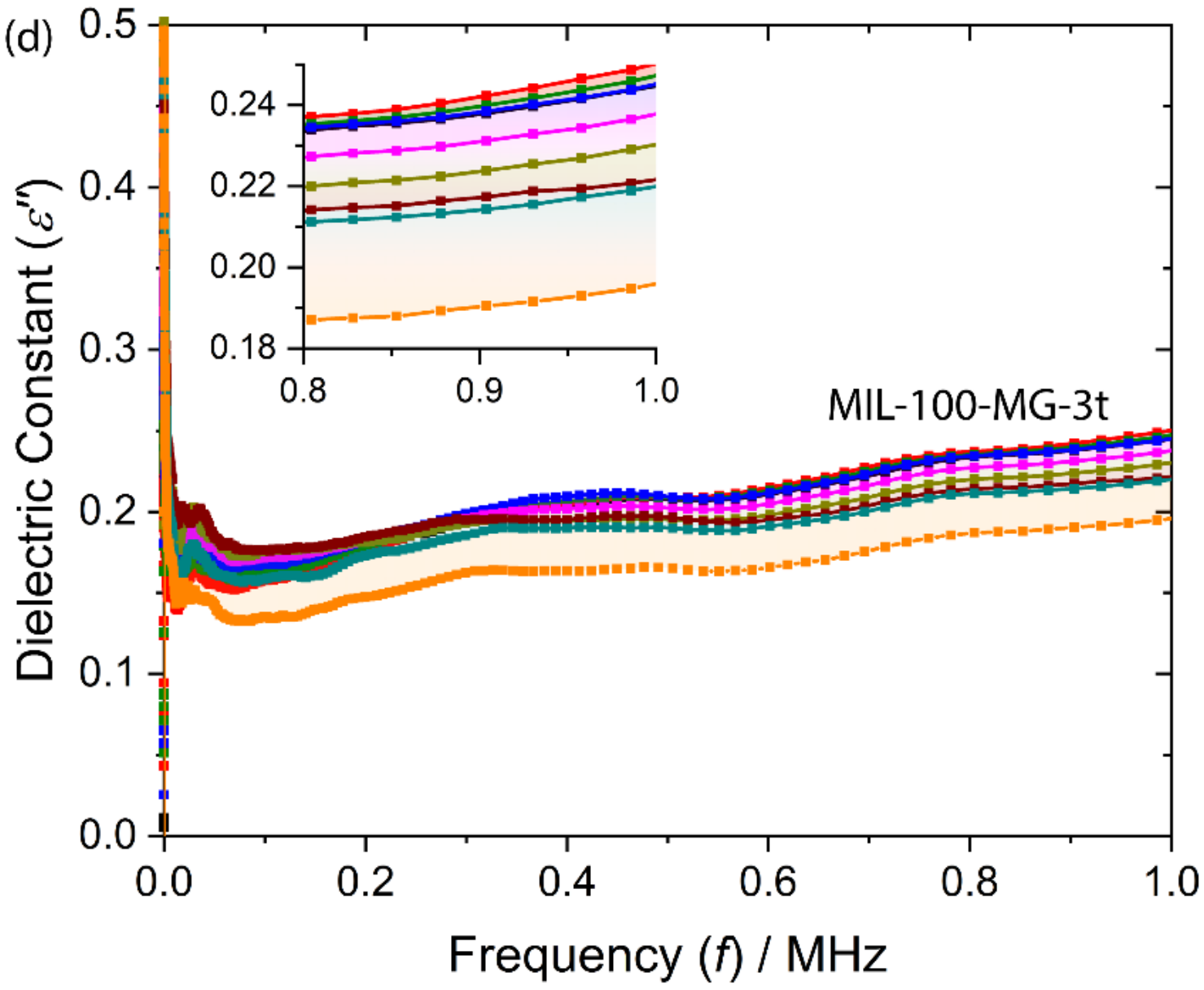
MIL-100-MG-3t
Dielectric Constant (ε'')
Frequency (f) / MHz

(e)

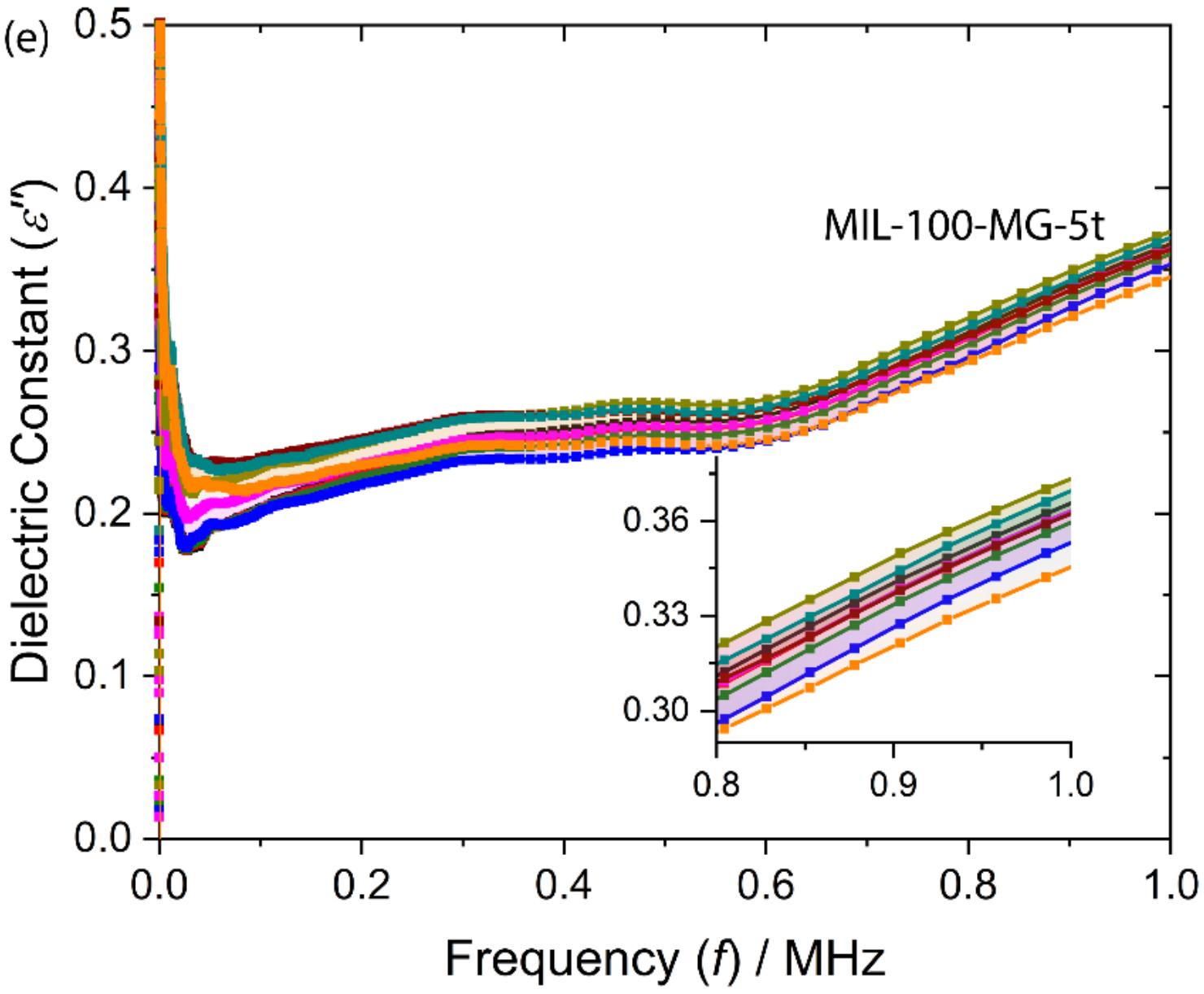
MIL-100-MG-5t
Dielectric Constant (ε'')
Frequency (f) / MHz

(f)

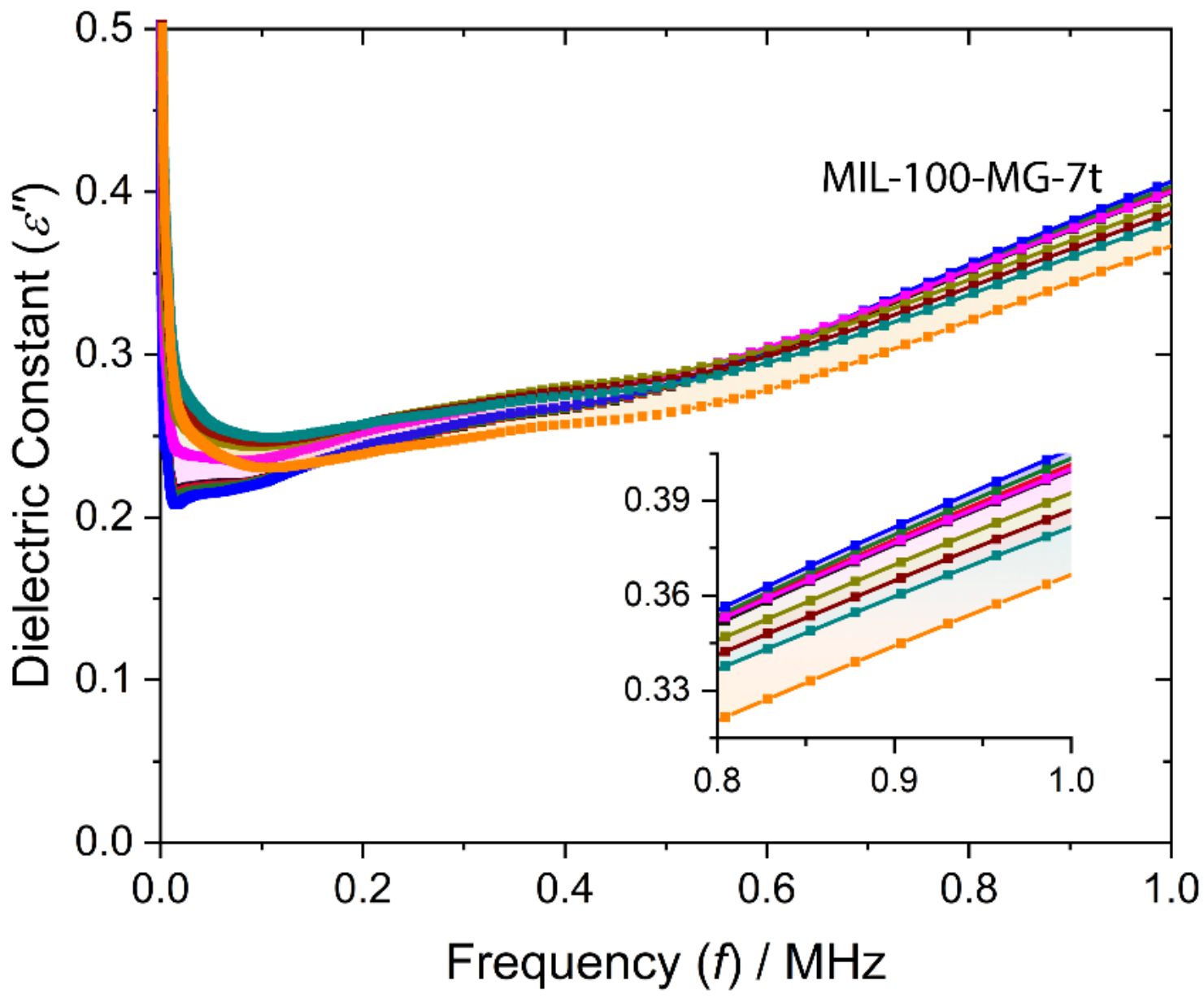
MIL-100-MG-7t
Dielectric Constant (ε'')
Frequency (f) / MHz

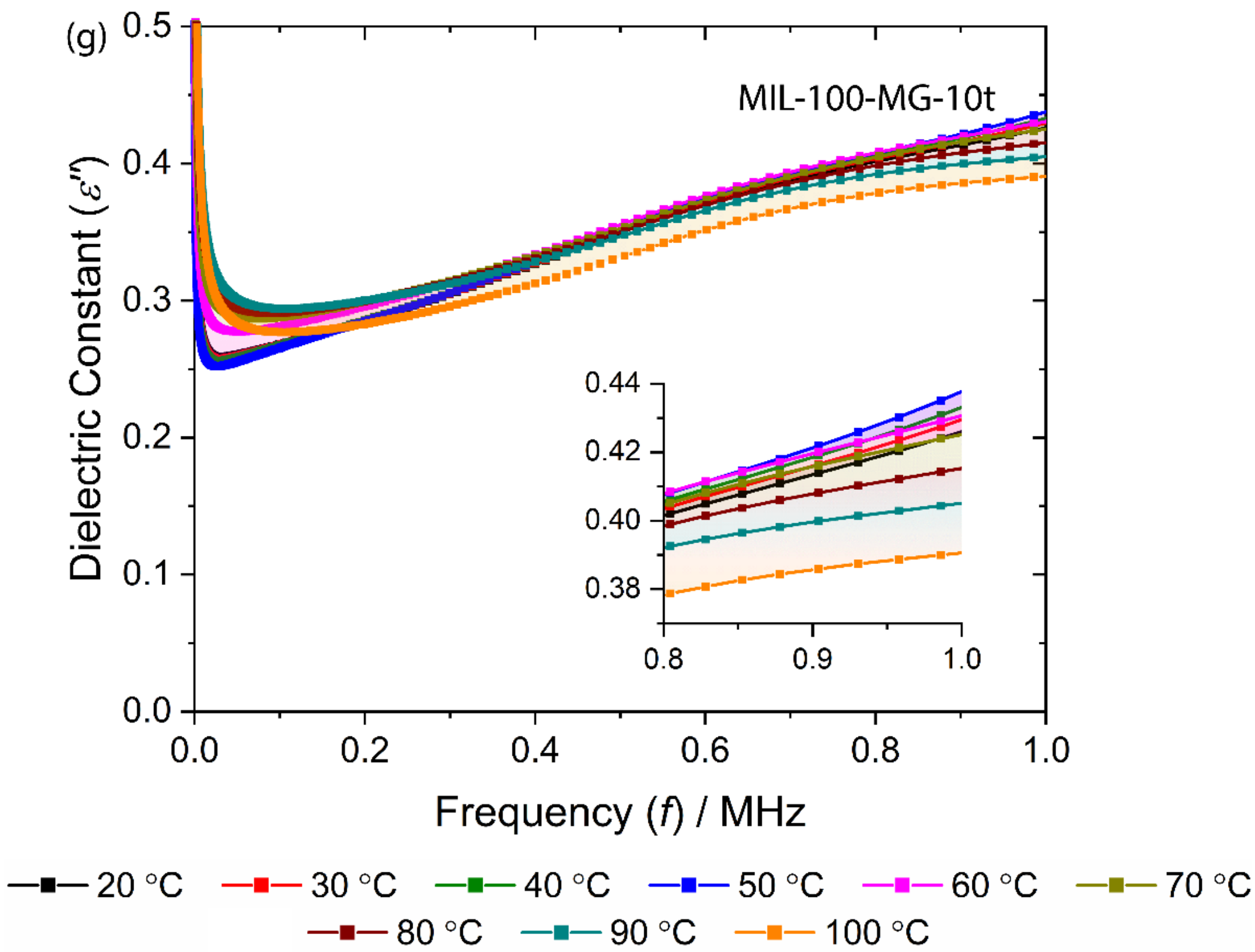

**Figure S8:** The imaginary part of dielectric constant for MIL-100-MG pellets prepared under a compression load of: (a) 0.5-ton, (b) 1-ton, (c) 2-ton, (d) 3-ton (e) 5-ton (f) 7-ton, and (g) 10-ton, respectively.

### 4.2.3 Dielectric Loss

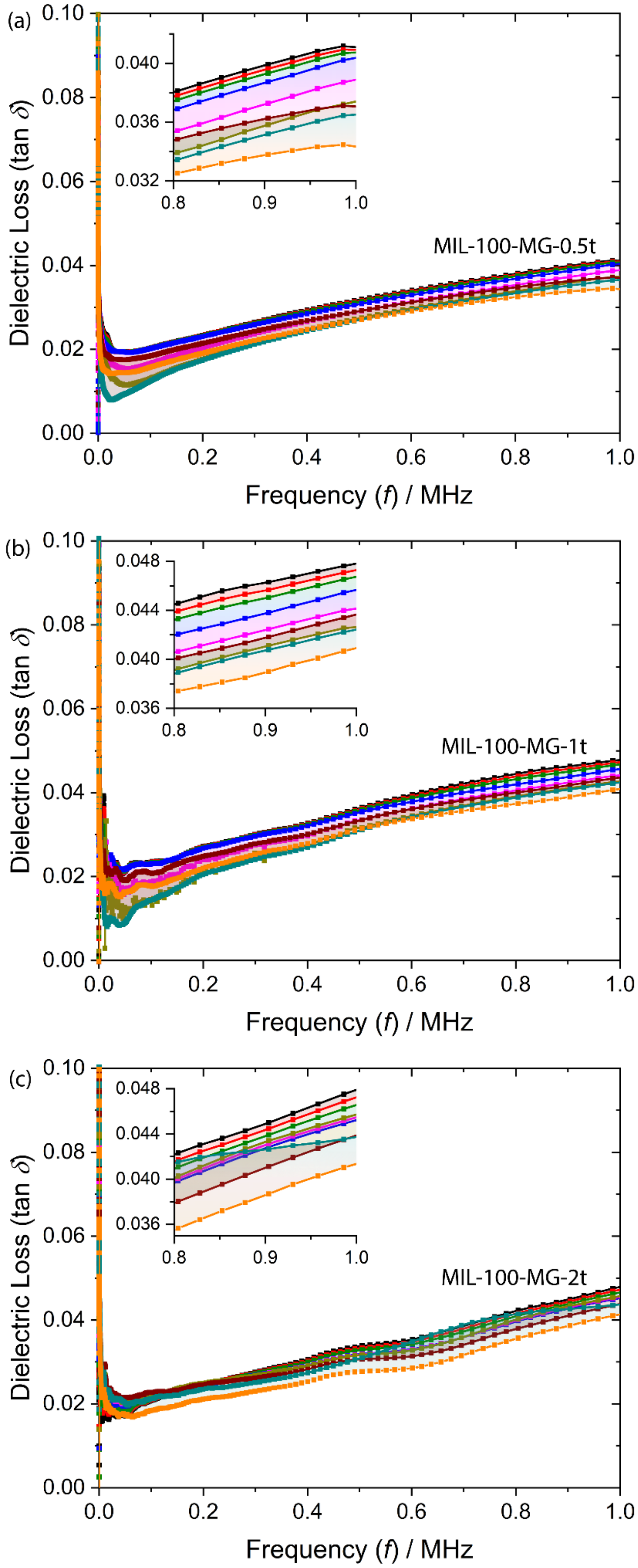

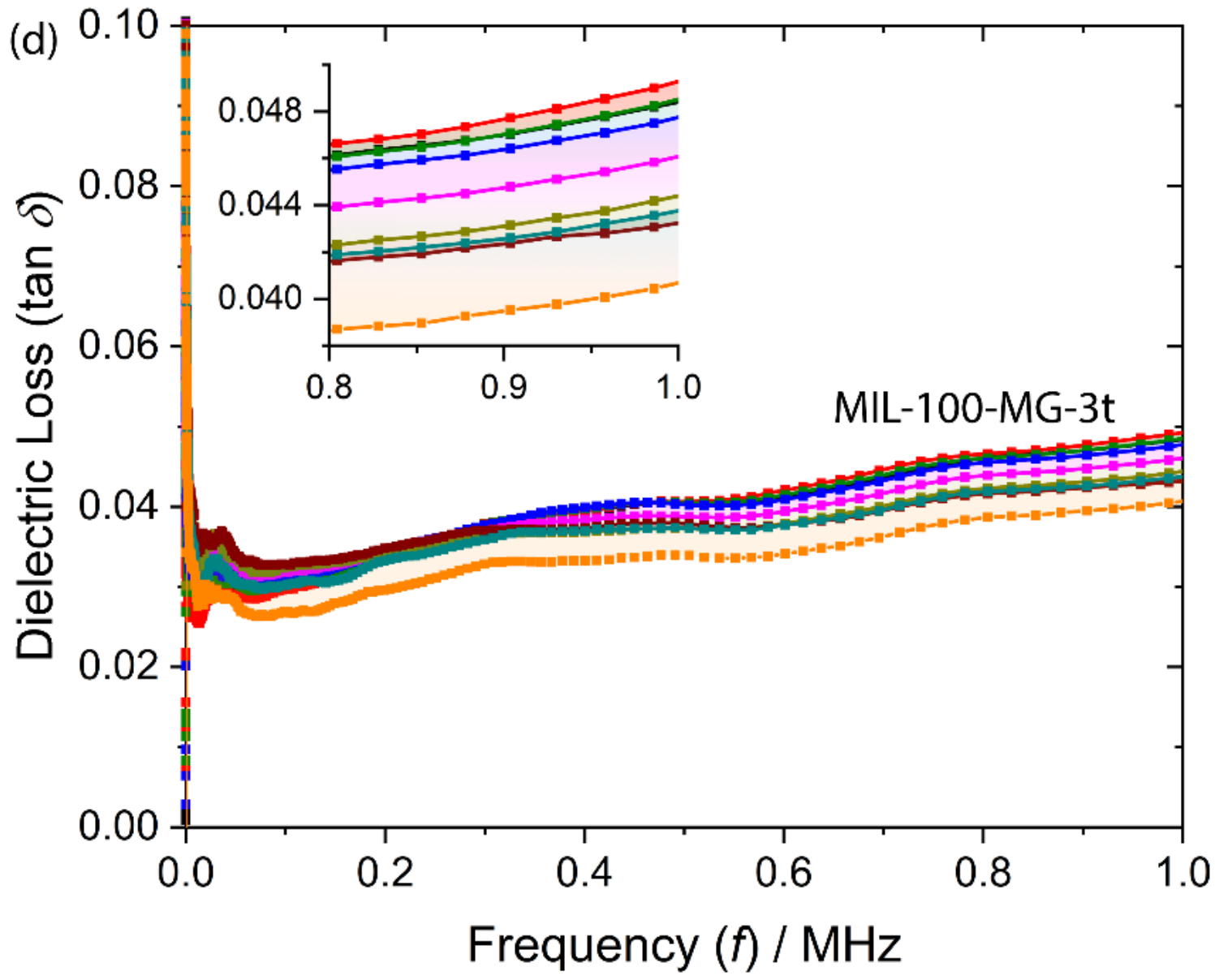
(d)
Dielectric Loss (tan δ)
Frequency (f) / MHz
MIL-100-MG-3t

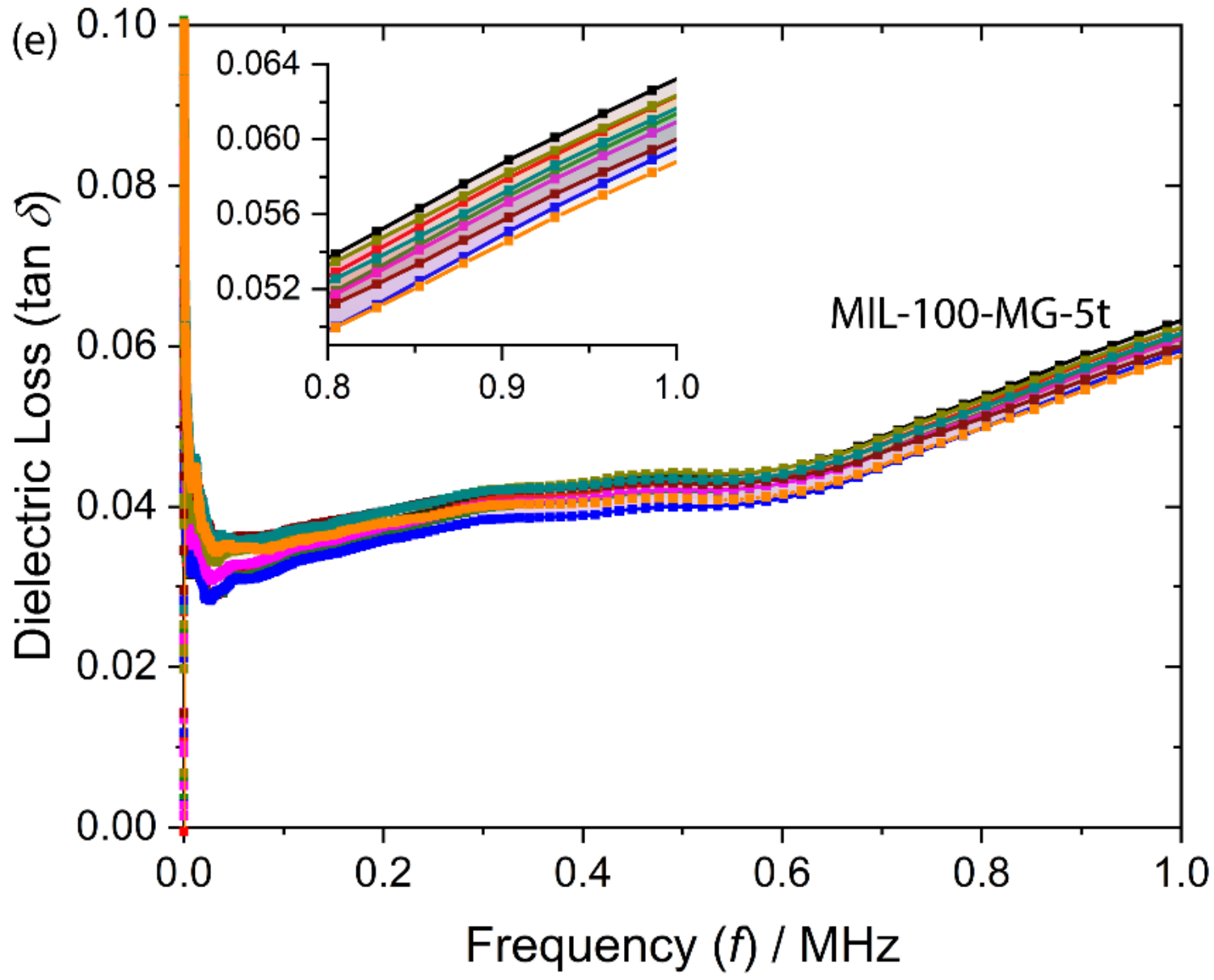
(e)
Dielectric Loss (tan δ)
Frequency (f) / MHz
MIL-100-MG-5t

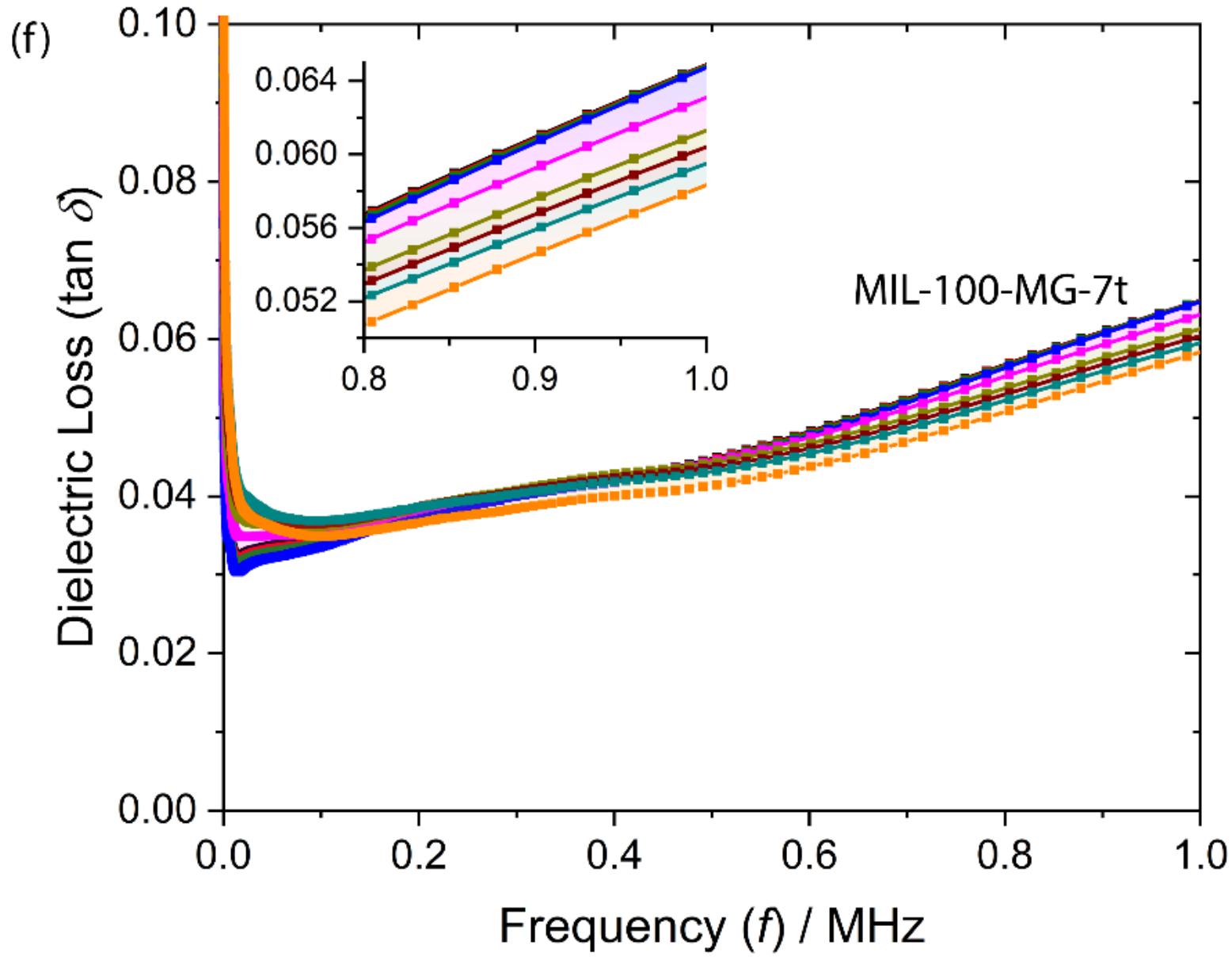
(f)
Dielectric Loss (tan δ)
Frequency (f) / MHz
MIL-100-MG-7t

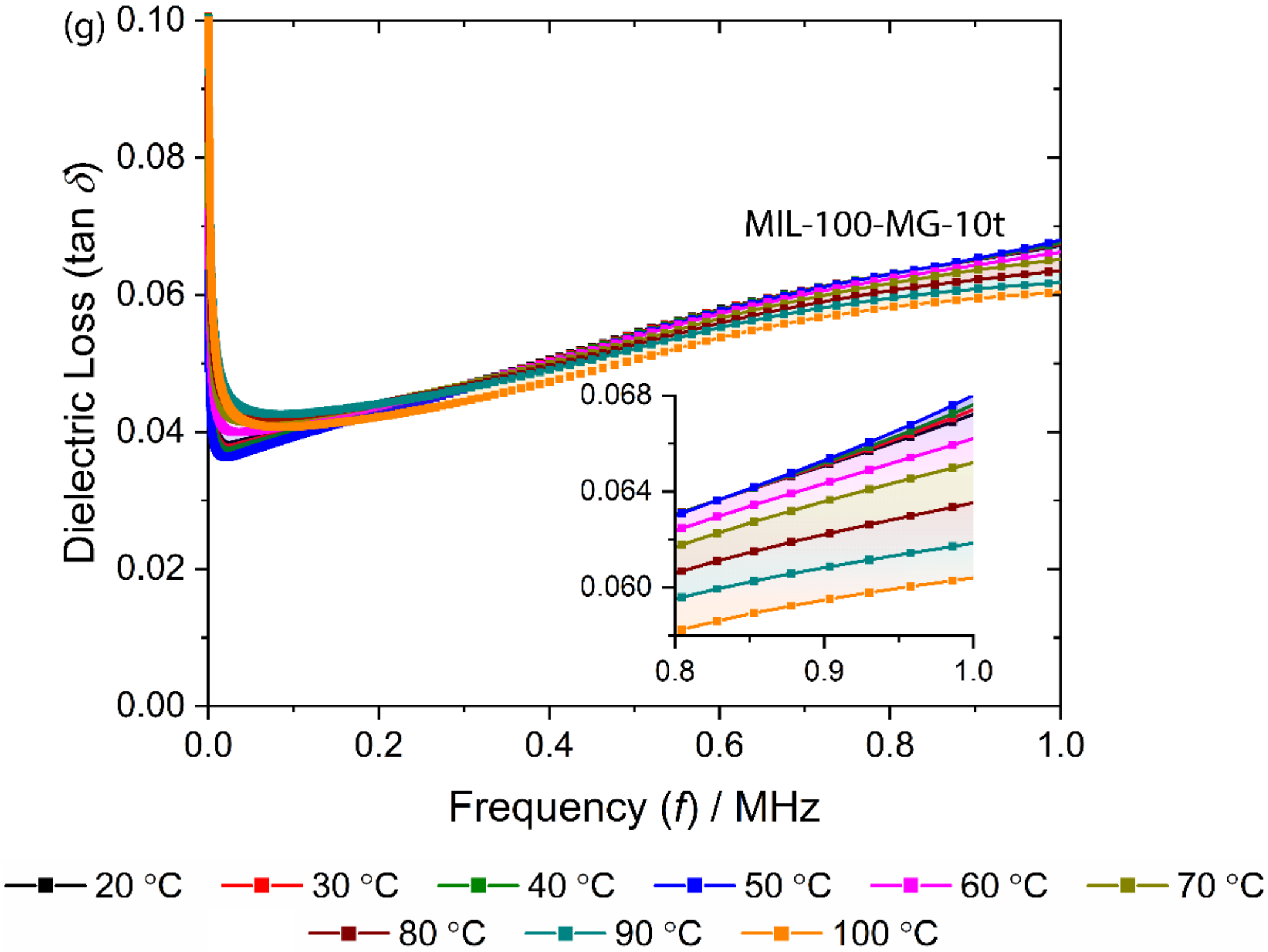

**Figure S9:** The dielectric loss for MIL-100-MG pellets prepared under a compression load of: (a) 0.5-ton, (b) 1-ton, (c) 2-ton, (d) 3-ton (e) 5-ton (f) 7-ton, and (g) 10-ton, respectively.

### 4.3 Comparative dieletric loss

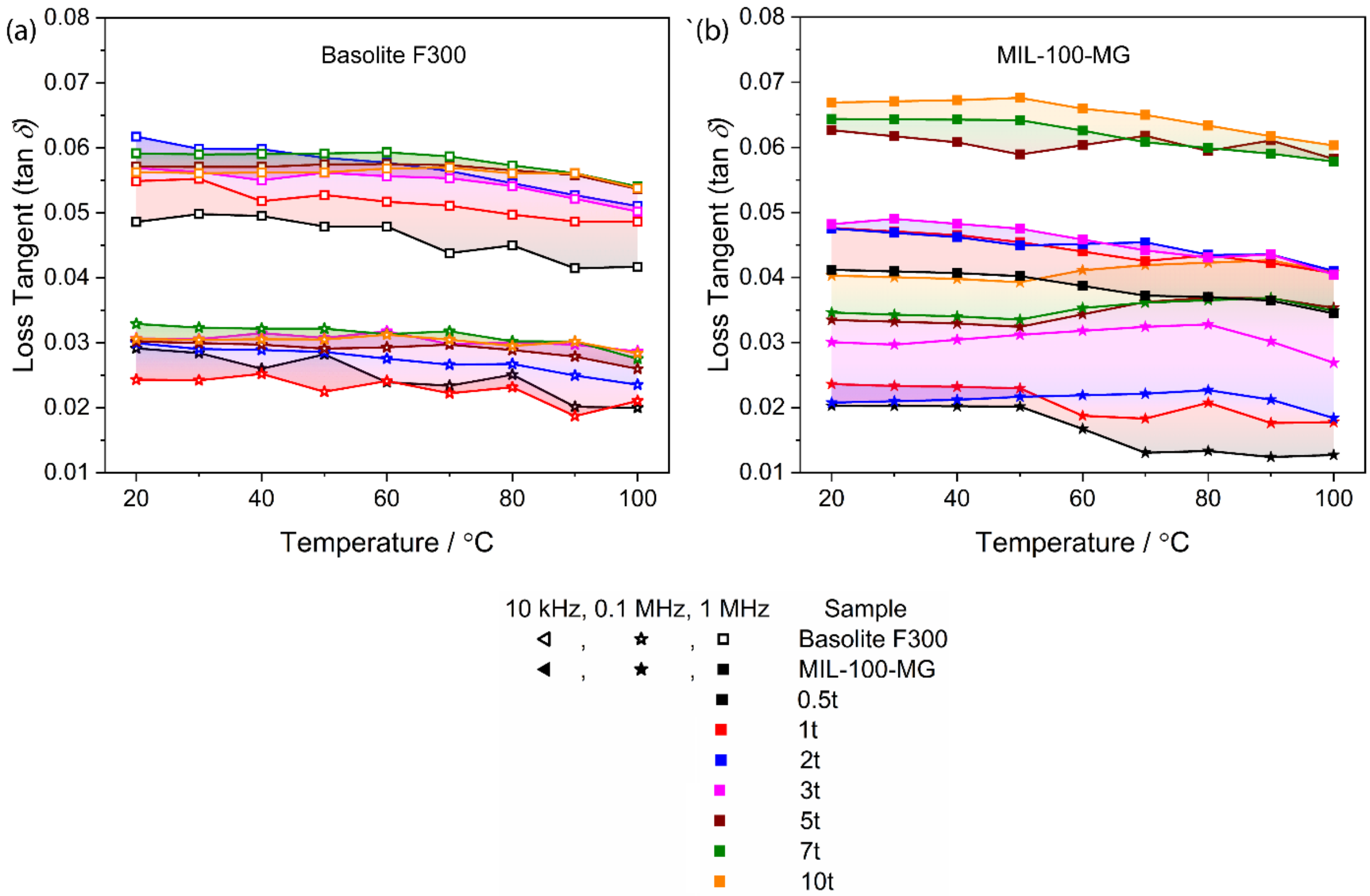

**Figure S10:** Dielectric loss of MOF pellets as a function of pelleting pressure and temperature: (a) Basolite F300 and, (b) MIL-100-MG pellets specific frequencies (0.01, 0.1, and 1 MHz).

## 5. Reflectivity spectra R(ω) in the far-IR and mid-IR regions

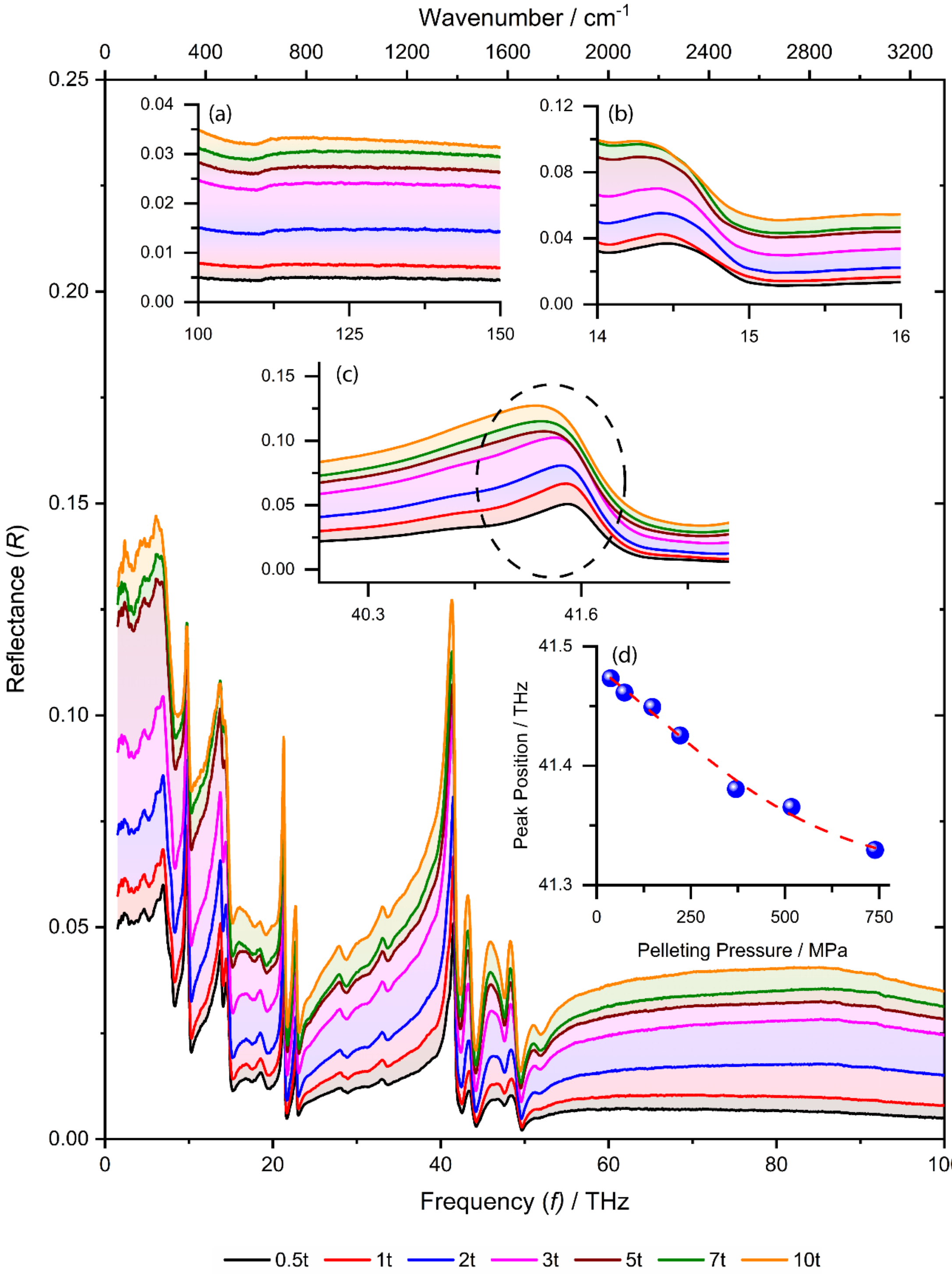

**Figure S11:** Reflectance spectra of MIL-100-MG pellets. Inset: (a) reflectivity spectra in the near-IR region, (b) joining of the far-IR and mid-IR spectra at 517 $cm^{-1}$ (~15.5 THz), (c) pelleting pressure-dependent redshift in transition mode of peaks obtained from the Gaussian peak fitting and (d) plot for pressure-dependent peak shift in peak positions.

## 6. Refractive index in THz region

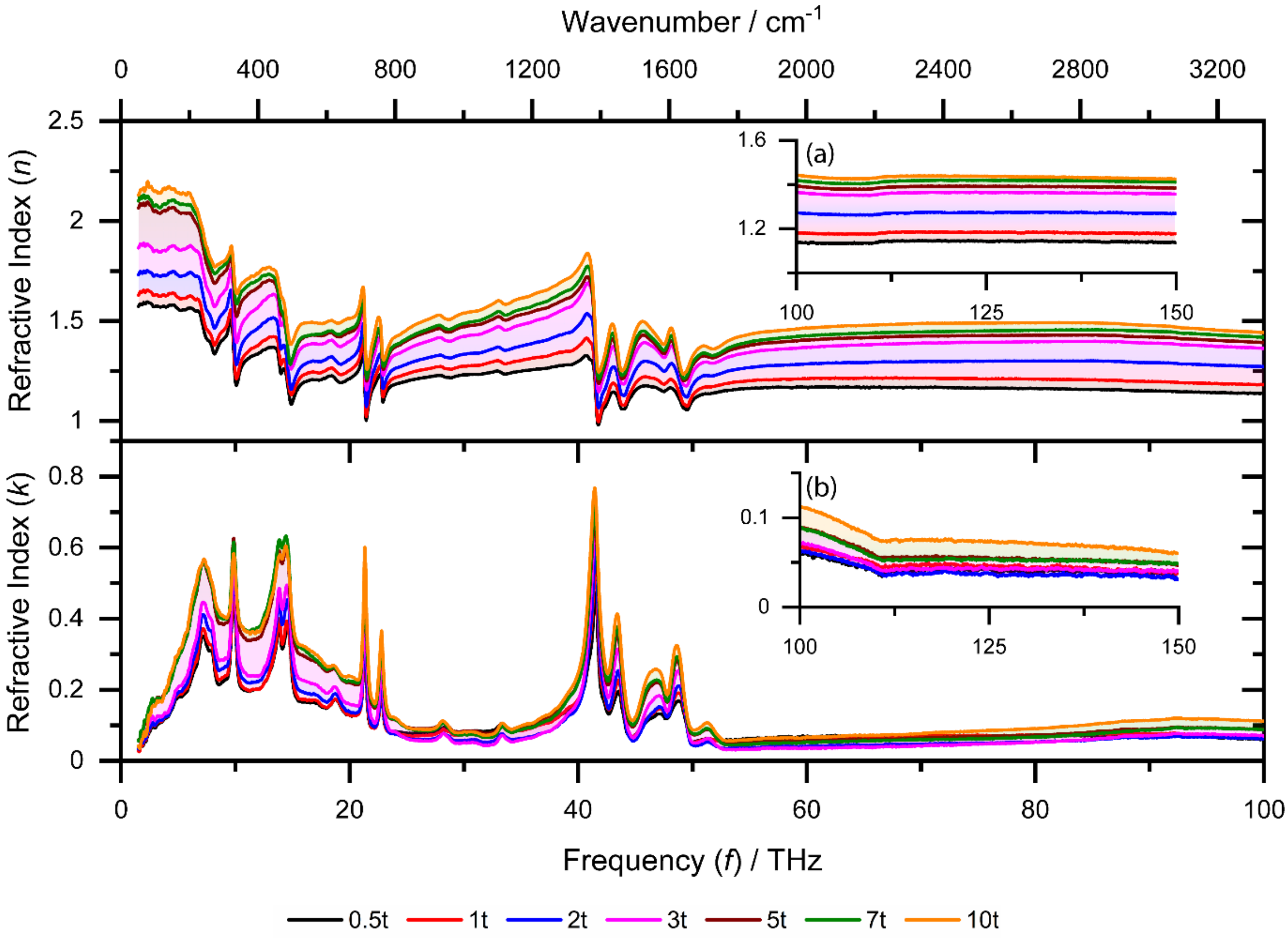

**Figure S12:** (a) Real and (b) imaginary parts of the refractive index of MIL-100 pellets in IR frequency range. Inset (c)-(d) shows the optically insensitivity of the framework in near-IR region.

## 7. The imaginary part of dielectric constant in the THz region

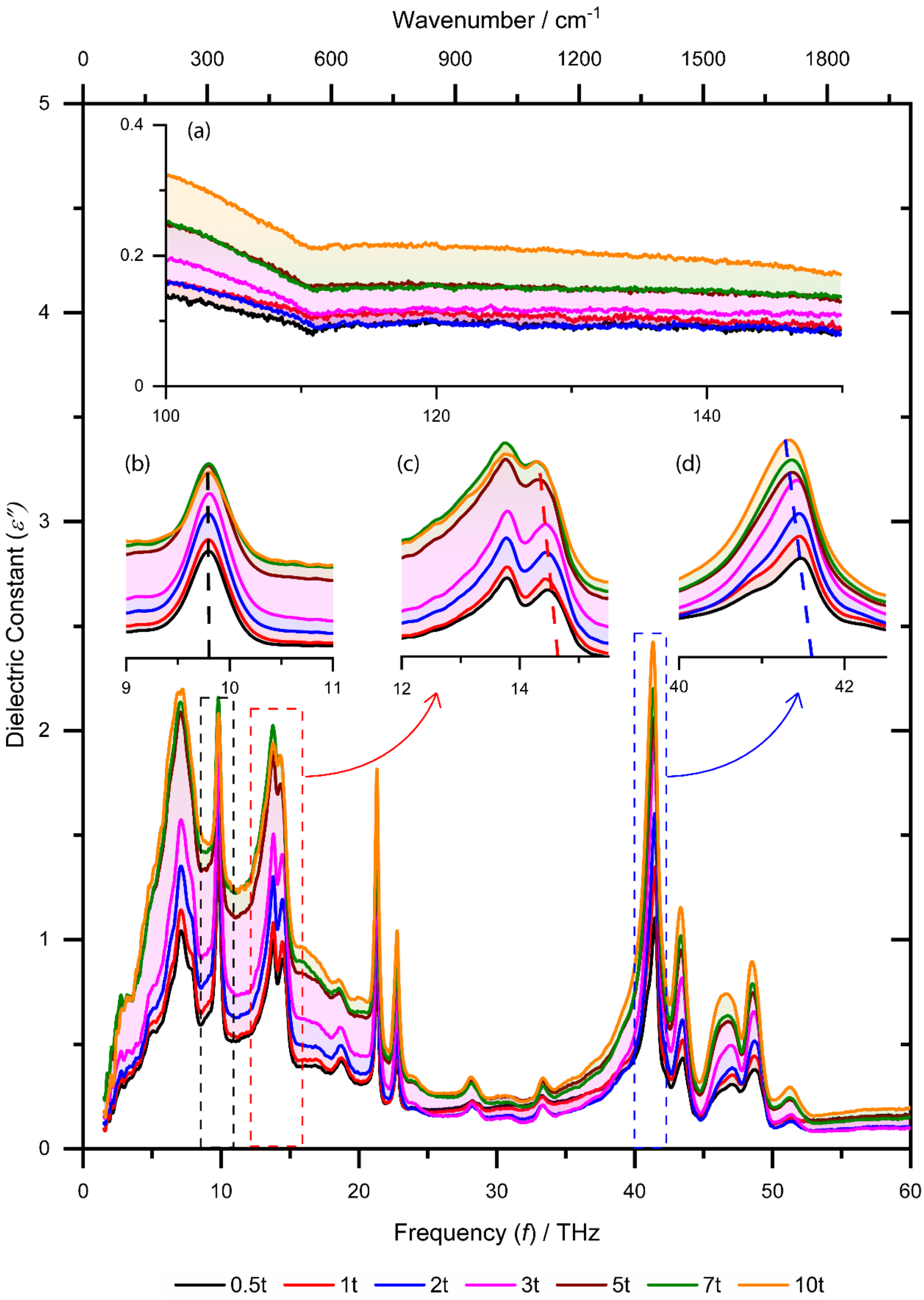

**Figure S13:** The imaginary part of dielectric constant ($\varepsilon''$) in the IR frequency range for MIL-100 pellets. Inset (a) shows the $\varepsilon''$ spectrum in the near-IR region. Inset (b) doesn't show any shift in the transition mode, whereas, in inset (c)-(d) the redshift is evident in the transition modes caused by the pelleting force-induced amorphization.

# 8. AC conductivity

## 8.1 Basolite F300

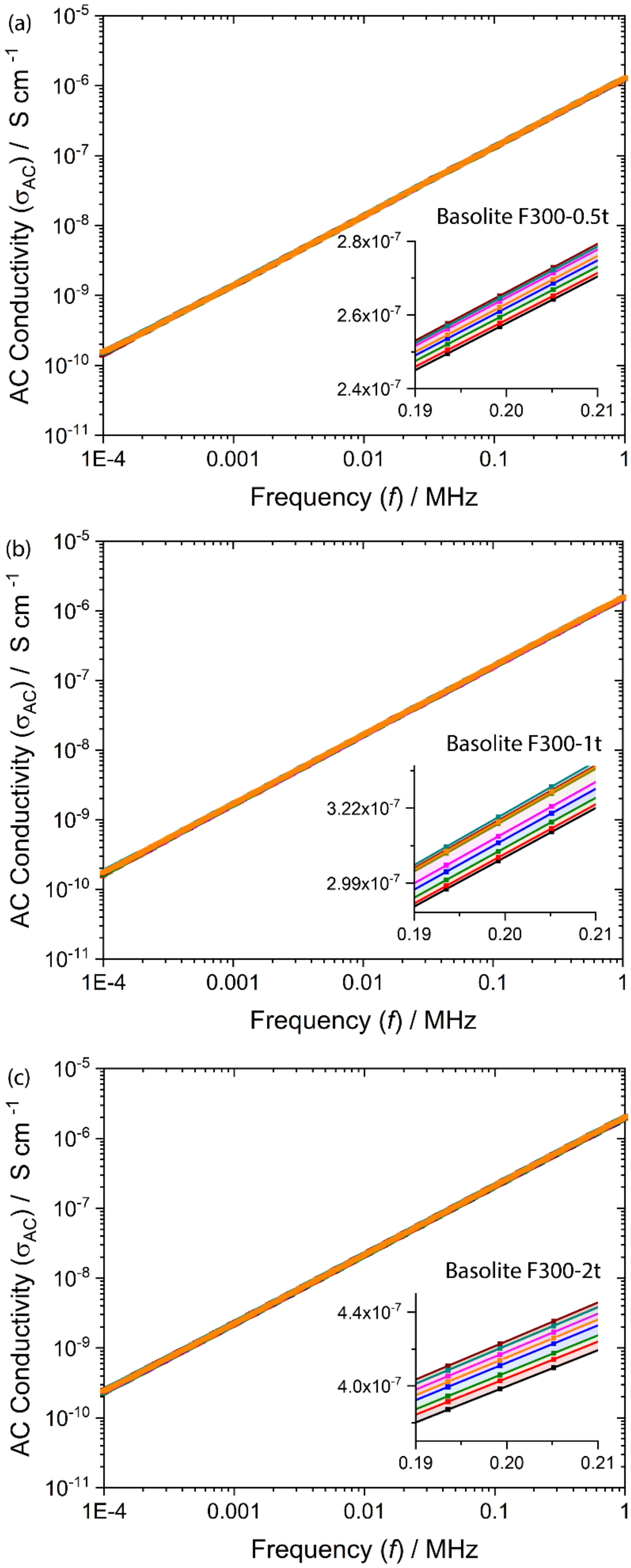

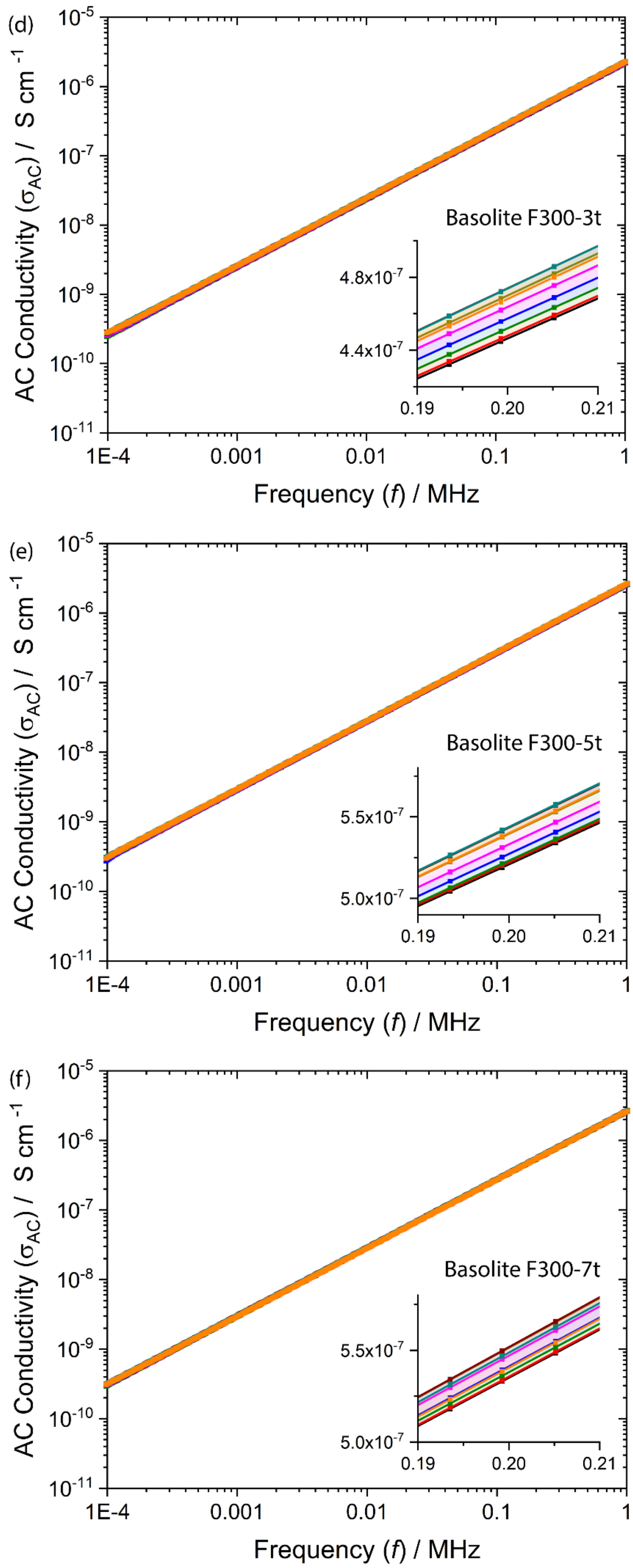
(d)
AC Conductivity ($\sigma_{AC}$) / S cm$^{-1}$
Frequency (*f*) / MHz
Basolite F300-3t
(e)
AC Conductivity ($\sigma_{AC}$) / S cm$^{-1}$
Frequency (*f*) / MHz
Basolite F300-5t
(f)
AC Conductivity ($\sigma_{AC}$) / S cm$^{-1}$
Frequency (*f*) / MHz
Basolite F300-7t

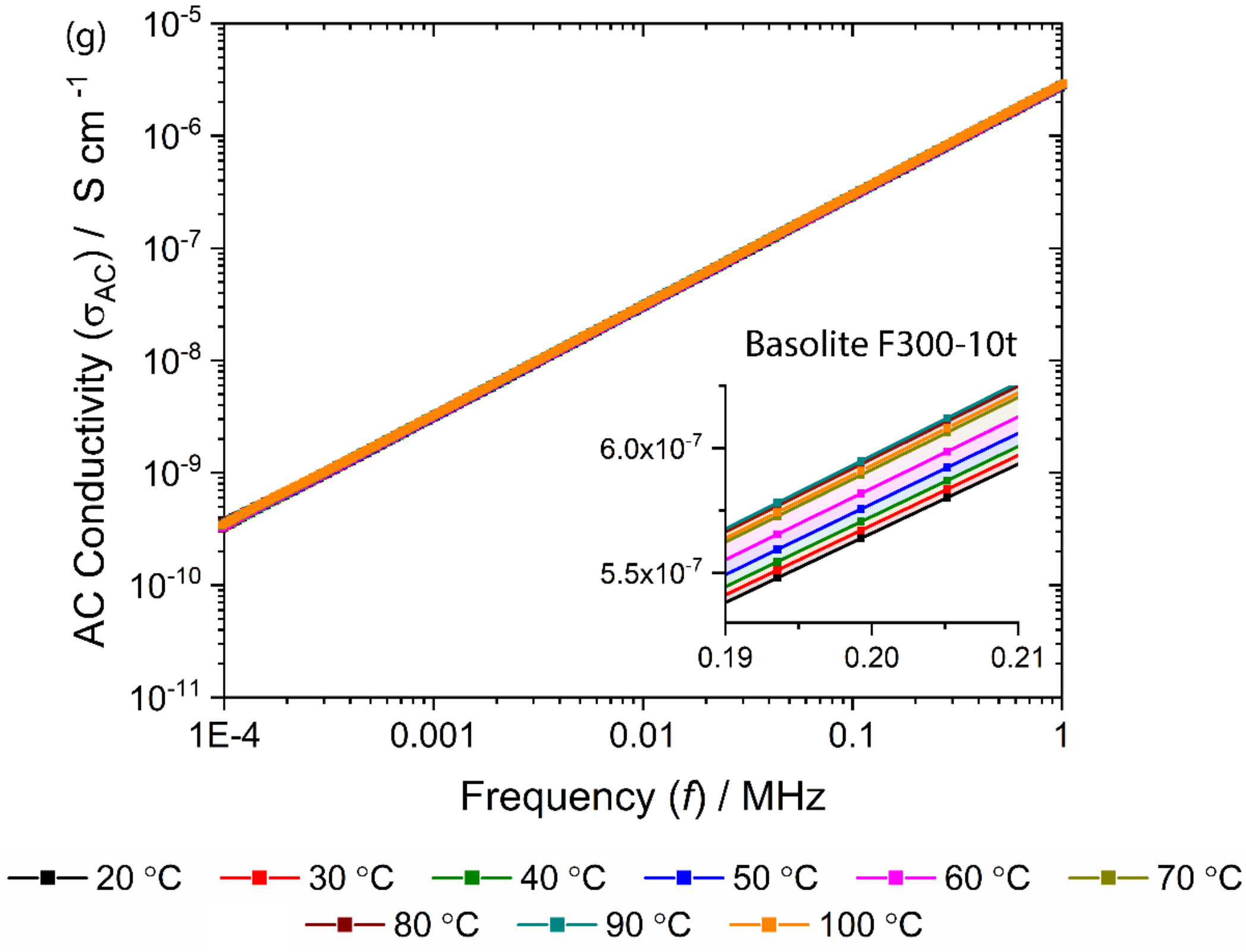

**Figure S14:** The AC conductivity of Basolite F300 pellets prepared under a compression load of: (a) 0.5-ton, (b) 1-ton, (c) 2-ton, (d) 3-ton (e) 5-ton (f) 7-ton, and (g) 10-ton, corresponding to the pressure of 36.96, 73.92, 147.84, 221.76, 369.6, 517.44 and 739.20 MPa, respectively.

## 8.2 MIL-100-MG

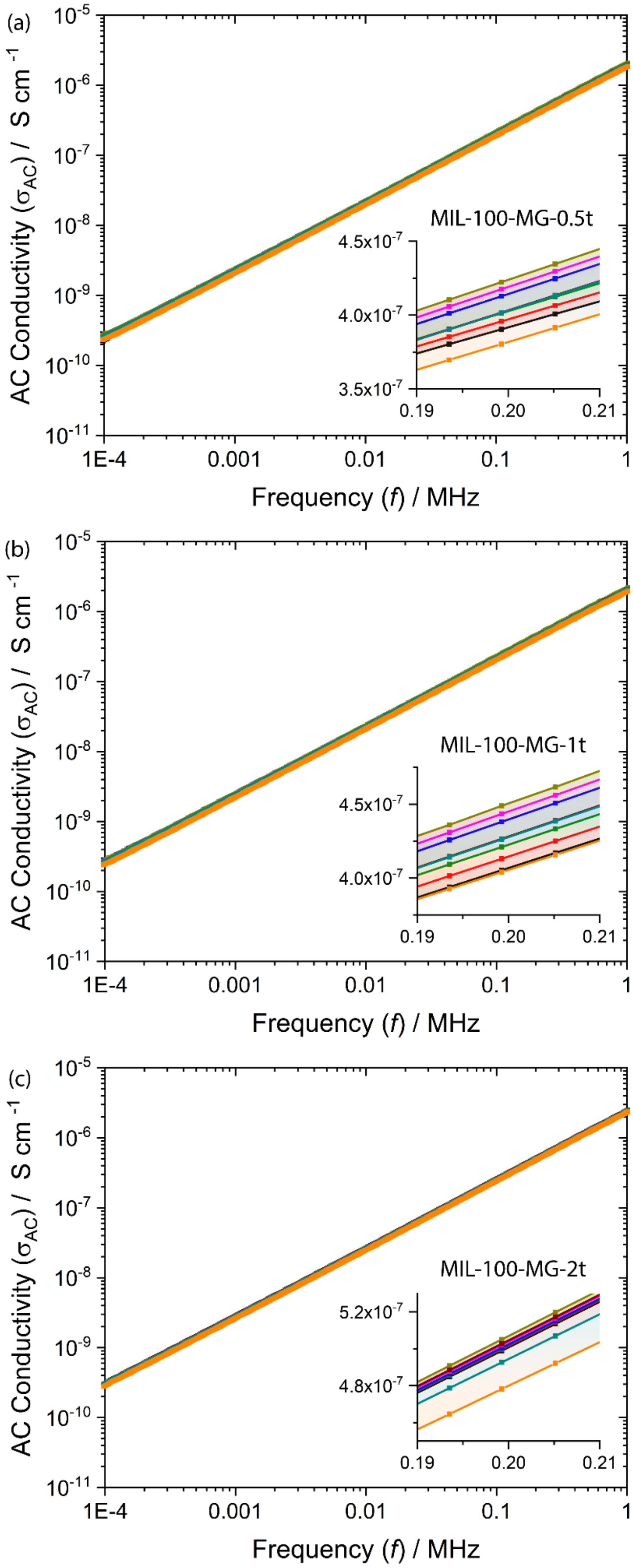
(a)
AC Conductivity (σAC) / S cm-1
Frequency (f) / MHz
MIL-100-MG-0.5t
(b)
AC Conductivity (σAC) / S cm-1
Frequency (f) / MHz
MIL-100-MG-1t
(c)
AC Conductivity (σAC) / S cm-1
Frequency (f) / MHz
MIL-100-MG-2t

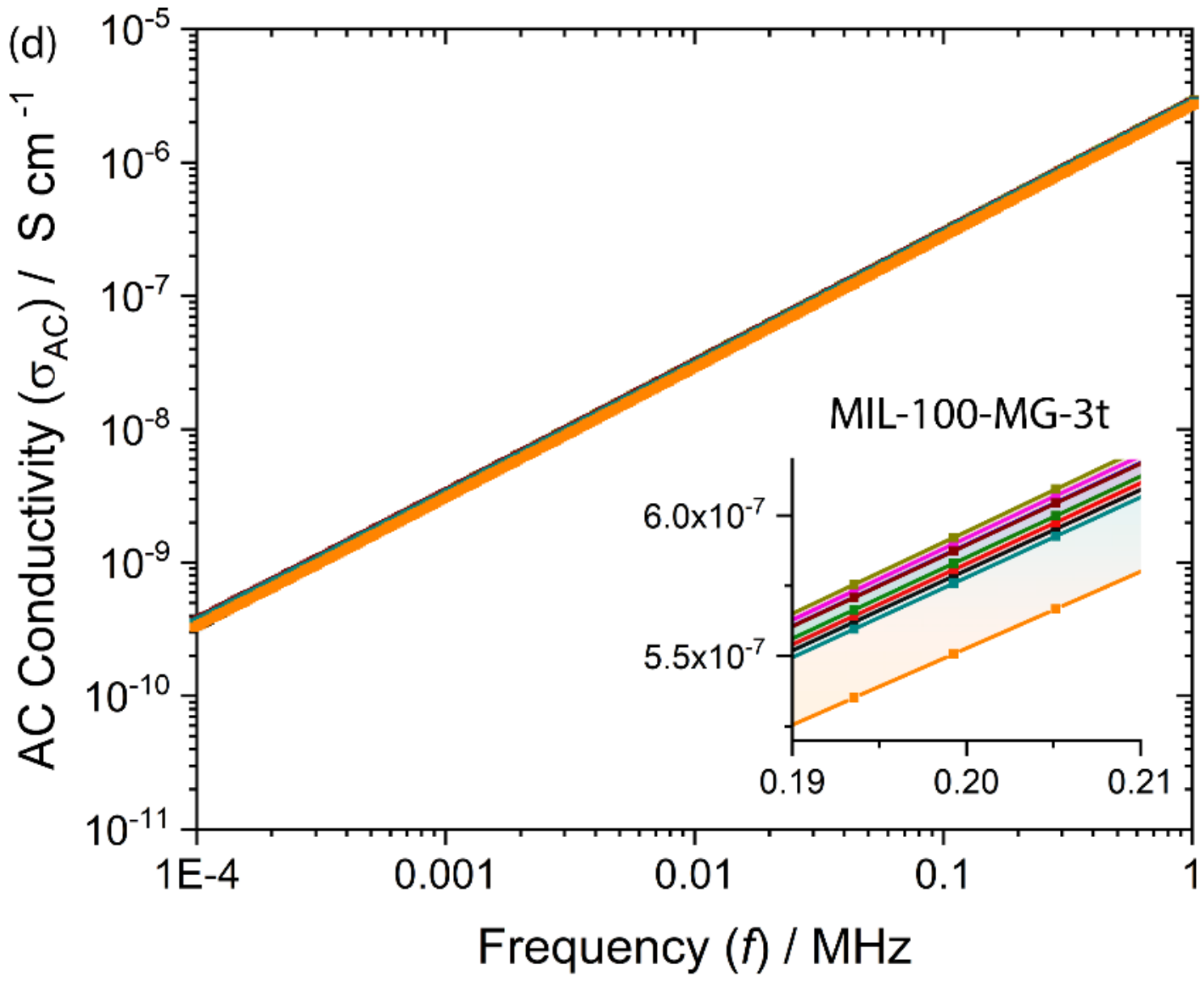
(d)
AC Conductivity ($\sigma_{AC}$) / S cm$^{-1}$
$10^{-5}$
$10^{-6}$
$10^{-7}$
$10^{-8}$
$10^{-9}$
$10^{-10}$
$10^{-11}$
1E-4
0.001
0.01
0.1
1
Frequency (*f*) / MHz
MIL-100-MG-3t
6.0x10$^{-7}$
5.5x10$^{-7}$
0.19
0.20
0.21

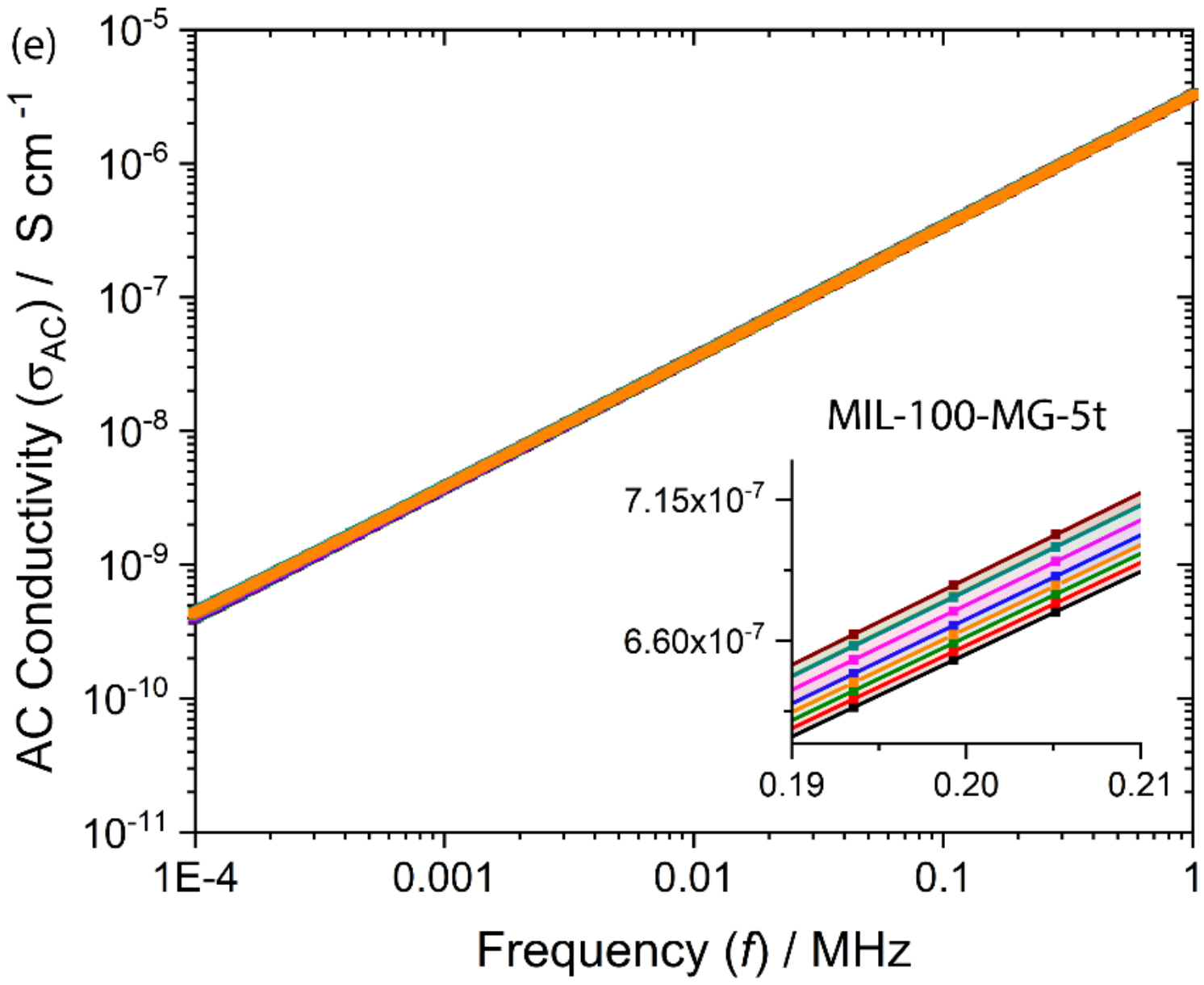
(e)
AC Conductivity ($\sigma_{AC}$) / S cm$^{-1}$
$10^{-5}$
$10^{-6}$
$10^{-7}$
$10^{-8}$
$10^{-9}$
$10^{-10}$
$10^{-11}$
1E-4
0.001
0.01
0.1
1
Frequency (*f*) / MHz
MIL-100-MG-5t
7.15x10$^{-7}$
6.60x10$^{-7}$
0.19
0.20
0.21

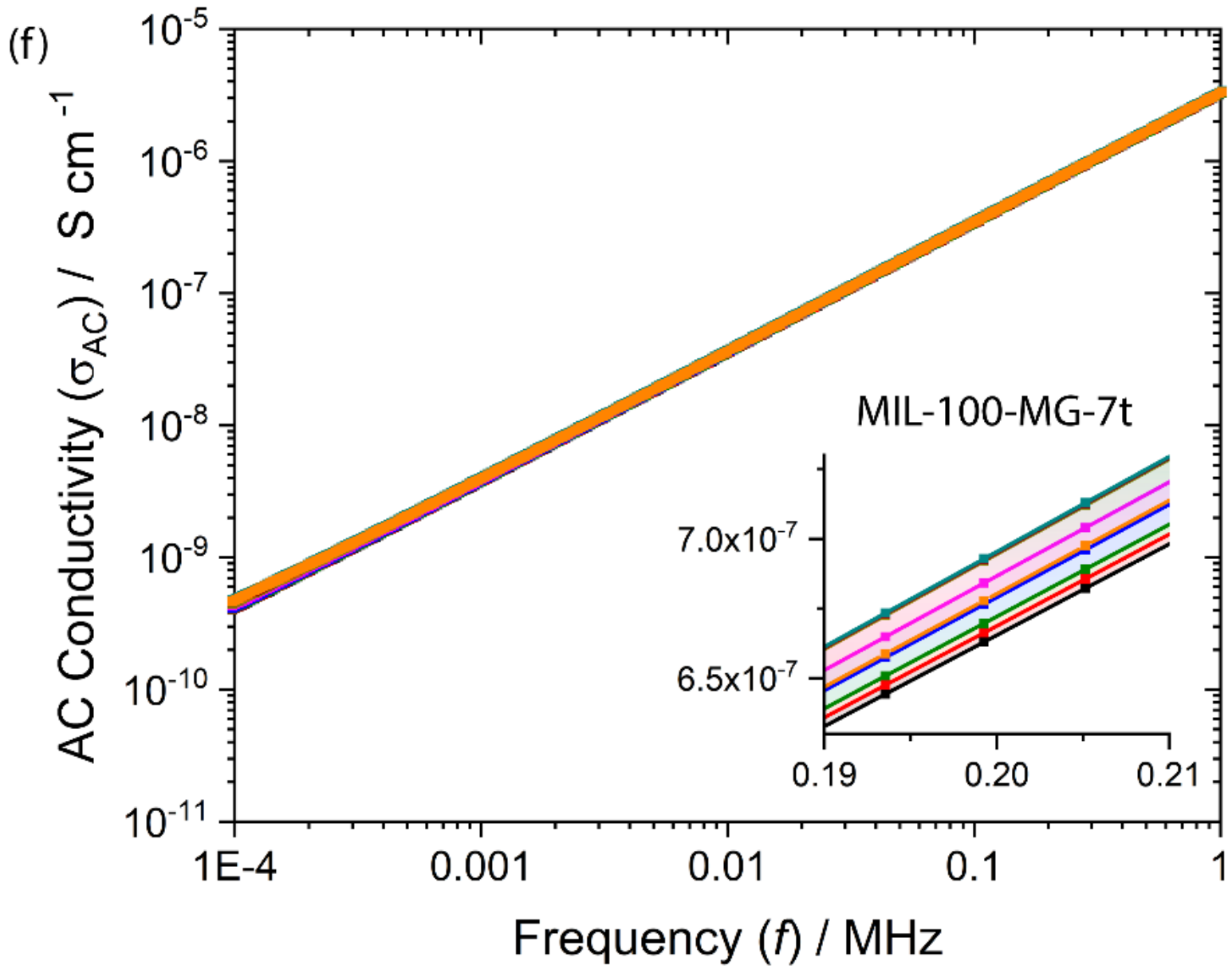
(f)
AC Conductivity ($\sigma_{AC}$) / S cm$^{-1}$
$10^{-5}$
$10^{-6}$
$10^{-7}$
$10^{-8}$
$10^{-9}$
$10^{-10}$
$10^{-11}$
1E-4
0.001
0.01
0.1
1
Frequency (*f*) / MHz
MIL-100-MG-7t
7.0x10$^{-7}$
6.5x10$^{-7}$
0.19
0.20
0.21

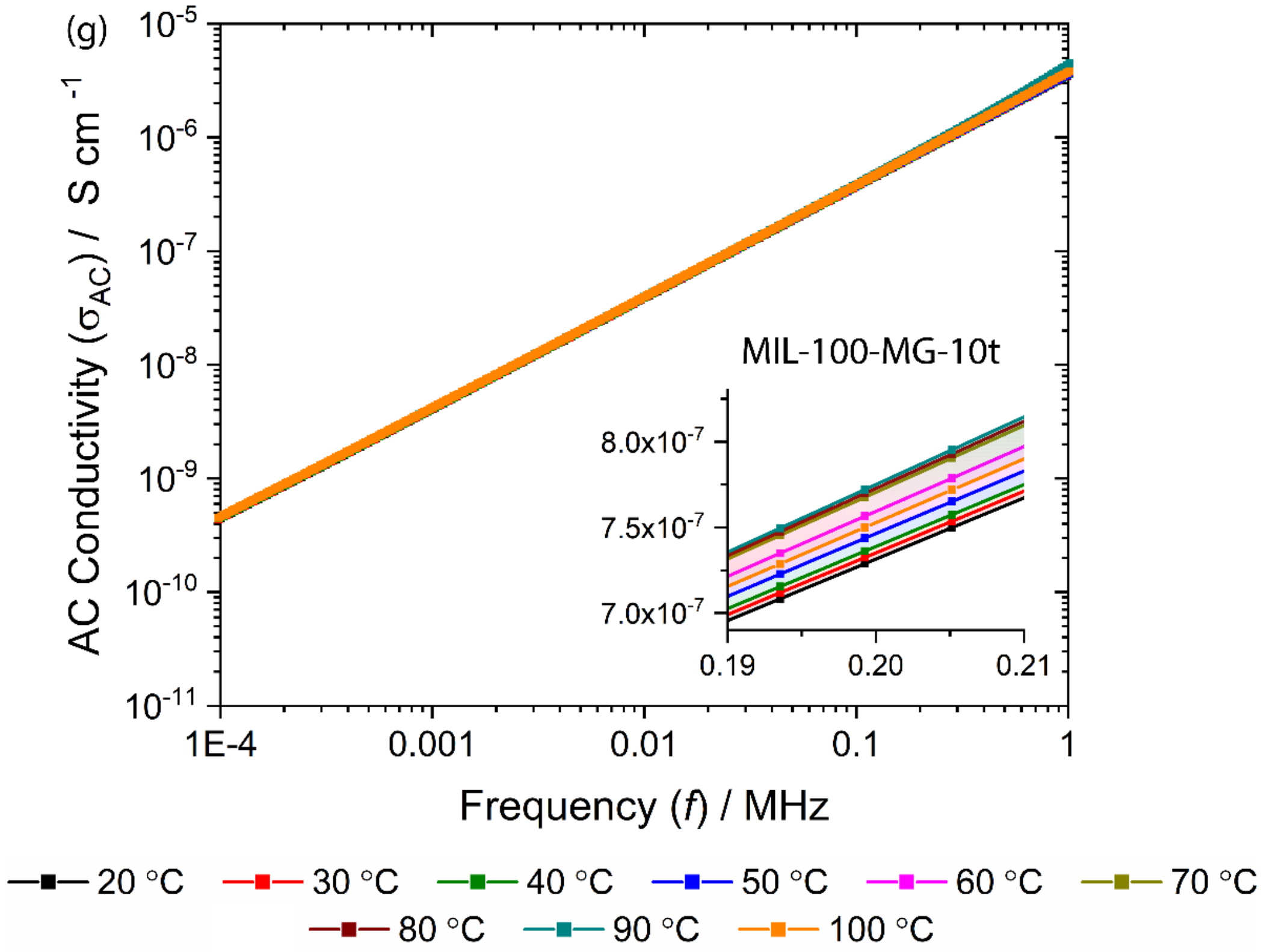

**Figure S15:** The AC conductivity of MIL-100-MG pellets prepared under a compression load of: (a) 0.5-ton, (b) 1-ton, (c) 2-ton, (d) 3-ton (e) 5-ton (f) 7-ton, and (g) 10-ton, corresponding to the pressure of 36.96, 73.92, 147.84, 221.76, 369.6, 517.44 and 739.20 MPa, respectively.